\documentclass[3p,times,twocolumn,sort&compress]{elsarticle}

\usepackage{ecrc-mb}
\usepackage[british]{babel}
\usepackage{color}

\usepackage{widetext}

\volume{00}

\firstpage{1}

\journalname{Nuclear Physics B Proceedings Supplement}

\runauth{M. Beneke and M. Steinhauser}
\runtitle{Non-relativistic high-energy physics: top production and 
dark matter annihilation}
\runprep{TUM-HEP 1003/15, TTP14-040, SFB/CPP-14-106, 
15 January 2015}

\jid{nuphbp}

\jnltitlelogo{Nuclear Physics B Proceedings Supplement}

\usepackage{amssymb}
\usepackage[figuresright]{rotating}

\newcommand{\gsim}{\;\rlap{\lower 3.5 pt \hbox{$\mathchar \sim$}} \raise 1pt
 \hbox {$>$}\;}
\newcommand{\lsim}{\;\rlap{\lower 3.5 pt \hbox{$\mathchar \sim$}} \raise 1pt
 \hbox {$<$}\;}

\def\bm#1{\mbox{\boldmath$#1$\unboldmath}}
\newcommand{\bffmath}[1]{\mbox{\boldmath ${#1}$}}
\newcommand{\bff}[1]{{\bf {#1}}}
\newcommand{\bfm}[1]{\mbox{\boldmath$#1$}}

\begin{document}

\begin{frontmatter}



\dochead{}

% Prperint numbers:
% SFB/CPP-14-106
% TTP14-040
% TUM ...

\title{Non-relativistic high-energy physics: top production and 
dark matter annihilation}


\author{Martin Beneke}

\address{Physik Department T31, 
James-Franck-Stra\ss e~1, 
Technische Universit\"at M\"unchen,
85748 Garching, Germany}

\author{Matthias Steinhauser}

\address{Institut f{\"u}r Theoretische Teilchenphysik, Karlsruhe 
  Institute of Technology (KIT), 76128 Karlsruhe, Germany}

\begin{abstract}
\noindent
Non-relativistic physics is often associated with atomic physics and 
low-energy phenomena of the strong interactions between nuclei and 
quarks. In this review we cover three topics in contemporary high-energy 
physics at or close to the TeV scale, where non-relativistic dynamics 
plays an important if not defining role. We first discuss in detail the
third-order corrections to top-quark pair production in
electron-positron collisions in the threshold region, which plays a 
major role at a future high-energy $e^+ e^-$ 
collider. Threshold effects are also 
relevant in the production of heavy particles in hadronic collisions, 
where in addition to the Coulomb force soft gluon radiation contributes 
to enhanced quantum corrections. We review the joint resummation of 
non-relativistic and soft gluon effects for pair production of top 
quarks and supersymmetric particles to next-to-next-to-leading logarithmic 
accuracy. The third topic deals with pair annihilation of dark matter 
particles within the framework of the Minimal Supersymmetric Standard
Model. Here the electroweak Yukawa force generated by the exchange of 
gauge and Higgs bosons can cause large ``Sommerfeld'' enhancements 
of the annihilation cross section in some parameter regions.
\end{abstract}

\begin{keyword}
top quark production \sep perturbative QCD \sep (P)NRQCD \sep ILC \sep
LHC \sep dark matter


\end{keyword}

\end{frontmatter}

% \linenumbers

%- }}}


%- {{{ Introduction:

\section{Introduction}
\label{sec:1}

Non-relativistic particle 
physics is often associated with atomic physics, 
the interactions between nuclei, and the low-energy phenomena of the strong 
interactions of the charm and bottom quarks, which form quarkonium 
bound states. But non-relativistic dynamics also governs the interactions 
of weak-scale particles such as the top quark or the hypothetical 
supersymmetric partners of the Standard Model (SM) particles, when 
they are slowly moving. In this review we cover three topics in 
contemporary high-energy physics where the unique non-relativistic 
dynamics caused by instantaneous, potential interactions 
plays an important if not defining role: the production of top-quark 
pairs in electron-positron collisions in the threshold 
region, which can provide a measurement of the top-quark mass 
with unchallenged precision and could be realized at a future 
high-energy $e^+ e^-$ collider. The hadronic pair production of top 
quarks at Tevatron and LHC, or yet unobserved heavy particles at 
the LHC, where in addition to the Coulomb force soft gluon radiation contributes 
to enhanced quantum corrections, which should be summed. And finally, 
the pair annihilation of dark matter particles within the framework 
of the Minimal Supersymmetric Standard Model (MSSM). While the first 
two processes are determined by the long-range colour Coulomb force, 
the low-velocity annihilation 
of heavy neutralinos in the MSSM can be dramatically enhanced by 
the short-range electroweak Yukawa force generated by the exchange of 
gauge and Higgs bosons.

We discuss the phenomena and results but also emphasize theoretical 
methods and techniques. Since there is always a small relative 
velocity $v$ in the problem, non-relativistic systems involve (at 
least) the three scales $m$ (mass), $m v$ (momentum), $m v^2$ 
(energy), and it is appropriate to formulate effective Lagrangians 
to organize the calculation, even if the underlying theory 
--- QCD, the SM, or the MSSM --- is known. The three topics discussed  
here have further in common that the relevant Lagrangian couplings 
are weak at all three scales. This allows for a systematic treatment 
with approximations of well-defined parametric accuracy, even though 
ordinary perturbation theory in the coupling breaks down, as will 
be discussed. 

Each of the three topics provides its own specific challenges. In 
top-quark pair production near threshold in $e^+ e^-$ collisions, it is 
the high precision of the envisioned mass measurement, which demands 
calculations with unprecedented third-order accuracy in the 
non-relativistic approximation that can be achieved by a combination 
of effective field theory and multi-loop techniques. In hadronic 
production the required precision is less, but the coloured partonic 
initial state of quarks and gluons implies that soft-gluon resummation 
must be integrated into the non-relativistic expansion, and a more 
general treatment of colour is needed for final states which can be 
produced in various colour representations. Dark matter (DM) annihilation 
involves yet another issue. The non-relativistic ``Sommerfeld'' 
enhancements from electroweak gauge boson exchange are operative 
only, when the DM particles are in the TeV range, such that the 
inverse Bohr radius scale is of order of or larger than the mass of 
the electroweak gauge bosons. For such heavy, weakly interacting 
particles, the mass splitting between the members of the electroweak 
multiplet are small, so that co-annihilation effects at dark matter 
freeze-out are generic. While the non-relativistic treatment is only 
performed at leading-order, the complexity of the problem arises from 
a large number of interacting two-particle states, kinematically 
closed channels in the Schr\"odinger problem, which must be solved 
numerically, and the large number of final states that contribute to 
the pair annihilation cross section in a model such as the MSSM.

The outline of this review is as follows: in the next section we briefly 
summarize several methods and technical tools, which are necessary to 
perform the calculations. In
particular, we briefly describe the threshold expansion, introduce NRQCD and
PNRQCD, and mention a few details in connection to the underlying multi-loop
calculations. Section~\ref{sec:3} is devoted to the top-quark pair production
cross section near threshold in $e^+ e^-$ annihilation. We introduce the 
main ingredients in the third-order calculation and discuss the numerical
results.  In Section~\ref{sec:4} top-quark and supersymmetric pair 
production is considered at hadron colliders. After describing the 
joint soft and Coulomb resummation formalism, the top-pair invariant 
mass distribution near threshold is considered. We then summarize results 
on the resummed total top and SUSY pair production cross sections, and 
compare it to fixed-order results, and results with soft-gluon but without 
Coulomb resummation. Section~\ref{sec:5} leads through the effective field 
theory treatment of Sommerfeld enhancement in dark matter annihilation 
with a short discussion of a method to determine the enhancement reliably 
in situations with several channels without extreme degeneracies. We also 
highlight a few results for a wino-like dark matter particle and a series of 
models that interpolates from a Higgsino- to wino-like neutralino.
We conclude our review with a summary in Section~\ref{sec:sum}.

\section{Methods and techniques}
\label{sec:2}

Before addressing more specifically the three topics of this review, 
we summarize the main methods and techniques.

\subsection{Threshold expansion}

Close to the production threshold of two heavy particles (for simplicity, 
with equal masses) there are three characteristic scales: the mass of 
the particle, $m$, its three-momentum of
order $mv$, and the kinetic energy $\sqrt{s}-2m= mv^2$. For example, for 
top quarks we have $m\approx 170$~GeV, $mv\approx 20$~GeV and 
$mv^2\approx 2$~GeV, and thus a strong hierarchy among the scales. 
Furthermore, $mv^2\gg \Lambda_{\rm QCD}$, which means that the strong 
coupling $\alpha_s(\mu)$ is always in the perturbative regime. In this 
review, we only consider systems where this condition is satisfied.

The presence of the small parameter $v$ leads to a breakdown of the
standard perturbation expansion in $\alpha_s$ due to kinematic $1/v$ 
enhancements. A reorganized, non-relativistic expansion must be used, 
where both $\alpha_s$ and $v$ are simultaneously considered as small 
but $\alpha_s/v$ of order one. In terms of Feynman diagrams, a summation 
of certain terms to all orders in $\alpha_s$ is required. At the same 
time, since $v$ is small, one does not need the full dependence 
on $v$ of every fixed-order diagram. 

For a given Feynman diagram the expansion in $v$ can be 
constructed without first computing the full expression using 
the threshold expansion~\cite{Beneke:1997zp}. The method exploits
that every diagram can be decomposed into a sum of terms, 
for which each loop momentum is in one of the following four 
momentum regions:
\begin{eqnarray}\label{eq:scales}
  \mbox{hard (h)}: & \ell^0 \sim m, & \bm{\ell} \sim m \nonumber \\
  \mbox{soft (s)}: & \ell^0 \sim m v, & \bm{\ell} \sim m v \nonumber \\
  \mbox{potential (p)}: & \,\ell^0 \sim m v^2, & \bm{\ell} \sim m v \nonumber \\
  \mbox{ultrasoft (us)}: & \,\ell^0 \sim m v^2, & \bm{\ell} \sim m v^2\,\,
\end{eqnarray}
In each region the integrand of the Feynman diagram is expanded in the small
parameters of the respective region. Afterwards the loop integration over the
complete $d$-dimensional space-time volume is performed. The property of 
dimensional regularization that scaleless integrals vanish, guarantees 
that no double counting occurs. The arguments in favour of this 
construction provided in~\cite{Beneke:1997zp} rely on assumptions on 
the location of singularities in the Feynman integrand and examples. 
Further justification can be found in \cite{Jantzen:2011nz}.
It may happen that the separation of the integrand into regions generates
spurious ultraviolet and infrared divergences. However, they cancel 
in the sum of all contributions.

\subsection{Non-relativistic effective theory}

The procedure described in the previous subsection is essentially equivalent
to the explicit construction of effective non-relativistic Lagrangians 
within dimensional regularization, since the terms that arise in the 
threshold expansion can be interpreted as modified propagators and 
vertices. In a certain sense, the threshold expansion defines these 
Lagrangians in dimensional regularization 
by providing the precise rules for performing consistent matching 
calculations.

To be definite, we refer to a system of a heavy quark (Q) 
and antiquark interacting 
via gluons. The effective Lagrangians are constructed in two steps 
following the following scheme for integrating out momentum modes:\\[0.0cm]

\begin{displaymath}
  \begin{array}{cc}
    {\cal L}_{\rm QCD}\,[Q(h,s,p),\,g(h,s,p,us)] & \mu>m \\[0.3cm]
    \Big\downarrow & \\[0.3cm]
    {\cal L}_{\rm NRQCD}\,[Q(s,p),\,g(s,p,us)] &  m v < \mu < m \\[0.3cm]
    \Big\downarrow & \\[0.3cm]
    {\cal L}_{\rm PNRQCD}\,[Q(p),\,g(us)]  & \mu < m v\\[0.3cm]
  \end{array}
%  \label{eq::leff}
\end{displaymath}
In square brackets the modes of the heavy quarks ($Q$) and massless
particles ($g$) are displayed which are still contained in the effective
Lagrangian; the others are integrated out
when the energy scale $\mu$ is lowered as indicated on the right.

The first step leads to
NRQCD~\cite{Thacker:1990bm,Lepage:1992tx,Bodwin:1994jh}. Its Lagrangian has
the following structure
\begin{eqnarray}
  \label{nrqcd}
  {\cal L}_{\rm NRQCD} &=& {\cal L}_{\psi} + {\cal L}_{\chi}
  + {\cal L}_{\psi\chi} + {\cal L}_{g} + {\cal L}_{\rm light}
  \,,
\\[-0.2cm]
\nonumber
\end{eqnarray}
where explicit expressions of the individual contributions can be 
found in~\cite{Beneke:2013jia}.
The bilinear heavy-quark Lagrangians ${\cal L}_{\psi}$ and ${\cal L}_{\chi}$
contain the kinetic terms for the two-component quark and antiquark fields
$\psi$ and $\chi$ and the interactions with the chromomagnetic field
in an expansion up to including terms of order $1/m^3$.
${\cal L}_{\psi\chi}$ contains four-fermion quark-antiquark terms
and ${\cal L}_{g}$ is the pure gluon part, again expanded in $1/m$.
Finally, ${\cal L}_{\rm light}$ is the same as the light-quark Lagrangian in
full QCD. The Feynman rules derived from the Lagrangian~(\ref{nrqcd})
can, e.g., be found in~\cite{Beneke:2013jia}. All matching coefficients 
relevant for a third-order calculation can also be
found in~\cite{Beneke:2013jia}. A subtlety that is 
explained there in detail is that the short-distance coefficient functions 
of the operators in the NRQCD Lagrangian must be kept $d$-dimensional 
in dimensional regularization.

In the second matching step in the above scheme soft modes and potential 
massless modes are integrated out. It has been suggested in the context of the
threshold expansion in~\cite{Beneke:1997zp} and at the level of an
effective Lagrangian in~\cite{Pineda:1997bj}. The result is the so-called 
potential NRQCD (PNRQCD) Lagrangian developed in various forms 
in~\cite{Pineda:1997bj,Pineda:1997ie,Beneke:1998jj,Brambilla:1999xf,Beneke:1999qg}. It only contains ultrasoft light fields and potential heavy quarks, 
which simplifies the scaling in the velocity expansion. Note that since 
the components of the hard momentum integrated out in the first matching 
step are larger than those of the modes in the effective theory, 
the NRQCD Lagrangian is local. This can no longer be expected for 
PNRQCD. However, since only the three-momentum of the potential modes 
$Q(p)$ is of order of the soft scale $m v$, the non-locality of the 
PNRQCD Lagrangian refers only to space but not to time. The relevant 
terms in the PNRQCD Lagrangian are given by~\cite{Beneke:2013jia}
\begin{widetext}
  \begin{eqnarray}
    \lefteqn{ {\cal L}_{\rm PNRQCD} = \psi^\dag
      \Big(i\partial_0+g_s A_0(t,\bff{0})+\frac{\bffmath{\partial}^2}{2m}+ 
      \frac{{\bffmath\partial^4}}{8m^3}\,\Big)\psi +\chi^\dag
      \Big(i\partial_0+g_s A_0(t,\bff{0})-\frac{{\bffmath
          \partial^2}}{2m}-\frac{ {\bffmath\partial^4}}{8m^3}\Big) \, \chi
    }
    \nonumber \\ &&\mbox{}
    + \int d^{d-1} {\bf r} \, \Big[ \psi^\dag_a
    \psi^{}_b \Big](x+{\bf r}) \, V_{ab;cd} (r,{\bffmath \partial})\,
    \Big[\chi^\dag_c \chi^{}_d\Big](x)
    -g_s\psi^\dagger(x)\bff{x}\cdot\bff{E}(t,\bff{0})\psi(x)-
    g_s\chi^\dagger(x)\bff{x}\cdot\bff{E}(t,\bff{0})\chi(x)\,,
    \label{eq:pnrqcd}
  \end{eqnarray}
\end{widetext}
\noindent
where 
\begin{equation}
  V_{ab;cd} (r,{\bffmath \partial})=T^A_{ab} T^A_{cd}V_0 (r)+
  \delta V_{ab;cd} (r,{\bffmath \partial})
  \,.
\label{eq:potV}
\end{equation}
$V_0=-\alpha_s/r$ is the tree-level colour Coulomb potential.  The
first two terms of ${\cal L}_{\rm PNRQCD}$ contain the kinetic terms
including the relativistic corrections and the third term is
responsible for heavy-quark potential interactions.  Note that the
heavy-quark potentials generated in the matching to PNRQCD enter this
term as short-distance coefficients of the spatially non-local 
but temporally local, i.e. instantaneous, PNRQCD interactions.
Ultrasoft interactions between the heavy quarks and the gluon field
are contained in the last two terms of~(\ref{eq:pnrqcd}) and the 
$A_0$ terms in the first line.  The latter do not contribute to colour-singlet production of a $Q\bar {Q}$ pair, in which case ultrasoft effects 
enter for the first time at third order in non-relativistic perturbation
theory.

Perturbation theory in PNRQCD closely resembles quantum-mechanical
perturbation theory, since the leading colour-Coulomb interaction is
part of the unperturbed theory. Thus, the propagator of PNRQCD
includes the leading Coulomb interaction exactly, which effects the
required resummation of conventional perturbation theory to all
orders. The PNRQCD Feynman rules can again be found 
in~\cite{Beneke:2013jia}.

\subsection{\label{sub::multiloop} Multi-loop calculations}

The computation of matching coefficients with high precision involves 
fixed-order multi-loop calculations. For the topics reviewed in this 
article, the highest order is demanded in the calculation of the
three-loop corrections to the static potential and the matching coefficient
between QCD and NRQCD of the vector current.
For calculations of this type it is necessary to have a high level
of automation, which reaches from the generation of the Feynman diagrams
to the final summation and expansion in $\epsilon$. 
The scheme used for the three-loop vector-current 
matching coefficient requires particularly little
manual interaction and is described in detail in~\cite{Marquard:2009bj}.

For the generation of the contributing Feynman diagrams we use the {\tt
  Fortran} program {\tt QGRAF}~\cite{Nogueira:1991ex}. The output is passed
via {\tt q2e}~\cite{Harlander:1997zb,Seidensticker:1999bb}, which transforms
Feynman diagrams into Feynman amplitudes, to {\tt
  exp}~\cite{Harlander:1997zb,Seidensticker:1999bb} that generates {\tt
  FORM}~\cite{Vermaseren:2000nd,Kuipers:2012rf} code. At the same time a topology is assigned to each
diagram.  The {\tt FORM} code is processed to perform
traces, apply projectors, and map the resulting scalar expressions to
functions, which have a one-to-one relation to the topologies.  They contain
the exponents of the involved propagators as arguments. At this point one has in
general a large number of different functions. Thus, in a next step one
passes them to a program which implements the Laporta
algorithm~\cite{Laporta:2001dd} and performs a reduction to a small number of
so-called master integrals.  The latter have to be computed using analytical or
numerical methods.  In the case of the three-loop corrections to the static
potential all but three coefficients in the $\epsilon$ 
expansion could be computed analytically.
For the three-loop corrections to the matching coefficient a significant
number of integrals have been evaluated numerically using the program {\tt
  FIESTA}~\cite{Smirnov:2008py,Smirnov:2009pb,Smirnov:2013eza}, which is based
on sector decomposition.

\section{\boldmath Top-quark pair production near threshold 
in $e^+ e^-$ annihilation}
\label{sec:3}

Top-quark pair production near threshold in $e^+ e^-$ annihilation 
provides a unique possibility to
measure a number of parameters with high precision. Among them are the 
top-quark mass and width, the strong coupling constant, and the 
top-quark Yukawa coupling. In particular the mass determination 
attracts lot of attention, since the
current uncertainty can be improved by about an order of magnitude, and  at
the same time the renormalization scheme is precisely defined.  To reach this
goal precise predictions are needed including both QCD and electroweak higher
order corrections.  The top-antitop system near threshold involves 
only scales in the perturbative regime~\cite{Fadin:1987wz}, which allows 
for systematic approximations. 
Next-to-next-to-leading order (NNLO) QCD corrections have
been computed at the end of the 1990s by several independent
groups~\cite{Yakovlev:1998ke,Melnikov:1998pr,Hoang:1998xf,Hoang:1999zc,Nagano:1999nw,Penin:1998ik,Beneke:1999qg}. The results are summarized and 
compared in~\cite{Hoang:2000yr}.  The inclusion of the NNLO corrections 
led to a significant shift of the cross section, and the theoretical 
uncertainty estimated through scale dependence remained larger than 
naively expected. The prospect of an accurate mass measurement at a 
future high-energy $e^+ e^-$ collider therefore makes it mandatory to 
push the theoretical calculation to the third order.

In this section we sketch the computation of the cross section
$\sigma(e^+e^-\to t\bar{t}+X)$ close to threshold to third order in 
non-relativistic 
perturbation theory. Normalized to the cross section for the 
production of $\mu^+\mu^-$ pairs in the high-energy limit, 
$\sigma_0 = 4\pi\alpha^2/(3s)$, it
is given by
\begin{eqnarray}
  R &=& \frac{\sigma(e^+ e^- \to t\bar{t} +X)}{\sigma_0}
  \nonumber\\
  &=& 12\pi e_t^2\,\,\mbox{Im}\left[ \Pi^{(v)}(q^2 = s+i\epsilon)\right]
  \,,
  \label{eq::R}
\end{eqnarray}
where $e_t=2/3$ is the top-quark electric charge, and 
the vector polarization function near threshold has the form
\begin{eqnarray}
  \label{eq:pitoNRQCD}
  \Pi^{(v)}(q^2)
  &=& \frac{N_{c}}{2m_t^{2}}\,c_v 
  \left[c_v-\frac{E}{m_t}\,\left(c_v+\frac{d_v}{3}\right)\right]
  \nonumber\\&&\mbox{}\times
  G(E) + \ldots\,.
\end{eqnarray}
For simplicity of notation, we restrict this expression to the 
terms from a virtual photon, neglecting the $Z$-boson contribution. The 
full expressions are given in~\cite{Beneke:2013jia}.
$c_v$ and $d_v$ are NRQCD matching coefficients with perturbative 
expansions in $\alpha_s$, which are defined through the
expansion of the vector current $j^{(v)\,\mu}$ in terms of non-relativistic
fields given by
\begin{eqnarray}
  \label{eq:QCDVectorCurrent}
  j^{(v) \,i}=c_v\, \psi^\dag\sigma^i\chi
  + \frac{d_v}{6m_t^2}\,\psi^\dag\sigma^i\,{\bf D^2}\chi
  +\ldots.
\end{eqnarray}
The terms proportional to $E=\sqrt{s}-2m_t$ in 
(\ref{eq:pitoNRQCD}) arise from expanding the prefactor $1/q^2=1/s$ and 
from the $1/m_t^2$ suppressed current in~(\ref{eq:QCDVectorCurrent}), 
whose matrix element can be reduced to the one of the leading 
current by equation-of-motion relations. Thus, the central quantity 
in the non-relativistic representation is the two-point function
\begin{eqnarray}
  \lefteqn{G(E) = \frac{i}{2 N_c (d-1)} 
    \int d^{d} x\, e^{iEx^0}\,} 
  \nonumber \\&&\mbox{}  
  \langle 0| \,T(\,
  [\chi^{\dag}\sigma^i\psi](x)\,
  [\psi^{\dag}\sigma^i\chi](0))
  |0\rangle_{| \rm NRQCD}
  \label{eq:G}
\end{eqnarray}
of non-relativistic currents, which has to be computed within NRQCD. 

In the following subsections we discuss various ingredients which are needed
for the third-order prediction to $\sigma(e^+e^-\to
t\bar{t}+X)$. In particular, we describe in Section~\ref{sub::c3} the
calculation of the three-loop corrections to the matching coefficient $c_v$ and
in Section~\ref{sub::a3} the three-loop corrections to the static
potential.  The latter is an important ingredient for the non-relativistic
correlator which is discussed in Section~\ref{sub::Green}.
Ultrasoft corrections appear for the first time at third order
and are discussed in Section~\ref{sub::US} before presenting results
for S-wave QCD contribution to the 
third-order cross section in Section~\ref{sub::R}.
Section~\ref{sub::further}
contains further results, in particular a discussion of Higgs boson
contributions at order $\alpha\alpha_s$, of P-wave effects
and of non-resonant contributions.

%- {{{ cv:

\subsection{\label{sub::c3}\boldmath Matching of the vector current, $c_v$}

One ingredient of the N$^3$LO corrections to top-quark threshold
production is the three-loop matching coefficient between QCD and NRQCD.
It is defined via the equation~\cite{Beneke:1997jm}
\begin{equation}
  \label{eq:newmatch}
  Z_{2,{\rm QCD}}\,\Gamma_{\rm QCD} =
  c_v\,Z_{2,{\rm NRQCD}}\,Z_J^{-1}\,\Gamma_{\rm NRQCD} \,,
\end{equation}
where $Z_{2,{\rm QCD}}$ and $Z_{2,{\rm NRQCD}}$ are the on-shell wave function
renormalization constants in QCD and NRQCD, respectively.
$\Gamma_{\rm QCD}$ and $\Gamma_{\rm NRQCD}$ represent the amputated, bare 
electromagnetic current vertex functions evaluated for on-shell 
heavy quarks directly at threshold, that is, with zero relative momentum.
It is understood that they are
expressed in terms of the renormalized QCD coupling and the pole mass. 
One furthermore chooses for the photon momentum $q^2 = 4 m^2$.
In this kinematic configuration $\Gamma_{\rm QCD}$ corresponds to
the hard part of the vertex function in full QCD. Moreover, we have 
$Z_{2,{\rm NRQCD}}=1$ and only tree-level contributions to $\Gamma_{\rm
  NRQCD}$, since loop corrections reduce to scaleless integrals in dimensional
regularization. The renormalization constant $Z_J$ takes care of the
infrared divergences of the renormalized on-shell QCD vertex leading to a
finite result for $c_v$.

One- and two-loop QCD corrections to $c_v$ have been known since more than 
fifteen years and have been computed in~\cite{Kallen:1955fb}
and~\cite{Czarnecki:1997vz,Beneke:1997jm,Kniehl:2006qw}, respectively.  The
so-called renormalon contribution, consisting of the one-loop diagram with
arbitrarily many massless quark loop insertions in the gluon propagator, has
been computed in~\cite{Braaten:1998au}.  More recently, the fermionic
three-loop term became available~\cite{Marquard:2006qi,Marquard:2009bj} and
the gluonic corrections have been computed in~\cite{Marquard:2014pea}.
One-loop electroweak corrections in the SM are available
from~\cite{Guth:1991ab} and corrections of ${\cal O}(\alpha\alpha_s)$ have
been considered in~\cite{Eiras:2006xm,Kiyo:2008mh}.  Supersymmetric one-loop
corrections to $c_v$ have been computed in~\cite{Kiyo:2009ih}.

%%%%%%%%%%%%%%%%%%%%%%%%%%%%%%%%%%%%%%%%%%%%%%%%%%%%%%%%%%%%%%%%%%%%
\begin{figure}[t]
\vspace*{0.2cm}
\hskip1cm
  \includegraphics[width=5cm]{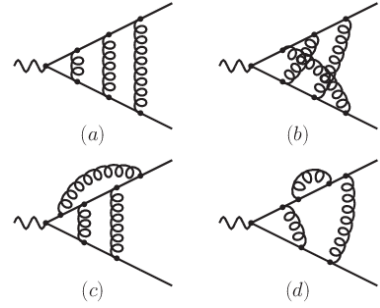}
\caption{\label{fig::sample_c3} Feynman diagrams contributing to $c_3$.
  Straight and curly lines denote heavy quarks with mass $m$ and gluons,
  respectively.}
\end{figure}
%%%%%%%%%%%%%%%%%%%%%%%%%%%%%%%%%%%%%%%%%%%%%%%%%%%%%%%%%%%%%%%%%%%%

Sample Feynman diagrams contributing to $c_v$ at the three-loop order 
are shown in
Figure~\ref{fig::sample_c3}.  After generating all relevant Feynman amplitudes
and mapping them to scalar functions we use the program 
{\tt crusher}~\cite{crusher} for
the reduction to about 100 master integrals. Some of them are known 
analytically~\cite{Marquard:2006qi,diss_piclum} but the major part is
evaluated numerically using the program {\tt
  FIESTA}~\cite{Smirnov:2008py,Smirnov:2009pb,Smirnov:2013eza}, which provides
for each coefficient in the $\epsilon$ expansion 
an uncertainty originating from the
Monte-Carlo integration. When summing the individual contributions these
uncertainties are added in quadrature.  Such an approach requires strong
checks which are described in detail
in~\cite{Marquard:2009bj,Marquard:2014pea}.  A very powerful check is provided
by the change of the master integral basis. We employ the integral tables
generated during the reduction procedure in order to re-express the master
integrals, which are not known analytically, through different, in general
more complicated ones. This transformation is done analytically for general
space-time dimension $d$. In a next step the new master integrals are again
evaluated with {\tt FIESTA} and inserted into the new expression for $c_v$. For
the final result we find excellent agreement within the uncertainties which is
a strong check on the correctness of the result.

In the following we present the result
specifying $N_c=3$ and $\mu=m$
\begin{eqnarray}
  \lefteqn{c_v \approx 1 - 2.667 \frac{\alpha_s^{(n_l)}}{\pi}
  + \left(\frac{\alpha_s^{(n_l)}}{\pi}\right)^2\left[
    - 44.551 + 0.407 n_l \right]}
  \nonumber\\&&\mbox{} 
  + \left(\frac{\alpha_s^{(n_l)}}{\pi}\right)^3\Big[
    -2091(2)  +120.66(0.01)\, n_l 
  \nonumber\\&&\mbox{} 
    -0.823\, n_l^2
    \Big]
  + \mbox{singlet terms}
  \,.
  \label{eq::cv3num}
\end{eqnarray} 
Results for individual colour coefficients and general values of $\mu$ can be
found in~\cite{Marquard:2006qi,Marquard:2009bj,Marquard:2014pea}.

%- }}}
%- {{{ a3:

\subsection{\boldmath Coulomb potential, $a_3$\label{sub::a3}}

In this subsection we summarize the potentials needed for top-quark pair
production where our special emphasis lies on the three-loop corrections to
the static potential.  In momentum space the potential can be written in terms
of operators expanded in $1/m$. Its colour-singlet projection reads
\begin{eqnarray}
  \lefteqn{V({\bf p},{\bf p}^{\prime}) =
  -{\cal V}_C(\alpha_s)\frac{4\pi C_{F}\alpha_s}{{\bf
      q}^2}  }
  \nonumber\\&&\mbox{}
  + {\cal V}_{1/m}(\alpha_s) \frac{\pi^2 (4\pi)C_F\alpha_s} {m|{\bf q}|}  
  + {\cal O}(1/m^2)
  \,,
  \label{eq:pot0}
\end{eqnarray}
where for brevity only the first two terms are shown; 
the $1/m^2$ operators can be found in~\cite{Beneke:2013jia}.
The coefficients of the operators are expanded in $\alpha_s$ where to N$^3$LO
the potential coefficient 
${\cal V}_C(\alpha_s)$ is needed to three-, ${\cal V}_{1/m}(\alpha_s)$
to two-loop and the coefficient of the $1/m^2$ term to one-loop accuracy. Note
that there is no tree-level contribution of order $1/m^3$.

The $1/m^2$ coefficients are known since long (see~\cite{Beneke:2013jia}
for a detailed discussion). Also ${\cal V}_{1/m}$ has been computed more then
ten years ago~\cite{Kniehl:2001ju}, however, only up to the constant term in
the $\epsilon$ expansion.  Since the potential loop-momentum integrals 
which have to be
solved to obtain the corrections to the Green function $G(E)$ are divergent, 
the ${\cal  O}(\epsilon)$ term of the two-loop correction to ${\cal V}_{1/m}$
is also required. It can be found in the Appendix of~\cite{Beneke:2014qea}.
The three-loop correction to ${\cal V}_C$, commonly denoted $a_3$, 
has been computed by two independent
groups~\cite{Smirnov:2008pn,Smirnov:2009fh,Anzai:2009tm}. The corresponding
results shall be discussed in more detail in the following.

\begin{figure}[t]
  \centering
  \includegraphics[width=\linewidth]{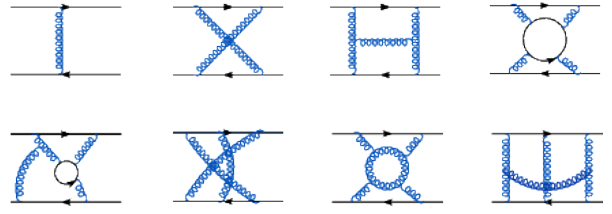}
  \caption{\label{fig::diags_a3}Sample Feynman diagrams contributing to the
    static potential at tree-level, one-, two- and three-loop order.
    Solid and curly lines represent quarks and gluons, respectively.
    In the case of closed loops the quarks are massless; the external 
    quarks are heavy and treated in the static limit.
    }
\end{figure}

\begin{figure}[t]
\vspace*{0.2cm}
  \centering
  \includegraphics[width=0.9\columnwidth]{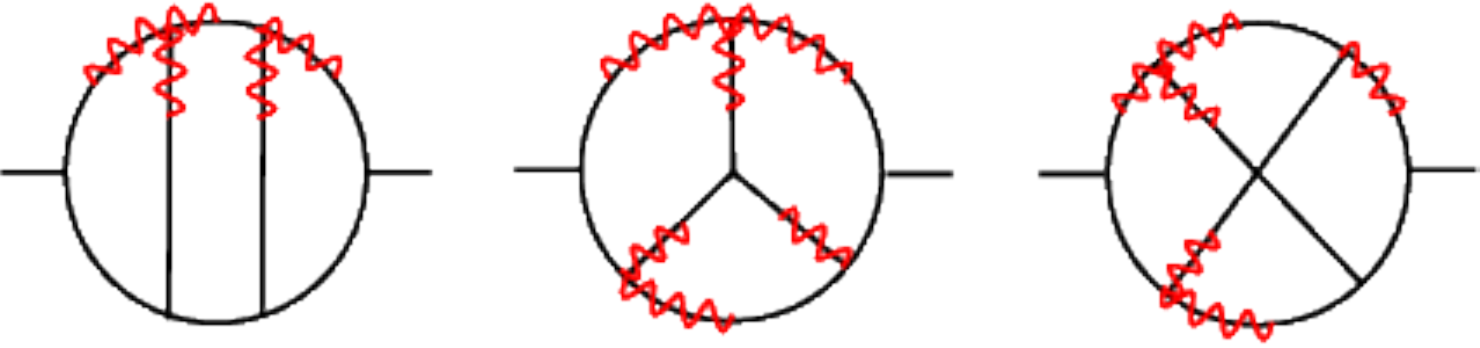}
  \caption{\label{fig::BasicGraphs}
    Families of Feynman integrals needed for the calculation of $a_3$. The
    solid lines correspond to relativistic massless propagators and the zigzag
    lines represents static propagators.}
\end{figure}

In Figure~\ref{fig::diags_a3} sample Feynman diagrams contributing to 
${\cal V}_C$ are shown. The external quarks are considered as heavy and 
thus for them the static approximation is applied. 
The only momentum scale in the problem is given
by the momentum transfer $q$ between the quark and antiquark which
has only non-vanishing space-like
components. Furthermore, the static heavy-quark propagators have the
form $1/(-v\cdot p - s_i i0)$ with $v\cdot p=p_0$ and $s_i=\pm 1$ and thus do
not depend on $q$.  As a consequence, any Feynman integral that contributes to
$a_3$ can be mapped to one of the three graphs shown in
Figure~\ref{fig::BasicGraphs} where solid lines stand for usual massless
propagators of the form $1/(-p^2-i0)^{a_i}$ and zigzag lines stand for static
propagators which might also be raised to an integer index.  In case the latter
type of propagators is absent the integrals reduce to usual massless two-point
functions, which can be treated with the help of {\tt
  MINCER}~\cite{Larin:1991fz}. Note, however, that the presence of the static
lines significantly increases the complexity of the problem.

\begin{figure}[t]
  \centering
\vspace*{0.2cm}
  \includegraphics[width=0.9\columnwidth]{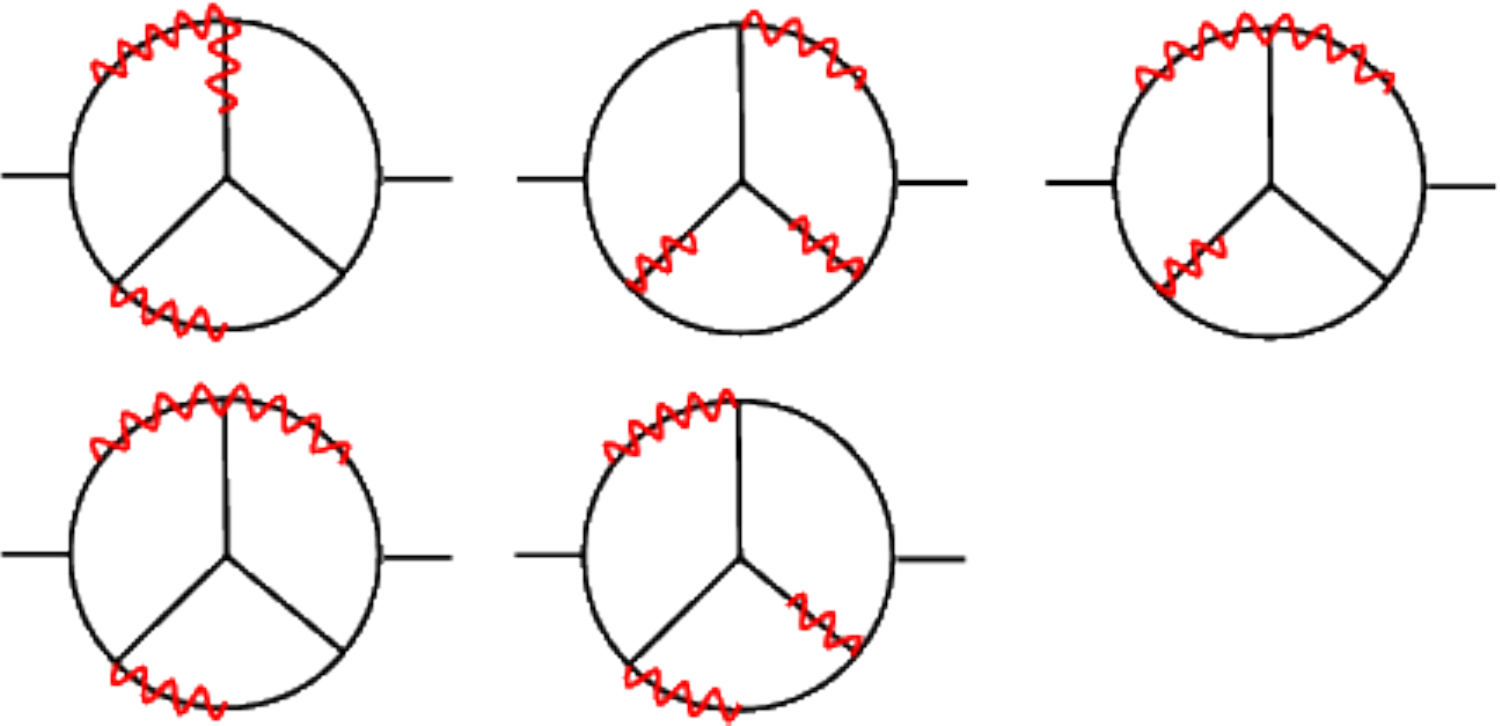}
  \caption{\label{fig::Merc}
    Families with twelve indices (eleven propagators plus one scalar product)
    after applying partial fractioning to the ``Mercedes'' graph of
    Figure~\ref{fig::BasicGraphs}.} 
\end{figure}

In general the integrals involve up to fifteen propagators (including an
irreducible numerator). In order to simplify the reduction problem we apply in
a first step partial fraction identities to arrive at various subfamilies of
integrals with at most three linear propagators.  Thus, any resulting integral
is labelled by twelve indices one of which corresponds to an irreducible
numerator and three indices correspond to linear propagators.
As an example we show in Figure~\ref{fig::Merc} the resulting families for the 
``Mercedes'' graph of Figure~\ref{fig::BasicGraphs}.

The next step is a reduction of all the Feynman integrals to master integrals
by solving integration-by-parts relations~\cite{Chetyrkin:1981qh}.  This is
done with the help of {\tt FIRE} (Feynman Integral
REduction)~\cite{Smirnov:2008iw,Smirnov:2013dia,Smirnov:2014hma}.

In total there are 41 master integrals which contribute to the three-loop
static potential. Ten integrals have no static lines and are thus known since
long~\cite{Chetyrkin:1980pr,Kazakov:1983ns,Bekavac:2005xs}.  Fourteen master
integrals contain a massless one-loop insertion which can easily be integrated
in terms of $\Gamma$ functions using standard formulae. As a result one
obtains two-loop integrals where one of the indices has a non-integer exponent
involving the space-time parameter $\epsilon = (4-d)/2$. Results for these
integrals are shown in~\cite{Smirnov:2010gi}.  The 17 remaining integrals
are more complicated and can only be computed in an expansion in
$\epsilon$. Explicit results can be found in~\cite{Smirnov:2010zc} (see
also~\cite{Anzai:2012xw}). Actually, all but three coefficients in the 
$\epsilon$ expansion could be computed analytically. The three numerical 
results are known to a
sufficiently high precision for all foreseeable phenomenological applications.

An important check for the correctness of the calculation is the use of a
general QCD gauge parameter~\cite{Smirnov:2009fh}.  Note that the
computational price one has to pay is quite high: a rough estimate of the
complexity based on the number of integrals which have to be reduced to
masters shows that the linear $\xi$ term is about seven times and the $\xi^3$
term even 18 times more complicated than the Feynman gauge expression. Let us
mention that nevertheless all occurring integrals could be reduced with the
help of {\tt FIRE}.

We write the perturbative expansion of ${\cal V}_{C}$ in the form
\begin{eqnarray}
  {\cal V}_{C}(\alpha_s)&=&{\cal V}_{C}^{(0)}
  +\frac{\alpha_s}{4\pi}{\cal V}_{C}^{(1)}
  +\bigg(\frac{\alpha_s}{4\pi}\bigg)^{\!2}{\cal V}_{C}^{(2)}
  \nonumber\\&&\mbox{}
  +\bigg(\frac{\alpha_s}{4\pi}\bigg)^{\!3}{\cal V}_{C}^{(3)}
  + O(\alpha_s^4)\,.
  \label{eq:calVexp}
\end{eqnarray}
Up to two-loop order the coefficients ${\cal V}_{C}^{(i)}$ are finite,
however, at three loops an infrared divergence 
occurs~\cite{Appelquist:1977es},
which is related to ultraviolet divergences in the calculation of the
ultrasoft corrections. It is convenient to subtract these
divergences in $V({\bf p},{\bf p}^{\prime})$ and add them back to the
ultrasoft calculation. In the case of the Coulomb potential 
the infrared divergence is cancelled after adding the $\overline{\rm MS}$
counterterm
\begin{eqnarray}
  \delta V_{C,c.t.} &=& \frac{\alpha_s C_F}{6\epsilon}
  C_A^{3}\frac{\alpha_s^{3}}{{\bf q}^2},
\label{eq:counterC}
\end{eqnarray}
to $V({\bf p},{\bf p}^{\prime})$ in~(\ref{eq:pot0}).
One finally arrives at
\begin{eqnarray}
  {\cal V}_{C}^{\,(0)} &=&1,
  \nonumber\\
  {\cal V}_{C}^{\,(1)} &=&
  \bigg[\bigg(\frac{\mu^2}{\bff{q}^2} \bigg)^{\!\epsilon} -1
  \bigg]\, \frac{\beta_0}{\epsilon}\,
  +\bigg(\frac{\mu^2}{\bff{q}^2} \bigg)^{\!\epsilon}\, a_{1}(\epsilon),
  \nonumber\\
  {\cal V}_{C}^{\,(2)} &=& a_2+ 
  \left(2a_1\beta_0+\beta_1\right)\ln\frac{\mu^2}{\bff{q}^2}
  +\beta_0^2\ln^2\frac{\mu^2}{\bff{q}^2},
  \nonumber\\
  {\cal V}_{C}^{\,(3)} &=& a_3 
  + 
  \left(2a_1\beta_1+\beta_2+3a_2\beta_0+8\pi^2C_A^3\right)
  \nonumber\\[-.5em]
  \nonumber\\&&\mbox{}\times
  \ln\frac{\mu^2}{\bff{q}^2}
  +\left(\frac{5}{2}\beta_0\beta_1+3a_1\beta_0^2\right)
  \ln^2\frac{\mu^2}{\bff{q}^2}
  \nonumber\\&&\mbox{}
  +\beta_0^3\ln^3\frac{\mu^2}{\bff{q}^2}
  \,.
  \label{eq:vcoulombN3LO}
\end{eqnarray}
Note the $\ln\mu^2$ term in ${\cal V}_{C}^{\,(3)}$ not associated with the
QCD beta function is the remnant of the infrared divergence.  We refrain from
providing explicit results for $a_1$ and $a_2$ which can be found in the
literature~\cite{Fischler:1977yf,Billoire:1979ih,Peter:1996ig,Schroder:1998vy,Kniehl:2001ju},
even including higher order terms in
$\epsilon$~\cite{Smirnov:2008pn,Beneke:2013jia}.
The result for $a_3$ reads~\cite{Smirnov:2008pn,Smirnov:2009fh,Anzai:2009tm}
\begin{widetext}
\begin{eqnarray}
  a_3 &=&
  -\left(\frac{20}{9}n_f T_F\right)^3
%  \nonumber \\
%  &&
  +\bigg[
  C_A\left( \frac{12541}{243}
    +\frac{64\pi^4}{135}
    +\frac{368}{3}\zeta_3\right)
  +C_F\left( \frac{14002}{81}
    -\frac{416}{3}\zeta_3\right)
  \bigg] (n_f T_F)^2 
  \nonumber \\
  &&\mbox{}
  +\bigg[\,
  \,\left(-709.717\right) C_A^{\,2}
  +\left(-\frac{71281}{162}+264\zeta_3+80\zeta_5\right) C_A C_F
%  \nonumber\\
%  &&\hspace{1cm}
  +
  \left(\frac{286}{9}+\frac{296}{3}\zeta_3-160\zeta_5\right) C_F^{\,2}
  \bigg]\,n_f T_F
  \nonumber \\
  &&\mbox{}
  +\bigg[ 
  -56.83(1)\,
  \biggl(\frac{d_F^{abcd}d_F^{abcd}}{2N_A\,T_F}\biggr)\,
  \bigg] n_f
%  \nonumber \\
%  &&
  +\bigg[\,
  502.24(1)\, C_A^{\,3}
  +\,\left(-136.39(12)\right)\, 
  \biggl(\frac{d_A^{abcd}d_F^{abcd}}{2N_A\,T_F}\,\biggr)
  \bigg]\,,
  \label{eq:a3}
\end{eqnarray}
with
\begin{eqnarray}
\hspace*{2cm}  \frac{d_F^{abcd}d_F^{abcd}}{2N_A\,T_F} =
  \frac{N_c^4-6N_c^2+18}{96N_c^2}\,,\qquad
%  \nonumber\\
  \frac{d_A^{abcd}d_F^{abcd}}{2N_A\,T_F} =
  \frac{N_c\left(N_c^2+6\right)}{48}\,.
\end{eqnarray}
\end{widetext}
\noindent
As mentioned above, some of the ingredients entering $a_3$ are only known
numerically; they enter four of the colour factors in~(\ref{eq:a3}) and
are taken from~\cite{Smirnov:2009fh,Smirnov:2010zc}.  Also parts of
the N$^4$LO Coulomb potential are known~\cite{Brambilla:2006wp}, but not
needed for the third-order cross section calculation.
The two- and three-loop corrections to the static potential of heavy
quarks in the colour-octet state have been computed in~\cite{Kniehl:2004rk}
and~\cite{Anzai:2013tja}, respectively.

\subsection{\label{sub::Green}Third-order potential corrections to the 
PNRQCD correlation function}

Once the matching coefficients of NRQCD and PNRQCD are determined, 
the correlation  function $G(E)$ defined in (\ref{eq:G}) has to be 
calculated to the third order in PNRQCD perturbation theory. 
Since the leading-order Coulomb 
potential $V_0(r)$ in (\ref{eq:potV}) is part of the unperturbed 
Lagrangian, the position-space propagator is the matrix element 
$G_0(\bff{r},\bff{r}^\prime;E)=
\langle \bff{r}| \hat{G}_{0}(E)|\bff{r}'\rangle$ of the Green 
operator $\hat{G}_{0}(E) = [H_0-E-i\epsilon]^{-1}$. The propagator 
propagates a heavy quark-antiquark {\em pair}, and $\bff{r}$ 
refers to the separation of the heavy quark-antiquark fields. Since one 
usually works in the centre-of-mass (cms) frame, the cms motion is 
irrelevant. The unperturbed theory is not free, but it is still 
exactly solvable, since $H_0$ is the Hamiltonian of the Coulomb 
problem. An explicit expression for the Coulomb Green function 
in terms of Laguerre polynomials $L_s^{(2l+1)}$ 
and a partial-wave decomposition is \cite{Voloshin:1979uv,Voloshin:1985bd} 
\begin{widetext}
\begin{equation}
G_0^{(1,8)}(\bfm{r},\bfm{r}^{\,\prime}\!,E) = 
\frac{m y}{2\pi} e^{-y (r+r^\prime)} \sum_{l=0}^\infty \,(2l+1) 
(2 y r)^l (2 y r^\prime)^l
P_l\left(\frac{\bfm{r}\cdot\bfm{r}^{\,\prime}}
{r r^\prime}\right) \sum_{s=0}^\infty \frac{s! L_s^{(2l+1)}(2 yr) 
L_s^{(2l+1)}(2 yr^\prime)}{(s+2 l+1)! (s+l+1-\lambda)},
\label{eq:CoulombGreen}
\end{equation}
where $y=\sqrt{-m (E+i\epsilon)}$ and 
$\lambda=(m \alpha_s)/(2 y)\times 
\{C_F \,\mbox{(singlet)}; C_F-C_A/2\, \mbox{(octet)}\}$ for the 
propagation in a colour singlet and octet state, respectively. 
Further useful representations are summarized in~\cite{Beneke:2013jia}.
The PNRQCD perturbation expansion of $G(E)$ to third order reads 
\begin{equation}
\label{greenexpand}
G(E) = G_0(E) + \delta_1 G(E) + \delta_2 G(E) + \delta_3 G(E) 
+ \ldots
\end{equation}
with $G_0(E) = \langle \bff{0}| \hat G_{0}(E) |\bff{0}\rangle = 
G_0^{(1)}(0,0;E)$, and 
\begin{eqnarray}
  \delta_1 G(E) &=& \langle \bff{0}| \hat G_0(E)i\delta V_1 
  i\hat G_0(E)|\bff{0}\rangle,\\[0.2cm]
  \delta_2 G(E) &=& \langle \bff{0}| \hat G_0(E)i \delta V_1 
  i\hat G_0(E) i\delta V_1 i\hat G_0(E)|\bff{0}\rangle 
  + \langle \bff{0}| \hat G_0(E)i\delta V_2 
  i\hat G_0(E)|\bff{0}\rangle,\\[0.2cm]
  \delta_3 G(E) &=& \langle \bff{0}| \hat G_0(E)i \delta V_1 
  i\hat G_0(E) i \delta V_1 
  i\hat G_0(E) i\delta V_1 i\hat G_0(E)|\bff{0}\rangle 
  + 2 \langle \bff{0}| \hat G_0(E)i \delta V_1 
  i\hat G_0(E) i\delta V_2 i\hat G_0(E)|\bff{0}\rangle 
  \nonumber\\ && \mbox{}
  + \langle \bff{0}| \hat G_0(E)i\delta V_3 
  i\hat G_0(E)|\bff{0}\rangle 
  + \, \delta^{us}G(E)\,.
  \label{masterthirdorder}
\end{eqnarray} 
\end{widetext}
\noindent
The corrections consist of potential contributions from single and 
multiple insertions of perturbation potentials $\delta V_k$ of 
order $\mathcal{O}(v^k)$ relative to the leading-order Coulomb 
potential, which arise from radiative corrections to the Coulomb potential 
and further non-Coulomb potentials such as ${\cal V}_{1/m}$ in 
(\ref{eq:pot0}). Starting from the third order, there is an ultrasoft 
contribution, $\delta^{us}G(E)$,  
which is briefly discussed in Section~\ref{sub::US}.

The first term on the right-hand side of (\ref{masterthirdorder}) 
represents the triple insertion of the one-loop correction $\delta V_1$ 
to the Coulomb potential. The presence of four propagators of the 
form (\ref{eq:CoulombGreen}) turns this into a complicated expression. 
However, all the integrations are ultraviolet-finite, and this allows 
the triple insertion to be converted into sums that can be evaluated 
numerically as done in \cite{Beneke:2005hg}. Instead of computing 
order by order multiple insertions of $\delta V_1$, one can also include 
$\delta V_1$ into the unperturbed Lagrangian and solve for the Green 
function of the associated Schr\"odinger problem numerically, which 
sums the insertions of $\delta V_1$ to all orders. The comparison of the 
two approaches performed in \cite{Beneke:2005hg} shows that good 
convergence of the PNRQCD expansion to the exact result 
is only achieved when the 
renormalization scale $\mu$ is larger than 30 GeV. Somewhat surprisingly 
PNRQCD perturbation theory already breaks down at scales larger than the 
natural scale of the inverse Bohr radius, but works when $\mu$ is 
chosen closer to the hard scale. The explicit verification of this 
scale choice by comparison with the exact result, which is only 
possible for the Coulomb potential, lends support to the scale choice 
that will be made in the evaluation of the full third-order cross 
section below.

The second and third term on the right-hand side of 
(\ref{masterthirdorder}) are associated with the single insertion of 
third-order potentials, such as the three-loop correction to the 
Coulomb potential discussed in the previous subsection, and the double 
insertion of a second-order potential together with the one-loop correction 
to the Coulomb potential. While algebraically simpler than the triple 
insertion, they turn out to be more difficult to evaluate. The reason 
is that potentials more singular than $1/r$ as $r\to 0$ cause ultraviolet 
divergences of the potential integrals, 
which have to be consistently calculated and factorized in 
dimensional regularization, so that the divergences cancel with the 
matching coefficients and the finite term is correctly computed. 
At the same time, a calculation of the all-order diagrams summed by 
PNRQCD perturbation theory in $d$ dimensions is not possible, since 
the propagator (\ref{eq:CoulombGreen}) is known only in four dimensions.

The calculation of these non-Coulomb potential terms is described 
in detail in \cite{paperII}. (The corresponding result 
for the bound-state 
parameters has already been presented in \cite{Beneke:2007gj} and for the 
Green function in \cite{Beneke:2008ec}, though no details were given 
in this work.) Here we discuss as an example the single insertion of 
the $1/r^2$ potential to illustrate the strategy of the calculation. 
In momentum space the $d$-dimensional $1/r^2$ potential ${\cal V}_{1/m}$ 
is of the form 
$w(\epsilon)/({\bf q}^2)^{1/2+a \epsilon}$, where $a$ is an integer. 
A single insertion is an integral 
\begin{widetext}
\begin{equation}
\hspace*{2cm}I[x+a\epsilon]= \int\prod_{i=1}^4\Bigg[\frac{d^{d-1}{\bf
p}_i}{(2\pi)^{d-1}}\Bigg]\,\tilde{G}_0({\bf p}_1,{\bf
p}_2;E)\frac{1}{({\bf q}_{23}^2)^x}\bigg(\frac{\mu^2}{{\bf
q}_{23}^2}\bigg)^{\!a\epsilon}
 \tilde{G}_0({\bf p}_3,{\bf p}_4;E),
\label{singleinsertionI}
\end{equation}
where ${\bf q}_{ij}={\bf p}_i-{\bf p}_j$.
Some of the potential insertions are multiplied by a divergent
coefficient function $w(\epsilon)$. This means that one should calculate
$I[x+a\epsilon]$ to order $\epsilon$. However, these
divergent coefficient functions always appear in conjunction with a 
counterterm with a slightly different momentum dependence, such that 
the potential expanded in $\epsilon$ is finite. We therefore consider 
the expression 
\begin{equation}
\hspace*{2cm}\frac{1}{({\bf q}^2)^x} 
\left[\bigg(\frac{{\mu }^2}{\bf{q}^2} \bigg)^{\!a \epsilon }\,w(\epsilon)-\frac{w^{(1/\epsilon)}}{\epsilon}\right]
\qquad\mbox{with}\qquad
w(\epsilon)=\frac{w^{(1/\epsilon)}}{\epsilon}+w+
w^{(\epsilon)}\epsilon+\mathcal{O}(\epsilon^2),
\end{equation}
and define the corresponding counterterm-including single-insertion 
function as 
\begin{eqnarray}\label{eq:J-definition1}
\hspace*{2cm}J[x+a\epsilon;w(\epsilon)]&=&
\frac{1}{\epsilon}w^{(1/\epsilon)}\bigg(I[x+a\epsilon]-I[x]\bigg)
+\bigg(w+ w^{(\epsilon)}\epsilon\bigg)I[x+a\epsilon].
\end{eqnarray}
\end{widetext}
\noindent
The advantage of this expression is that it avoids the need to 
calculate in $I[x+a\epsilon]$ the $a$-independent $\mathcal{O}(\epsilon)$ 
terms, since they drop out in the difference in brackets in the first 
term. These terms would indeed be diificult to obtain. The second term is 
multiplied by a finite series, hence the $\mathcal{O}(\epsilon)$ term of 
$I[x]$ is never required to obtain the finite part of 
$J$ as $\epsilon\to 0$.

The single insertion of the $1/r^2$ potential generates ultraviolet 
$1/\epsilon$ poles from the integration 
over the potential loop momenta, which are related to the 
singularities in the dimensionally regulated hard current matching 
coefficients, and which have to be properly factorized. 
Power counting shows that 
the insertion has an overall
divergence coming from diagrams with less then two gluon exchanges,
and a vertex subdivergence, when there is no gluon exchange between 
the external vertex and the potential insertion. To accomplish the
correct factorization, we divide the integral into four different parts 
(and show the corresponding diagrams below):
\begin{widetext}
\begin{equation}
\hspace*{2.8cm}I[\frac{1}{2}+a\epsilon] = I_a[\frac{1}{2}+a\epsilon] + 2
I_b[\frac{1}{2}+a\epsilon] + 2 I_c[\frac{1}{2}+a\epsilon] +
I_d[\frac{1}{2}+a\epsilon].
\end{equation}
%%%%%%%%%%%%%%%%%%%%%%%%%%%%%%%%%%%%%%%%%%%%%%%%%%%%%%%%%%%%%%%%%%%%
%\begin{figure}[t]
  \begin{center}
    \includegraphics[width=0.68\linewidth]{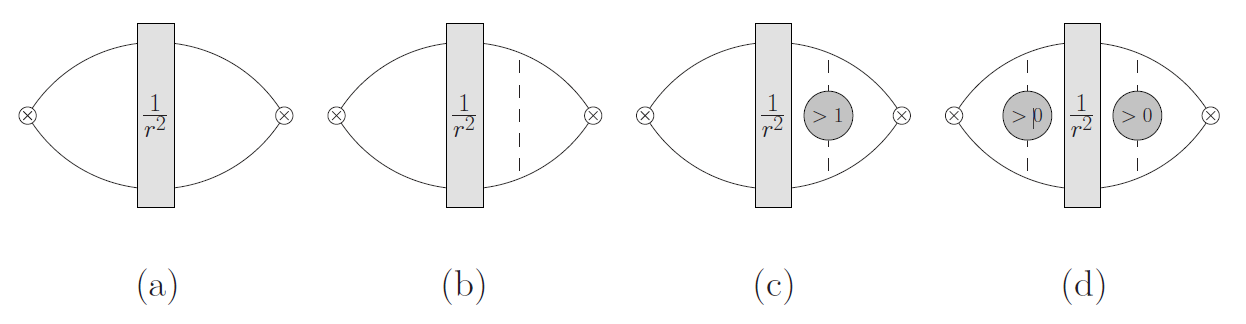}
\end{center}
%\end{figure}
%%%%%%%%%%%%%%%%%%%%%%%%%%%%%%%%%%%%%%%%%%%%%%%%%%%%%%%%%%%%%%%%%%%%
\end{widetext}
\noindent
Similar notation applies to the counterterm-including insertion 
function $J$. The first two terms map to two- and three-loop 
diagrams, which carry an overall
divergence and a divergence in the vertex subgraph(s) without gluon 
exchanges. The fourth term represents a sum of diagrams to all orders, 
but is finite. The third term has a divergence in the left vertex 
subgraph and all-order summation to the right of the potential insertion, 
where the notation ``$\,>\!1$'' refers to all ladder diagrams summed by 
$\hat{G}_0(E)$ with more than one gluon exchange. 

The calculation of the first two parts is
straightforward, since it involves only ordinary, solvable, dimensionally 
regularized two- and three-loop diagrams. 
Nevertheless, the divergent part must be properly factorized. To be 
explicit, the two-loop integral for diagram (a) evaluates to 
\begin{widetext}
\begin{eqnarray}
\hspace*{1cm}I_a[\frac{1}{2}+u]
&\!\!=\!\!&(\mu^2)^u\!\int\prod_{i=1}^4\Bigg[\frac{d^{d-1}{\bf
p}_i}{(2\pi)^{d-1}}\Bigg]\frac{\tilde{G}_0^{(0ex)}({\bf p}_1,{\bf
p}_2;E)\tilde{G}_0^{(0ex)}({\bf p}_3,{\bf p}_4;E)}{({\bf
q}_{23}^2)^{\frac{1}{2}+u}}
\nonumber\\
&\!\!=\!\!&\frac{m^2 \sqrt{-m E}}{64\pi^3} \left(-\frac{m E}{\mu^2}\right)^{\!-u} 
\!\left(-\frac{m E}{4\pi}\right)^{\!-2\epsilon}\,
\frac{\Gamma(1-\epsilon-u)\Gamma(\epsilon+u)^2
\Gamma(-\frac{1}{2}+u+2\epsilon)}{\Gamma(\frac{3}{2}-\epsilon)
\Gamma(2\epsilon+2 u)}\,,\hspace*{1.1cm}
\label{eq:i1halfu}
\end{eqnarray}
from which $J_a[1/2+a\epsilon;w(\epsilon)]$ follows according to 
the definition (\ref{eq:J-definition1}). The overall 
$\sqrt{-m E}$ factor indicates that the divergence of this 
integral persists in the imaginary part of the correlation function and 
is proportional to $G_0^{(0ex)}(E)$. Since the divergent part of 
the hard matching coefficient $c_v$ multiplies the $d$-dimensional 
correlation function, we must write the pole part 
of $J_a[\frac{1}{2}+a\epsilon;w(\epsilon)]$ in such a way that it 
multiplies the $d$-dimensional expression 
for $G_0^{(0ex)}(E)$, 
which is given by ($\tilde{\mu}^2 = \mu^2 e^{\gamma_E}/(4\pi)$)
\begin{eqnarray}
\hspace*{1cm}&& G_0^{(0ex)}(E) = (\tilde{\mu}^2)^\epsilon\!
\int\frac{d^{d-1}\bff{p}}{(2\pi)^{d-1}}
\frac{-1}{E -\frac{\bff{p}^2}{m}}  
= \frac{m\sqrt{-m E}}{8\pi^{3/2}} \left(-\frac{m E}{\mu^2}\right)^{\!-\epsilon} 
\!e^{\epsilon\gamma_E}\,\Gamma(-\frac{1}{2}+\epsilon).
\end{eqnarray} 
This results in ($L_\lambda = -\frac{1}{2}\ln(-4 m E/\mu^2)$)
\begin{eqnarray}
\hspace*{1cm}J_a[\frac{1}{2}+a\epsilon;w(\epsilon)] &=&-\frac{a m
w^{(1/\epsilon)}}{2\pi^2(a+1)\epsilon^2}\,G_0^{(0ex)}(E)
+\frac{m\Big(2a(\ln 2 -1) w^{(1/\epsilon)}+w\Big)}
{2\pi^2(a+1)\epsilon}\,G_0^{(0ex)}(E)
\nonumber \\
&&+ \,\frac{m^3C_F\alpha_s}{8\pi^3(a+1)\lambda} \Bigg[a
w^{(1/\epsilon)}\bigg(\ln^22-2\ln2-\frac{\pi^2}{24}(2a+5)-2a-2(a+1)L_{\lambda}
 \nonumber\\ 
&&
-\,(a+1)L_{\lambda}^2\bigg)+w\bigg(\ln
2-(a+1)L_{\lambda}-2-a\bigg)-\frac{1}{2}w^{(\epsilon)}\Bigg],
\label{eq:Jab}
\end{eqnarray}
\end{widetext}
\noindent
where the $1/\epsilon$ pole parts now multiply the $d$-dimensional 
zero-exchange Coulomb Green function.

Contributions such as part c, which have subdivergences and represent 
all-order graphs are the most complicated ones, since $G_0^{(>1ex)}(E)$ 
that appears to the right of the insertion is not known in $d$-dimensions. 
The strategy consists of isolating the subdivergence at the integrand 
level, so that it can be factorized without ever requiring an explicit 
representation of $G_0^{(>1ex)}(E)$. The finite remainder can then be 
evaluated in four dimensions. This strategy always works, because the
divergences can only arise from subdiagrams with a finite number of 
loops. This must be so, since the divergences must cancel with 
infrared divergences in the matching coefficients, which are computed 
in fixed-order perturbation theory. On the technical level, the finiteness 
of ladder diagrams with potential insertions once there are sufficiently 
many ladder rungs follows from the fact that every rung reduces the degree of 
divergence. For explicit results for the other parts of 
$J[\frac{1}{2}+a\epsilon;w(\epsilon)]$ we refer to~\cite{paperII}. 
An important consistency check is that the factored 
divergent parts of $J_a$, $2J_b$ 
and $2J_c$ are all the same. This is necessary for the sum of all
contributions to add to a term proportional to the full Green function
$G_0(E)=G_0^{(0ex)}(E)+G_0^{(1ex)}(E)+G_0^{(>1ex)}(E)$. That is, 
for the sum of all parts we have 
\begin{widetext}
\begin{eqnarray}
\hspace*{1cm}J[\frac{1}{2}+a\epsilon;w(\epsilon)]
&\!\!=\!\!&-\frac{am w^{(1/\epsilon)}}{
2\pi^2(a+1)\epsilon^2}\,G_0(E)+\frac{m\Big(2a(\ln 2 -1)
w^{(1/\epsilon)}+w\Big)}
{2\pi^2(a+1)\epsilon}\,G_0(E)+\mathcal{O}(\epsilon^0) ,
\end{eqnarray}
\end{widetext}
\noindent
as is required for cancelling the divergent part with the hard matching 
coefficient multiplying $G_0(E)$. The method also applies to double 
insertions, where some expressions require the factorization of a $1/\epsilon$ 
pole multiplying the single insertion of the NLO Coulomb potential. 
In the end, one verifies the cancellation of all poles in the sum of 
all terms, and evaluates the remainder numerically.

\subsection{\label{sub::US}Ultrasoft correction}

A contribution from the ultrasoft loop momentum region has to be taken into
account for the first time at the third order in perturbation theory. The 
ultrasoft correction $\delta^{us}G(E)$ to (\ref{masterthirdorder}) is 
technically and conceptually the most complicated contribution to the 
third-order correction to $G(E)$, since its (sub) divergence structure is 
rather involved and the integrals are more difficult. The correction 
to the Green function relevant to the top-pair cross section was 
computed in~\cite{Beneke:2008cr}, with the subtraction scheme  
already developed in~\cite{Beneke:2007pj} to determine the bound-state 
residues. In this subsection we want to
summarize the most important features.

The ultrasoft interaction terms in the PNRQCD
Lagrangian~(\ref{eq:pnrqcd}) relevant to third order are given by 
\begin{eqnarray}
  && g_s\psi^\dagger(x) \big[A_0(t,\bff{0})
  -\bff{x}\cdot\bff{E}(t,\bff{0})\big]\psi(x) 
  \nonumber\\ && \mbox{}
  + g_s\chi^\dagger(x)\big[A_0(t,\bff{0})
  -\bff{x}\cdot \bff{E}(t,\bff{0})\big]\chi(x).  \label{usterms}
\\[-0.45cm] \nonumber 
  \label{usterms}
\end{eqnarray}
The leading $A_0(t,\bff{0})$ couplings can be removed by a redefinition 
of the heavy-quark fields with 
a time-like Wilson line, which modifies the production current. 
The Wilson lines cancel for colour-singlet currents, hence the 
$A_0(t,\bff{0})$ couplings have no effect on top-pair production in $e^+ e^-$ 
collisions. Note that $\bff{x} \sim 1/v$ and $g_s
\bff{E} \sim v^{9/2}$ for ultrasoft gluon fields and thus the 
remaining chromoelectric
dipole interaction is suppressed by $v^{3/2}$
relative to the kinetic term in the action. Two ultrasoft
interaction vertices are required to form an ultrasoft loop, from which it
follows that the leading ultrasoft contribution arises only
at the third order.

Employing the Feynman rule for the $\bff{x}\cdot\bff{E}(t,\bff{0})$ vertex, 
the ultrasoft correction to the Green function can be expressed in the form
\begin{widetext}
  \begin{eqnarray}
    \delta^{us} G(E)&\!\!=\!\!& ig_s^2 C_F\!\int d^3{\bf r} \,d^3{\bf r}'\!\int
    \!\frac{d^4{k}}{(2\pi)^4}\, \Bigg[\frac{k_0^2\,{\bf r}\cdot {\bf r}'- ({\bf
        r}\cdot {\bf
        k})({\bf r}'\cdot {\bf k})}{k^2+i\varepsilon}
%    \nonumber\\ && \times\,
    G^{(1)}_0(\bff{0},{\bf r};E)G^{(8)}_0({\bf r},{\bf r}';E-k_0)
    G^{(1)}_0({\bf r}',\bff{0};E)\Bigg],\, 
%    \nonumber\\[-0.0cm]
    \label{eq:gfcorr}
  \end{eqnarray}
\end{widetext}
\noindent
with the understanding that one picks up only the pole at
$k^0=|\bff{k}|-i\epsilon$ in the gluon propagator.  
However, this expression cannot be used in practice, because it is 
ultraviolet divergent. Instead, one reverts the derivation of the 
chromoelectric dipole interaction and uses the Feynman 
rules for the threshold-expanded NRQCD momentum-space diagrams 
together with dimensional regularization. 

The ultrasoft correction $\delta^{us} G(E)$ 
has ultraviolet divergences from the integral over the three-momentum 
$\bff{k}$ of the ultrasoft gluon, and from the subsequent potential 
loop integrations. The former take 
the form of a single insertion of a third-order potential and 
of a one-loop correction to the coefficient $d_v$ of the 
$\mathcal{O}(v^2)$ suppressed derivative current. 
Note that ultraviolet divergence has to be regularized in 
dimensional regularization to be consistent with the calculation of
potential insertions and hard-matching coefficients. For this reason it is
convenient to introduce
appropriate subtraction terms to cancel the ultrasoft subdivergences, which 
leads to~\cite{Beneke:2007pj,Beneke:2013jia}
\begin{widetext}
  \begin{eqnarray}
    \delta^{us} G(E)&=& \big[\tilde\mu^{2\epsilon}\big]^2\int
    \frac{d^{d-1}\bffmath{\ell}}{(2\pi)^{d-1}} \frac{d^{d-1}\bffmath{\ell
        ^\prime}}{(2\pi)^{d-1}} \bigg\{\delta d_v^{\rm div}\,(-1)
    \frac{\bffmath{\ell}^2+\bffmath{\ell}^{\prime\,2}} {6 m^2}\,
    G^{(1)}_0(\bffmath{\ell},\bffmath{\ell}^\prime;E)
    \nonumber\\%[0.2cm]
    && \mbox{}
    +\,\big[\tilde\mu^{2\epsilon}\big]^2 \int \frac{d^{d-1}{\bf
        p}}{(2\pi)^{d-1}} \frac{d^{d-1}{\bf p^\prime}}{(2\pi)^{d-1}}
    \,G^{(1)}_0(\bffmath{\ell},{\bf p};E) \,i\,\Big[\delta U -\delta 
    V_{c.t.}\Big] \,iG^{(1)}_0({\bf p^\prime},\bffmath{\ell^\prime};E)
    \bigg\}\,,
    \label{eq:defUS}
  \end{eqnarray}
\end{widetext}
\noindent 
where $\delta V_{c.t.}$ represents the potential subtraction, and $\delta U$
is the ultrasoft insertion (containing the octet Green function).  The first
line of~(\ref{eq:defUS}) is related to the renormalization of the 
$\mathcal{O}(v^2)$ suppressed vector current.
In fact, $\delta d_v^{\rm div}$ contains the infrared divergence that 
was subtracted to obtain the finite expression for $d_v$ 
in~(\ref{eq:QCDVectorCurrent}). 
The divergences in the second line are associated with 
potential insertions. The counterterm (\ref{eq:counterC}) that 
is needed to make the three-loop Coulomb potential finite after 
coupling renormalization is contained in  $\delta V_{c.t.}$ above. 
The remaining divergences of (\ref{eq:defUS}) from the potential 
loop integrations are associated with 
the three-loop hard matching coefficient $c_v$ and isolated as 
explained in~\cite{Beneke:2007pj}. The final result is then evaluated 
by a combination of analytical and numerical methods.

\subsection{\label{sub::R}Third-order cross section}

In this subsection we put all third-order corrections together 
and briefly discuss their numerical effect on the production cross
section of top-quark pairs (see also~\cite{Beneke:2015}).  We restrict
ourselves to pure QCD corrections and furthermore only consider S-wave
contributions.  

The (normalized) total cross section is given by~(\ref{eq::R}).
We parametrize $R$ in terms of the potential subtracted (PS) 
mass~\cite{Beneke:1998rk}. This avoids the infrared sensitivity of the 
pole mass~\cite{Beneke:1994sw,Bigi:1994em}, which would otherwise 
prohibit a mass determination with accuracy below $\Lambda_{\rm QCD}$. 
The PS mass is related to the pole mass via
\begin{equation}
  m_t^{\rm PS} = m_t - \delta m_t(\mu_f),
  \label{eq::mPS_OS}
\end{equation}
where $\delta m_t(\mu_f)$ is obtained from the static potential
solving the following integral
\begin{equation}
  \delta m_t(\mu_f) =
  -\frac{1}{2} 
  \int_{|\bff{q}\,|<\mu_f} \frac{{\rm d}^3\bff{q}}{(2\pi)^3}
  V(\bff{q}\,)
  \,.
\end{equation}
The factorization scale $\mu_f$ is part of the definition of 
the PS mass. In the following 
we set it to $\mu_f=20$~GeV. To obtain $\delta m_t(\mu_f)$ to
N$^3$LO~\cite{Beneke:2005hg} the three-loop coefficient of the static
potential (cf. Section~\ref{sub::a3}) is needed.

We are now in the position to numerically evaluate~(\ref{eq::R}) where our
default input parameters are given by
\begin{eqnarray}
  \alpha_s(M_Z) &=& 0.1184\,, \nonumber \\
  m_t^{\rm PS} &=& 171.3~\mbox{GeV}\,, \nonumber \\
  \Gamma_t &=& 1.4~\mbox{GeV}\,.
  \label{eq::input}
\end{eqnarray}
As the central value for the renormalization scale we adopt $\mu=80$~GeV.

%%%%%%%%%%%%%%%%%%%%%%%%%%%%%%%%%%%%%%%%%%%%%%%%%%%%%%%%%%%%%%%%%%%%%
\begin{figure}[t]
\vspace*{0.2cm}
  \centering
\hskip-0.1cm
  \includegraphics[width=\columnwidth]{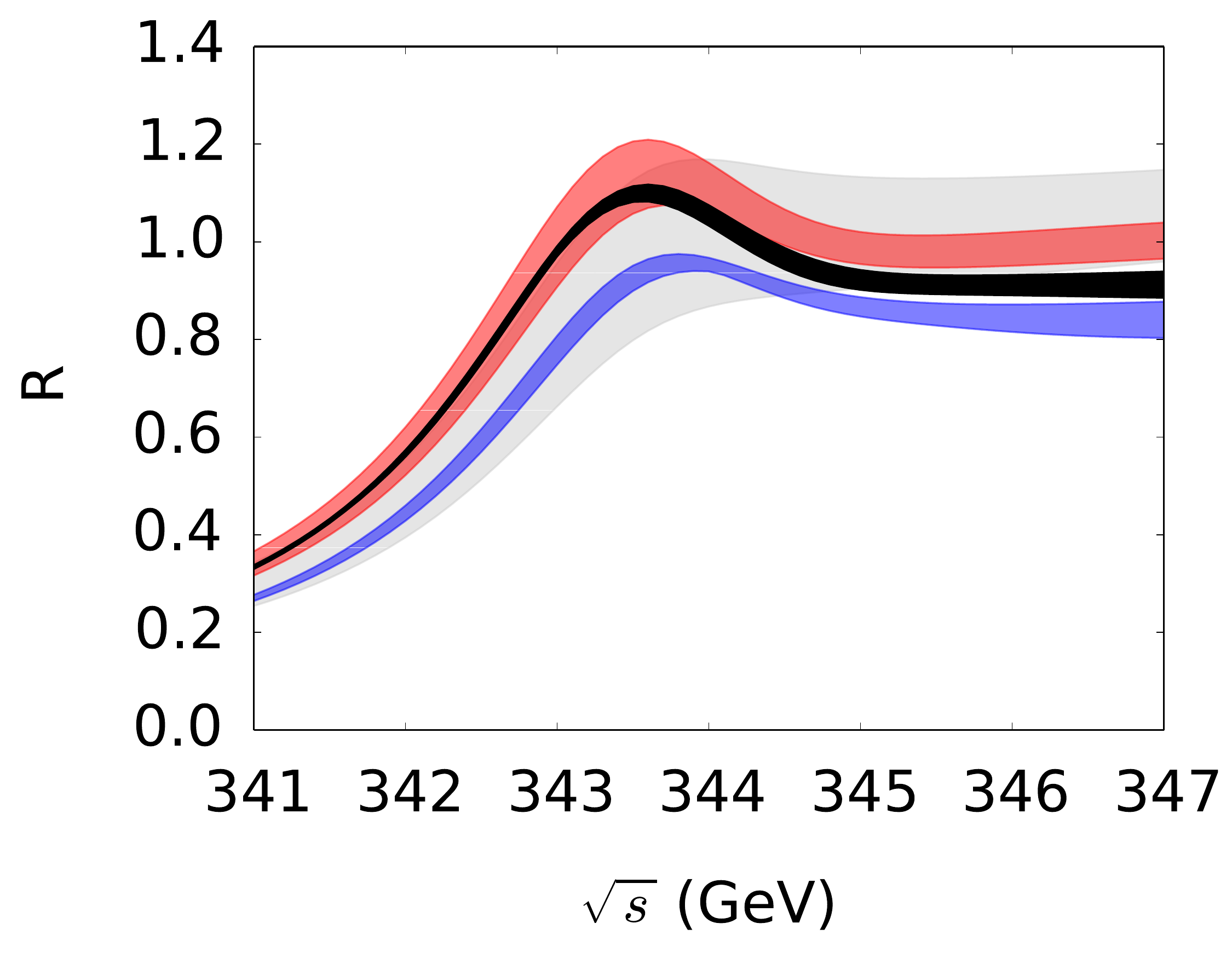}
\vspace*{-0.2cm}
  \caption{$R$ as a function of the $e^+e^-$ centre-of-mass energy in the
    threshold region. LO, NLO,
    NNLO and N$^3$LO results are shown in grey, blue (lowest band at 
$\sqrt{s}=347\,\mbox{GeV}$), red and black where the
    bands indicate the uncertainty due to scale variation.}
  \label{fig::Rs}
\end{figure}
%%%%%%%%%%%%%%%%%%%%%%%%%%%%%%%%%%%%%%%%%%%%%%%%%%%%%%%%%%%%%%%%%%%%%

In Figure~\ref{fig::Rs} the normalized total cross section is shown as a
function of $\sqrt{s}$ in the threshold region. The bands are obtained by
simultaneous variation of the renormalization and factorization scale between
$50$ and $350$~GeV.  After the inclusion of the third-order corrections one
observes a dramatic stabilization of the perturbative prediction, in
particular in and below the peak region, where most of the sensitivity to 
the top-quark mass comes from. In fact, the N$^3$LO band is
entirely contained within the NNLO one in this region.  
This is different above the peak position where a sizable negative 
correction is observed when going from NNLO to
N$^3$LO. For example, 2~GeV above the peak this amounts to about $-10\%$.

%%%%%%%%%%%%%%%%%%%%%%%%%%%%%%%%%%%%%%%%%%%%%%%%%%%%%%%%%%%%%%%%%%%%%
\begin{figure}[t]
  \centering
\hskip-0.4cm  
\includegraphics[width=\columnwidth]{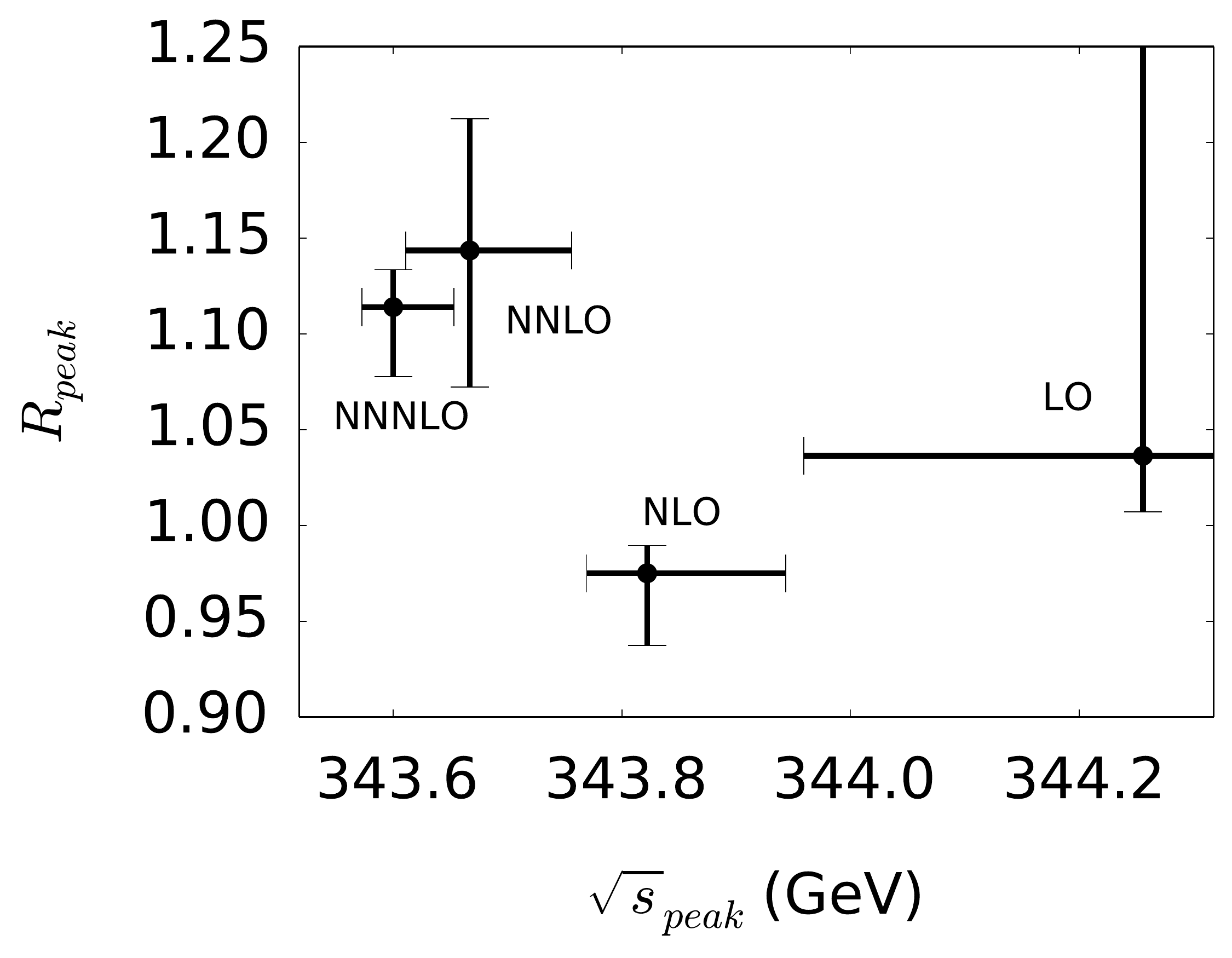}    
\vspace*{-0.2cm}
  \caption{\label{fig::peak_hei_pos}LO, NLO, NNLO and N$^3$LO results for the
    height (vertical axis) and position (horizontal axis) of the peak. The error bands
    originate from scale variation and uncertainties in $\alpha_s$.}
\end{figure}
%%%%%%%%%%%%%%%%%%%%%%%%%%%%%%%%%%%%%%%%%%%%%%%%%%%%%%%%%%%%%%%%%%%%%

It is interesting to have a closer look at the position and height of the
peak.  Figure~\ref{fig::peak_hei_pos} shows the peak height and its position at
LO, NLO, NNLO and N$^3$LO where the error bars reflects the uncertainty due to
the scale and $\alpha_s$ variation, which are added in quadrature.  In the
position of the peak one observes a relatively big jump from LO to NLO of
about 400~MeV, and from NLO to NNLO of approximately 150~MeV which reduces
to only 50~MeV from NNLO to N$^3$LO.  Furthermore, the NNLO and N$^3$LO
uncertainty bands show a significant overlap.  The uncertainty of the peak
position amounts to about 70~MeV, 60~MeV and 30~MeV at NLO, NNLO and N$^3$LO.
As far as the height is concerned there are big jumps from LO to NLO and NLO
to NNLO, however, the N$^3$LO correction stabilizes the perturbative series
and a shift below 3\% is observed.

%%%%%%%%%%%%%%%%%%%%%%%%%%%%%%%%%%%%%%%%%%%%%%%%%%%%%%%%%%%%%%%%%%%%%
\begin{figure}[t]
\vspace*{0.2cm}
  \centering
  \includegraphics[width=\linewidth]{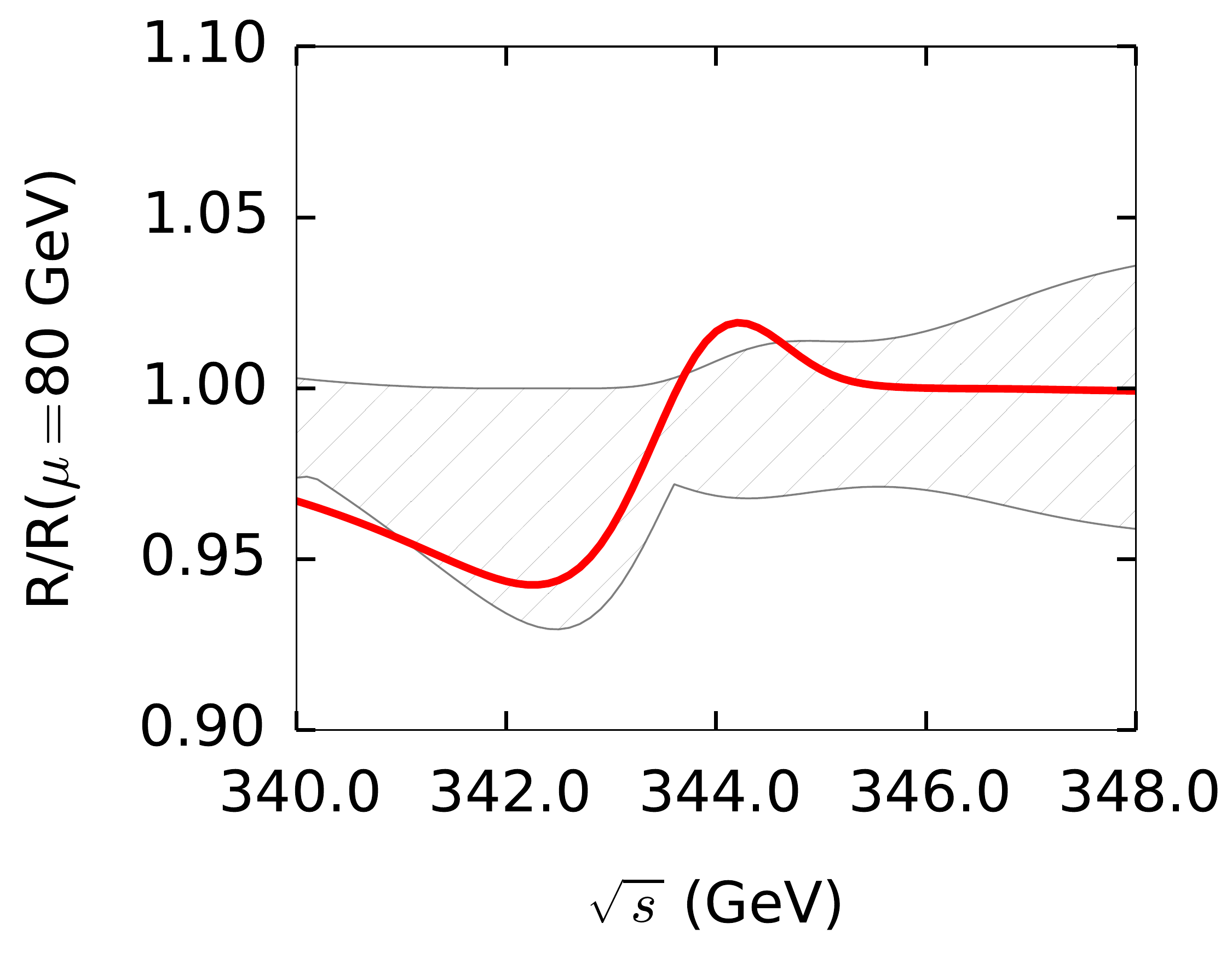}    
\vspace*{-0.2cm}
  \caption{\label{fig::mt_norm}N$^3$LO corrections to $R(s)$ normalized to its
    value for $\mu=80$~GeV where the band corresponds to scale variation
    between $50$ and $350$~GeV. For the (red) solid curve the input value 
    for $m_t^{\rm PS}$ in~(\ref{eq::input}) by $+50$~MeV.} 
\end{figure}
%%%%%%%%%%%%%%%%%%%%%%%%%%%%%%%%%%%%%%%%%%%%%%%%%%%%%%%%%%%%%%%%%%%%%

In Figure~\ref{fig::mt_norm} we study the sensitivity of the total 
cross section on the variation of the top-quark mass. This is done by 
normalizing the N$^3$LO band of Figure~\ref{fig::Rs} to the result 
obtained for $\mu=80$~GeV. The hatched band in Figure~\ref{fig::mt_norm} 
represents the theoretical uncertainty of this normalized quantity.  
The red curve in
Figure~\ref{fig::mt_norm} is obtained by adopting $\mu=80$~GeV and shifting
$m_t^{\rm PS}$ by $+50$~MeV. Above threshold basically no effect is
observed. However, the strong variation of the cross section below 
threshold implies a several percent effect even for 
such a small variation of  $m_t^{\rm PS}$, and 
provides a handle for the precise determination of the top-quark mass. From
Figure~\ref{fig::mt_norm} one may expect a precision of about 
$50\,\mbox{MeV}$, and it is obvious that the small theoretical uncertainty 
after including the third-order corrections to $R$ is crucial to 
reach this goal. Similar results are obtained for variations of 
other parameters
like the strong coupling constant and the top-quark width~\cite{Beneke:2015}.

Finally we mention that (almost complete) next-to-next-to-leading logarithmic 
predictions for the total cross section have been 
obtained~\cite{Pineda:2006ri,Hoang:2013uda}. In this approach, 
the logarithmically enhanced terms in $v$ of the third-order correction 
are included, but not the sizable ``constant'' terms. On the other hand, 
logarithms of $v$ are summed to all orders. At first
glance there appears to be good agreement with our N$^3$LO results. 
However, a detailed comparison of the different terms included in the two 
approaches is required to determine whether the agreement is 
more than coincidental.

\subsection{\label{sub::further}Further results}

With the third-order QCD corrections to the dominant S-wave production 
mode completed and uncertainties under good control, attention must be 
paid to other corrections that may be less challenging from the 
technical perspective, but important phenomenologically. There are 
QED and electroweak corrections, where we count $\alpha_{\rm em}\sim 
\alpha_{\rm EW} \sim \alpha_s^2\sim v^2$; Higgs boson contributions; 
production of $t\bar t$ in a P-wave state; non-resonant contributions 
to the physical final state $W^+W^- b\bar b$; and a consistent treatment 
of initial-state radiation with controlled dependence on the factorization 
scheme for the initial-state electron distribution function. 
We discuss some of these items below (see \cite{Beneke:2007zg,Actis:2008rb} 
for a general discussion of initial-state radiation in this context 
and some relevant formalism).

\subsubsection{P-wave contribution}

The P-wave contribution arises from the axial-vector coupling of the 
$Z$-boson to the $t\bar t$ pair. Since the imaginary part of the 
axial-vector polarization 
function $\Pi^{(a)}(q^2)$ is $\mathcal{O}(v^2)$ suppressed in the 
non-relativistic limit, a NLO calculation of $\Pi^{(a)}(q^2)$ in 
PNRQCD results in a third-order contribution to the total top-pair 
production cross section near threshold. The NLO calculation was performed 
in \cite{Beneke:2013kia}.\footnote{Some results for the P-wave Green 
function were already obtained 
in~\cite{Bigi:1991mi,Penin:1998ik,Penin:1998mx,Kuhn:1999hw}, but
none of these computations were performed in dimensional
regularization as required for consistency with other pieces 
of the calculation.} The NLO correction stabilizes the 
scale dependence, but the result is ambiguous until a scheme is defined 
to combine the resonant with non-resonant contributions, see 
Section~\ref{sub::non-res} below. However, the P-wave contribution is 
always a very small contribution, below 1\%, to the total cross section.

\subsubsection{Higgs contributions}

\newcommand{\yh}{y_H}
\newcommand{\yhoneb}{y_{H,1b}}

The Higgs contributions are of particular interest, since they 
provide sensitivity of the cross section $\sigma(e^+e^-\to t\bar{t}+X)$
to the top-quark Yukawa coupling to the Higgs boson. The dependence 
originates from Higgs exchange between the produced top quarks (potential 
region), and from radiative corrections to the production vertex 
(hard region). These effects are incorporated into the PNRQCD framework 
via modifications of the vector current which result from simultaneously 
integrating out the top quark and the Higgs boson, assuming  
$M_H\sim m_t$ for power counting. Furthermore, a new operator occurs 
in the effective theory, which in momentum space is given 
by~\cite{Eiras:2006xm}
\begin{eqnarray}
  \delta {\cal L}_H &=& \frac{\alpha \pi m_t^2}{s_W^2 M_W^2 M_H^2},
\label{eq::Hop}
\end{eqnarray}
where $s_W$ is the sine of the weak mixing angle and $M_W$ is the $W$
boson mass. In coordinate space, this is equivalent to a delta function 
potential that approximates the short-range Yukawa potential.
If we employ the counting rule $\alpha \sim \alpha_s^2$ it is easy to
see that $\delta {\cal L}_H$ gives contributions which are
parametrically of the same order as the ones from the third-order PNRQCD 
Lagrangian. 

We parametrize electroweak corrections to the matching
coefficient $c_v$ as
\begin{eqnarray}
  c_v &=& 1 
  + \frac{\alpha}{\pi s_W^2} c_v^{\rm ew}
  + \frac{\alpha\alpha_s}{\pi^2 s_W^2} C_F c_v^{\rm mix}
  + \ldots
  \,,
  \label{eq::cvdef}
\end{eqnarray}
where the the ellipses stand for QCD corrections which can be found
in~(\ref{eq::cv3num}).  The complete one-loop electroweak corrections to the
matching coefficient $c_v^{\rm ew}$ have been computed
in~\cite{Guth:1991ab}. The Higgs boson contribution can be cast in the form
\begin{eqnarray}
    c_v^{H, \rm ew} &=& \frac{m_t^2}{M_W^2}
  \Bigg[ \frac{3\yh^2-1}{12\yh^2}-\frac{2-9\yh^2+12\yh^4}{48\yh^4}
  \ln\yh^2 
  \nonumber\\&&\mbox{}
  - \frac{(-2+5\yh^2-6\yh^4)}{24\yh^2}
  \nonumber\\&&\mbox{}\times
  \frac{\sqrt{4\yh^2-1}}{\yh^2}
   \arctan{\sqrt{4\yh^2-1}}
  \Bigg]\, ,
  \label{eq::cv1exact}
\end{eqnarray}
where $\yh=m_t/M_H$ and $m_t$ is the on-shell top-quark mass.  Mixed
contributions of order $\alpha\alpha_s$ involving the Higgs boson have
been computed in~\cite{Eiras:2006xm} where various expansions
have been considered. For $M_H\approx 125$~GeV an expansion around
$m_t\approx M_H$ leads to the best results for $c_v^{H, \rm mix}$,
which can be written as
\begin{eqnarray}
  \lefteqn{c_{v,1b}^{H, \rm mix} =  \frac{m_t^2}{M_W^2} \Bigg [
  \frac{\pi^2}{8} \left ( 1-\yhoneb \right
  ) \ln{\frac{m_t^2}{\mu^2}} 
  -5.760}
  \nonumber \\ && \mbox{}
  + 5.533 \yhoneb   
  -0.171 \yhoneb^2
  +0.0124\yhoneb^3
  \nonumber \\ && \mbox{}
  +0.0304\yhoneb^4
  +0.0296 \yhoneb^5 +\ldots \Bigg ] 
  \,,
  \label{eq::cv2exp}
\end{eqnarray}
with $\yhoneb = (1-\yh^2)$.

%%%%%%%%%%%%%%%%%%%%%%%%%%%%%%%%%%%%%%%%%%%%%%%%%%%%%%%%%%%%%%%%%%
\begin{figure}[t]
  \vspace*{0.2cm}
  \hskip0.2cm
  \includegraphics[width=7cm]{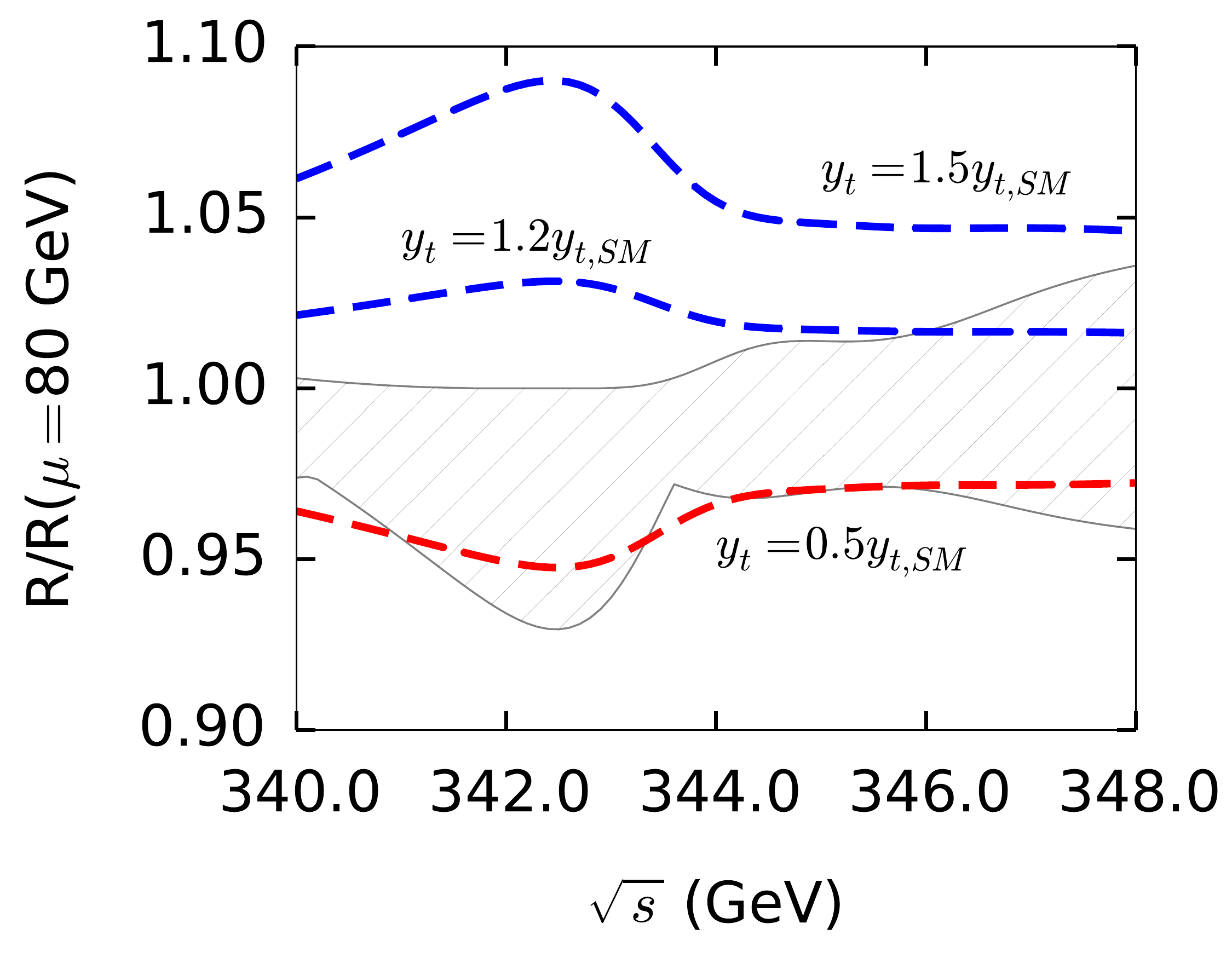}    
  \caption{\label{fig::yt_norm}N$^3$LO corrections to $R(s)$ normalized to its
    value for $\mu=80$~GeV where the band corresponds to scale variation
    between $50$ and $350$~GeV. For the dashed curves the input value
    for the top Yukawa coupling in the $t\bar t H$ vertex is rescaled 
    relative to the SM value by a factor 1.5, 1.2, 0.5 (from top to bottom) 
    while keeping the top-quark mass at its input 
    value~(\ref{eq::input}). The central value to which the 
    variation is normalized includes the SM Higgs contributions.} 
\end{figure}
%%%%%%%%%%%%%%%%%%%%%%%%%%%%%%%%%%%%%%%%%%%%%%%%%%%%%%%%%%%%%%%%%%

The PNRQCD computation of the Higgs contributions to  
$\sigma(e^+e^-\to t\bar{t}+X)$ has been performed 
in \cite{BenekePiclumRauhHiggs} and results in corrections 
at the $10\%$ level as expected from the discussion 
of the corrections to the bound-state residue~\cite{Eiras:2006xm}. 
Higgs contributions of order $\alpha\alpha_s$ clearly have to be incorporated 
in order to reach a precision at the few-percent level.

Figure~\ref{fig::yt_norm} shows the effect of a modification of the 
top-Yukawa coupling relative to its SM value of the top-pair 
production cross section near threshold in a format similar to 
Figure~\ref{fig::mt_norm}. Comparing the width of the uncertainty 
band to the shift of the curves provides a first estimate of how 
well the top-Yukawa coupling can be constrained at a future 
$e^+ e^-$ collider before the cms energy for open $t\bar t H$ 
production is reached \cite{BenekePiclumRauhHiggs}. 

\subsubsection{\label{sub::non-res}Non-resonant contributions}

The most important piece that is still missing for a realistic prediction 
of the total cross section is related to the fact that the top quark 
decays rapidly. The measurement of top-pair production near threshold 
really refers to the measurement of $e^+ e^- \to W^+W^- b\bar b$ 
with cms energy $\sqrt{s}$ near $4 m_t^2$. The pure QCD contribution, 
which we referred to above, is {\em defined} by the polarization function 
evaluated in the complex energy plane with the prescription 
$E\to E+i \Gamma_t$. 

The limitations of this approximation manifest themselves within 
the (NR)QCD calculation itself. The current correlation function $G(E)$ 
exhibits an uncancelled ultraviolet divergence from an overall 
divergence of the form $[\delta G(E)]_{\rm overall} 
\propto \alpha_s E/\epsilon$ in dimensional 
regularization~\cite{Beneke:2008cr}. Since $E$ acquires an 
imaginary part $\Gamma_t \sim m_t\alpha_{\rm EW}$ through the above 
prescription, the divergence 
survives in the cross section,
\begin{equation} 
\mbox{Im}\,[\delta G(E)]_{\rm overall} \propto 
m_t\times \frac{\alpha_s\alpha_{\rm EW}}{\epsilon}.
\end{equation}
The divergence 
appears first at NNLO (since at LO, $G(E) \sim v \sim \alpha_s$), 
and results in dependence on an additional regularization scale $\mu_w$ 
after subtraction. 
A consistent calculation therefore requires that one considers 
the process $e^+ e^- \to W^+W^- b\bar b$ within unstable-particle 
effective theory~\cite{Beneke:2003xh,Beneke:2004km} including the effects of off-shell top quarks and 
processes that produce the $W^+W^- b\bar b$ final state with no or only one 
intermediate top-quark line. The physical cross section is then the 
sum of two terms, 
\begin{eqnarray}
&& \sigma_{e^+ e^- \to W^+W^- b\bar b} = 
\underbrace{\sigma_{e^+ e^- \to [t\bar t]_{\rm res}}(\mu_w)}_{
\mbox{\footnotesize pure (NR)QCD}} 
\nonumber\\[0.2cm] &&
\hspace*{0.8cm}+ \,
\sigma_{e^+ e^- \to W^+W^- b\bar b{}_{\rm nonres}}(\mu_w). 
\end{eqnarray}
Both terms separately have a ``finite-width scale dependence'', 
and only the sum is well-defined. The calculation of the non-resonant 
part has been performed so far only at 
NLO \cite{Beneke:2010mp,Penin:2011gg}, including the 
possibility to apply invariant-mass cuts on the top decay products. 
Parts relevant to NNLO are known 
\cite{Penin:2011gg,Hoang:2004tg,Jantzen:2013gpa,Ruiz-Femenia:2014ava}, 
but the complete computation is still missing. For a further discussion 
of non-resonant contributions and unstable-particle effective theory 
we refer to the review \cite{BenekeB4} in this volume.

\section{Threshold resummation in hadronic collisions}
\label{sec:4}

The production of top-quark pairs, and pairs of heavy particles in general, 
in proton-proton (LHC) or proton-antiproton (Tevatron) collisions is 
different from $e^+ e^-$ annihilation in two important aspects. First, 
the centre-of-mass energy of the colliding partons is not fixed. 
Non-relativistic theory is strictly relevant only in the kinematic region 
of small invariant mass of the heavy-particle pair near the kinematic  
limit. Second, the initial-state particles are coloured, so that the 
heavy-particle pair can be produced in different colour states. This 
implies that soft gluon effects are much more important than in $e^+ e^-$ 
annihilation, both 
due to initial-state radiation, and due to the coupling to the coloured 
final state. (Following standard terminology, 
what we call ``soft'' in this section, refers to the ultrasoft region 
in previous sections.) In this section we discuss 
the theoretical formalism that brings together non-relativistic theory 
with the theory of soft-gluon resummation, the latter having been 
developed first for Drell-Yan 
production~\cite{Sterman:1986aj,Catani:1989ne} and then extended to di-jets 
or pairs of heavy particles~\cite{Kidonakis:1998nf,Laenen:1991af,Berger:1996ad,Catani:1996yz,Kidonakis:1996aq,Kidonakis:1997gm,Bonciani:1998vc}. 
We then consider the 
invariant-mass distribution of hadronically 
produced top pairs, and the total cross section for top and superpartner 
particle pairs.

\subsection{Joint soft and Coulomb resummation}

The theoretical quantities of interest are the hard-scattering 
cross sections $\hat{\sigma}$ for the partonic subprocesses 
\begin{equation}
\label{eq:subprocess}
p(k_1) p'(k_2)\rightarrow H(p_1)H'(p_2)+X
\end{equation}
with $pp'\in\{qq, q\bar q, \bar q\bar q, gg, gq,g\bar q\}$, 
when the partonic centre-of-mass energy $\hat{s}$ is close 
to $4 M^2 \equiv (m_H+m_{H^\prime})^2$, such that 
$\beta = (1-4 M^2/\hat{s})^{1/2}$ is 
small. The two heavy particles $H$, $H^\prime$ can be in arbitrary, 
not necessarily identical colour representations $R$, $R^\prime$.

The expansion of $\hat{\sigma}$ contains enhanced terms of 
the form $(\alpha_s \ln^2\beta\,)^n$, primarily (but not exclusively) 
related to soft gluon effects, and the ``Coulomb singularities''
$(\alpha_s/\beta)^n$, which both cause a breakdown of the 
perturbation expansion for small $\beta$. Although the power-like 
Coulomb singularities are formally stronger than the logarithmic 
soft-gluon effects, the importance of one or the other effect 
depends on the initial state (quark vs. gluon) and colour of the 
final state (for instance, singlet vs. octet). Joint soft and Coulomb 
resummation \cite{Beneke:2009rj,Beneke:2010da}
concerns the question whether both effects can be 
resummed simultaneously.

\subsubsection{Factorization}

This is a non-trivial issue, since the energy of soft gluons is of the same
order as the kinetic energy $M\beta^2$ of the heavy particles
produced. Soft-gluon lines may therefore connect without parametric
suppression to the heavy-particle propagators in between the Coulomb ladder 
rungs, as well as to the Coulomb gluons itself (by virtue of the gluon
self-coupling), impeding the standard factorization arguments that assume
either no couplings to the final state, or to the energetic initial-state
particles.

Factorization was investigated in \cite{Beneke:2009rj,Beneke:2010da}, 
where it was shown that the partonic cross section factorizes into three
separate contributions, related to hard, soft and Coulomb effects.  
The latter two are coupled only through a convolution in an energy 
variable $\omega$, which roughly speaking accounts for the fact that 
near threshold the production of the heavy particle pair is sensitive 
to the small amount of energy radiated into soft gluons. The factorization 
formula reads
\begin{widetext}
\begin{equation}
\label{eq:fact}
\hspace*{2.5cm}
\hat\sigma_{pp'}(\hat s,\mu)
= \sum_{i}H^{i}_{pp'}(M,\mu) \;\int d \omega\;
\sum_{R_\alpha} J_{R_\alpha}(E-\frac{\omega}{2})\,
W^{R_\alpha}_i(\omega,\mu)\, .
\end{equation}
\end{widetext}
\noindent 
Here $E=\sqrt{\hat s}-2 m_t$ is the energy relative to the production
threshold. Referring to \cite{Beneke:2010da} for the derivation, we explain 
here the elements of the formula and properties of the effective 
interactions that lead to this simple form. The formula as written applies 
to the total cross section. A similar result without convolution integral 
holds for the distribution in the $H H^\prime$ invariant mass.

The first factor $H^{i}_{pp'}(M,\mu)$ accounts for the short-distance 
production of $H H^\prime$. It depends on the masses of the heavy particles, 
the partonic initial state, but not on the small scales $E$ and $M\beta$ in 
the problem. The superscript $i$ refers to a colour decomposition to be 
discussed below. The function $J_{R_\alpha}(E)$ accounts for the 
non-relativistic effects from the potential region and coincides with 
the zero-distance Coulomb Green function $G(E)$, generalized to the 
case of unequal-mass particles and an arbitrary irreducible SU(3) 
representation. At leading order, this is easily accomplished by 
replacing $m$ by the reduced mass, and the colour-singlet coefficient 
$C_F$ of the Coulomb potential by $-D_{R_\alpha}$. $D_{R_\alpha}$ is 
determined by \begin{equation}
\label{eq:coulomb-coeff}
  {\bf T}^{(R)b}_{a_1c_1}{\bf T}^{(R')b}_{a_2c_2}P^{R_{\alpha}}_{c_1c_2a_3a_4}
  = D_{R_\alpha}P^{R_{\alpha}}_{\{a_1a_1a_3a_4\}},
\end{equation}
where $P^{R_{\alpha}}_{a}$ projects on the irreducible representation 
$R_\alpha$ in the tensor product $R\otimes R^\prime$, and $T^{(R)}$ are the 
SU(3) generators in representation $R$. Finally, 
$W^{R_\alpha}_i(\omega,\mu)$ is the soft-gluon function in the 
representation $R_\alpha$.

Eq.~(\ref{eq:fact}) holds up to corrections ${\cal O}(\beta^2)$ for the 
total cross section near threshold. Within this approximation the 
interactions of soft gluons with collinear and non-relativistic particles 
take a particularly simple form, which is essential for proving 
factorization. The coupling of soft gluons to an energetic initial-state 
quark with four-momentum in the light-like $n$ direction is given 
by the term
\begin{equation}
\bar{\xi}_c(x) \,i g_s n \cdot A_s(x_+) \,\frac{\not\!\bar n}{2}
\,\xi_c(x), 
\label{lagrangian}
\end{equation}
in the position-space SCET Lagrangian, which is equivalent to the 
eikonal approximation. Here $\xi_c$ denotes the 
$n$-collinear quark field, and $\bar{n}$ is another light-like vector 
with $n\cdot \bar{n}=2$. It is important that only the $n\cdot A_s$ 
component appears, and the soft field is evaluated at the point 
$x_+^\mu=(\bar n\cdot x/2)\,n^\mu\equiv
x_- n^\mu $ due to the light-cone 
multipole expansion \cite{Beneke:2002ph,Beneke:2002ni}. This allows 
the soft-gluon coupling to be removed from the Lagrangian by 
performing the field redefinition 
$\xi_c(x) = S^{(3)}_n(x_-) \xi_c^{(0)}(x)$ \cite{Bauer:2001yt}, 
where 
\begin{equation}
S^{(R)}_{n}(x) = \mbox{P} \exp 
\left[i g_s \!\int_{-\infty}^0 \!\!\!d t 
\,n \cdot  A^c_s(x+n t){\bf T}^{(R)c}\right]
\end{equation}
denotes a light-like, soft Wilson line for a particle in the representation 
$R$ of SU(3). Similar results apply to initial-state gluons. The leading 
soft-gluon interaction with the non-relativistic fields for $H$ and 
$H^\prime$, 
\begin{equation}
\label{eq:NRQCDsoft}
\psi^\dagger(x) \,i g_s A_s^0(x_0)\,\psi(x)
+\psi^{\prime\,\dagger}(x) \,i g_s A_s(x_0)\, \psi'(x),
\end{equation}
involves only the $A_0$ fields at point $(x_0,\bff{0}\,)$, and 
therefore can be removed 
by $\psi(x) = S^{(R)}_{w}(x_0) \psi^{(0)}(x)$, where $w=(1,\bff{0}\,)$ 
is a time-like four-vector, and 
\begin{equation}
S^{(R)\dagger}_{w}(x) =
\mbox{P}
 \exp \left[i g_s \!\!\int_{0}^\infty \!\!\!d t\, w \cdot A^c_s(x+w t)\,
   {\bf T}^{(R)c}\right]
\end{equation}
is the corresponding Wilson line. The decoupling of soft gluons from the 
Coulomb gluons in ladder diagrams now follows from the identity 
$S_w^{(R)\dagger} {\bf T}^{(R)a}S_w^{(R)}= S_{w;ab}^{(8)}{\bf T}^{(R)b}$, 
which in turn implies 
$\left[\psi^{\dagger}{\bf T}^{(R)a} \psi\right]\!(x+\bff{r}\,) = 
  S_{w;,ab}^{(8)} (x_0) 
 \left[\psi^{\dagger (0)}{\bf T}^{(R)b} \psi^{(0)}\right]\!(x+\bff{r}\,)$. 
Since the Wilson line in the adjoint representation is real and 
hermitian, and the argument is $x_0$ independent of $\bff{r}$, the 
Wilson line drops out from the Coulomb interaction
\begin{equation}
\int \!\!d^3 \vec{r}\, 
\left[\psi^{\dagger}{\bf T}^{(R)a} \psi\right]\!(x+\bff{r}\,)
\,\frac{\alpha_s}{r}
\left [\psi^{\prime\,\dagger}{\bf T}^{(R')a}\psi^\prime\right]\!(x),
\,
\end{equation}
when expressed in terms of the redefined fields \cite{Beneke:2010gi,Beneke:2010da}.

The soft gluons do not disappear completely. The short-distance production 
amplitude of the heavy-particle pair in the hard $2\to 2$ process 
is described by an operator of the 
form 
\begin{eqnarray}
\label{eq:opdef}
\mathcal{O}^{(0)}_{\{a;\alpha\}}(\mu) = 
\big[\phi_{c;a_1\alpha_1}\phi_{\bar c;a_2\alpha_2} 
\psi^\dagger_{a_3\alpha_3}\psi^{\prime \,\dagger}_{a_4\alpha_4}\big](\mu), 
&& \\[-0.2cm]\nonumber 
\end{eqnarray}
which is local (up to collinear Wilson lines $W_c$ in collinear fields 
such as $\phi_c$ for the initial-state partons). After squaring the 
amplitude and introducing the redefined fields, eight soft Wilson lines 
are left over. Since the effective Lagrangian, expressed in the redefined 
fields, does not contain soft interactions with the other fields at leading 
power, these Wilson lines can be collected into the soft function
\begin{widetext}
\begin{equation}
\label{eq:soft-general}
\hspace*{1.5cm}
\hat W^{\{k\}}_{\{ab\}}(z,\mu)
=\langle 0|\overline{\mbox{T}}[ 
S_{w,b_4 k_2} S_{w,b_3 k_1} S^\dagger_{\bar{n},jb_2} S^\dagger_{n, ib_1}](z)
\mbox{T}[S_{n,a_1i} S_{\bar{n},a_2j} S^\dagger_{w,k_3 a_3} 
S^\dagger_{w,k_4 a_4}](0)|0\rangle,
\end{equation}
\end{widetext}
\noindent
where $\mbox{T}$ ($\overline{\mbox{T}}$) denote (anti) time-ordering. 
Latin indices refer to SU(3) colour and the Wilson lines are understood 
to refer to the representations of the corresponding particles. Curly 
brackets $\{a\}$ denote a colour multi-index, here ${a_1 a_2 a_3 a_4}$. 

Further simplifications that lead to (\ref{eq:fact}) are closely connected 
to properties of the soft function \cite{Beneke:2009rj}. The physical 
intuition that soft gluons should couple only to the total colour charge 
when the heavy-particle pair is produced at rest at threshold, suggests that 
the gluon coupling to the pair can be described by 
a single soft function. Indeed, by virtue of 
\begin{equation}
\label{eq:combine-wilson}
C^{R_\alpha}_{\alpha a_1a_2}S^{(R)}_{v,a_1b_1}
S^{(R')}_{v,a_2b_2}=S^{(R_\alpha)}_{v,\alpha\beta}
C^{R_\alpha}_{\beta b_1b_2}, 
\end{equation}
where $C^{R_\alpha}_{\alpha a_1a_2}$ denotes the Clebsch-Gordan 
coefficient that couples the representations $R$, $R^\prime$ to the 
irreducible representation $R_\alpha$ in $R\otimes R^\prime$, 
the soft function (\ref{eq:soft-general}) can be related to a sum 
of simpler soft functions of the form
\begin{widetext}
\begin{equation}
\label{eq:soft-R}
\hspace*{2.5cm}
W^{R_\alpha}_{\{a\alpha ,b\beta \}}(z,\mu)\equiv
\langle 0|\overline{\mbox{T}}[S^{R_\alpha}_{v,\beta \kappa}
S^\dagger_{\bar{n},jb_2} S^\dagger_{n, ib_1} ](z)
\mbox{T}[S_{n,a_1i}S_{\bar{n},a_2j}S^{R_\alpha\dagger}_{v,\kappa\alpha}](0)|
0\rangle .
\end{equation}
\end{widetext}
\noindent
The Clebsch-Gordan coefficients can also be used to construct an 
orthonormal basis of colour tensors $c^{(i)}_{\{a\}}$ 
for the hard-scattering amplitude with 
initial-state partons in the representation $r\otimes r^\prime = 
\sum_\beta r_\beta$ and 
$H H^\prime$ in $R\otimes R^\prime = \sum_{\alpha} R_\alpha$, such that 
$\mathcal{A}_{pp'\{a\}}=\sum_i c^{(i)}_{\{a\}} \mathcal{A}_{pp'}^{(i)}$.
The index $i$ enumerates all pairs $P_i=(r_{\beta}, R_{\alpha})$ of equivalent 
representations $r_\beta$ and $R_\alpha$ that occur in the above 
decomposition of the tensor-product representations. For example, 
in top-pair production through gluon-gluon fusion, there are 
three combinations $P_i\in 
\{(1,1),\; (8_S,8),\;(8_A,8)\}$.
Colour conservation implies that 
\begin{equation}
\label{eq:prod-basis}
  c_{\{a\}}^{(i)}=\frac{1}{\sqrt{{\rm dim}(r_\beta)}}\,
  C^{r_\beta}_{\alpha a_1a_2} C^{R_{\alpha}\ast}_{\alpha a_3a_4}
\end{equation}
form an orthonormal basis. Defining 
\begin{equation}
\label{eq:c-ralpha}
c^{R_\alpha(i)}_{\{a\alpha\}}\equiv c^{(i)}_{\{a\}}C^{R_\alpha}_{\alpha a_3a_4}
\end{equation}
the soft function (\ref{eq:soft-R}) can be represented as a matrix in 
this basis:
\begin{equation}
\label{eq:soft-coulomb-2}
 W^{R_\alpha}_{ii'}(\omega,\mu)=  
c^{R_\alpha(i)}_{\{a\alpha\}} W^{R_\alpha}_{\{a\alpha,b\beta\}}(\omega,\mu)
c^{R_\alpha(i')\ast }_{\{b\beta\}}.
\end{equation}
The formula (\ref{eq:fact}) now follows from the fact that 
$ W^{R_\alpha}_{ii'}(\omega,\mu)$ can be shown to be a diagonal matrix to all 
orders in perturbation theory for all relevant cases \cite{Beneke:2009rj}.

The factorization formula (\ref{eq:fact}) justifies earlier treatments 
of threshold resummation for heavy particles, where soft-Coulomb 
factorization has been put in as an 
assumption~\cite{Kiyo:2008bv,Hagiwara:2008df,Kulesza:2009kq}, or the 
Coulomb-enhanced terms were not summed and technically considered as 
part of the hard function \cite{Bonciani:1998vc}. 
Soft-gluon resummation in heavy-particle production has been performed 
in Mellin-moment space in previous works. We therefore note that the 
convolution in (\ref{eq:fact}) turns into multiplicative 
soft-Coulomb factorization $\hat \sigma_{pp'}(N,\mu) \approx
\sum_{i}H^{i}_{p p^\prime}
\sum_{R_\alpha}\,J_{R_\alpha}(N)\,
W^{R_\alpha}_{i}(N,\mu)$ of the partonic cross sections in moment space.

%%%%%%%%%%%%%%%%%%%%%%%%%%%%%%%%%%%%%%%%%%%%%%%%%%%%%%%%%%%%%%%%%%%%%%
\begin{figure}[t]
\vskip0.2cm
\begin{center}
\includegraphics[width=6cm]{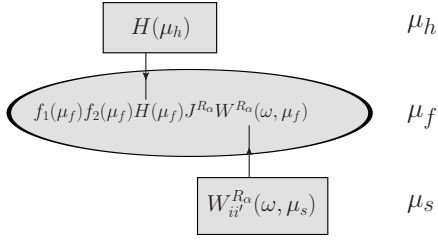}
\end{center}
\vspace*{-0.1cm}
\caption{\label{fig:running} 
Sketch of the resummation of soft gluon corrections using 
renormalization group equations. Figure from~\cite{Beneke:2010da}.}
\end{figure}
%%%%%%%%%%%%%%%%%%%%%%%%%%%%%%%%%%%%%%%%%%%%%%%%%%%%%%%%%%%%%%%%%%%%%%

Away from the production threshold large soft-gluon logarithms 
can appear in various other kinematic regions. When the relative velocity of 
the heavy particles is not small, Coulomb effects are not enhanced, and 
factorization takes the more standard form $H\cdot W$ without the 
$J$ function. However, the hard and soft functions now depend on the 
kinematic invariants of the general $2\to 2$ scattering process, as do 
the anomalous dimensions relevant to resummation. The soft function is 
no longer diagonal. Near threshold, interactions that change the 
colour-state of the heavy-particle pair appear only at the level of 
$\mathcal{O}(\beta^2)$ ($\mathcal{O}(\beta)$) corrections to the 
total cross section (amplitude). Some of these 
effects are discussed in \cite{Beneke:2010da}, but no calculation 
of $\mathcal{O}(\beta^2)$ corrections in non-relativistic perturbation 
theory has so far been performed for hadronic processes.

\subsubsection{Resummation}

In the present framework the resummation of ``Coulomb singularities''
$(\alpha_s/\beta)^n$ is automatic in the $J$ function. The 
logarithms $(\alpha_s \ln^2\beta\,)^n$ 
are summed by solving the renormalization group equations to 
evolve the hard functions $H$ from the scale $\mu_h\sim M$, and the 
soft functions $W$ from the scale $\mu_s \sim M\beta^2$ to a common 
scale $\mu_f$, chosen to be the factorization scale of the 
parton distributions, as illustrated in Figure~\ref{fig:running}.

For the systematics of the combined resummations
of the two types of 
corrections we count both $\alpha_s \ln\beta$ and $\alpha_s/\beta$
as quantities of order one and introduce a parametric representation of the
expansion of the cross section of the form
\begin{widetext}
\begin{eqnarray}
\label{eq:syst}
\hspace*{1.3cm}
\hat{\sigma}_{p p'} &\!\propto\!& \,\hat\sigma_{p p'}^{(0)}\, 
\sum_{k=0} \left(\frac{\alpha_s}{\beta}\right)^k \,
\exp\Big[\underbrace{\ln\beta\,g_0(\alpha_s\ln\beta)}_{\mbox{(LL)}}+ 
\underbrace{g_1(\alpha_s\ln\beta)}_{\mbox{(NLL)}}+
\underbrace{\alpha_s g_2(\alpha_s\ln\beta)}_{\mbox{(NNLL)}}+\ldots\Big]
\nonumber\\[0.2cm]
&& \,\times
\left\{1\,\mbox{(LL,NLL)}; \alpha_s,\beta \,\mbox{(NNLL)};
\ldots\right\}.
\end{eqnarray}
\end{widetext}
\noindent
The resummed
cross section at LL accuracy includes all terms of order
$1/\beta^{\,k}\times\alpha_s^{n+k}\ln^{2n}\beta$ relative to the Born cross
section near threshold.  Next-to-leading summation includes in
addition all terms of order $\alpha_s \ln\beta; \,\alpha_s^2
\{1/\beta\times\ln\beta, \ln^3 \beta\};\ldots$, while 
furthermore all terms $\alpha_s ; \,\alpha_s^2
\{1/\beta,\ln^{2,1} \beta\}; \ldots$ are included
in the NNLL approximation. With this counting, the formalism described 
above is limited to NNLL, since $\mathcal{O}(\beta^2)$ terms are 
required beyond this order. A caveat applying to current NNLL 
treatments should be mentioned. 
Higher-order non-relativistic 
potentials cause ultraviolet singularities in the 
Coulomb function $J$, which cancel with the hard functions, and 
cause non-relativistic logarithms beginning with $\alpha_s^2\ln\beta$ 
at NNLL order. Present NNLL results (as described below) 
sum soft logarithms to 
all orders by evolution of the soft function, but 
include the non-relativistic logarithms only 
at fixed order $\alpha_s^2\ln\beta$.

The renormalization group equations and the formula for the resummed 
cross section are very similar to those derived for Drell-Yan production 
within the SCET framework~\cite{Becher:2006nr,Becher:2006mr,Becher:2007ty}, 
with the $H H^\prime$ pair in representation $R_\alpha$ 
replacing the colour-singlet Drell-Yan pair or electroweak gauge boson.
The resummed partonic cross section then 
reads \cite{Beneke:2009rj,Beneke:2010da}
\begin{widetext}
\begin{equation}
\hat\sigma^{{\rm res}}_{pp'}(\hat s,\mu_f)=
\sum_{i} H^{i}_{pp^\prime}(\mu_h)  \;
 U_i(M,\mu_h,\mu_s,\mu_f) 
\int_0^\infty \!d \omega \;
\frac{ J_{R_{\alpha}}(E-\frac{\omega}{2})}{\omega} 
\left(\frac{\omega}{2M}\right)^{2 \eta}
\tilde{s}_{i}^{R_\alpha}
\left(2\ln\left(\frac{\omega}{\mu_s}\right)+\partial_\eta,\mu_s\right)
\frac{e^{-2 \gamma_E \eta}}{\Gamma(2 \eta)}
\label{eq:fact-resum}
\end{equation}
with $\eta = 2 a_{\Gamma}(\mu_s,\mu_f)$.
The summed logarithms are contained in the evolution function
\begin{eqnarray}
U_i(M,\mu_h,\mu_f,\mu_s )&=&
\left(\frac{4M^2}{\mu_h^2}\right)^{-2a_\Gamma(\mu_h,\mu_s)}\,
\left(\frac{\mu_h^2}{\mu_s^2}\right)^{\eta}
\times \,\exp\Big[4  (S(\mu_h,\mu_f)-S(\mu_s,\mu_f))
\nonumber\\[0.1cm]
&&
 -\,2a_i^{V}(\mu_h,\mu_s) +2 a^{\phi,r}(\mu_s,\mu_f)+
2 a^{\phi,r'}(\mu_s,\mu_f)\Big],
\label{eq:def-u}
\end{eqnarray}
\end{widetext}
\noindent
and 
\begin{equation}
\tilde{s}^{R_\alpha}_{i}(\rho,\mu) =\int_{0_-}^{\infty} d \omega e^{-s \omega}
\,W_{i}^{R_\alpha}(\omega,\mu)
\end{equation}
denotes the Laplace-transform of the $\overline{{\rm MS}}$-renormalized soft
function with respect to $s =
1/(e^{\gamma_E} \mu e^{\rho/2})$. The sum over the final-state
representations $R_\alpha$ in the factorization
formula~(\ref{eq:fact}) has disappeared in the colour
basis~(\ref{eq:prod-basis}), since there is a unique final-state 
representation for each term in the sum over $i$. The resummed 
partonic cross section is then matched to the fixed-order cross 
section at the highest available order (NNLO for top quarks, NLO 
for supersymmetric particles), and integrated with the parton 
luminosity. The presence of the Coulomb functions introduces some 
subtleties in the convolution with the parton densities, which are 
discussed in~\cite{Beneke:2011mq}.

For the definitions of the quantities appearing in (\ref{eq:fact-resum}),  
(\ref{eq:def-u}) we refer to \cite{Beneke:2010da}, and mention only 
briefly the necessary ingredients for NNLL resummation. The hard functions 
are process-specific and must be obtained from the one-loop 
$2\to 2$ production cross sections directly at threshold, separated into 
irreducible colour (and if necessary, spin) representations. The soft 
function is also needed at one-loop, and has been computed for 
arbitrary colour representations in~\cite{Beneke:2009rj}. The 
single-particle and cusp anomalous dimension must be used at the two-loop  
and three-loop order, respectively, and are the same as for the 
Drell-Yan process. The only new anomalous dimension is the  
anomalous dimension $\gamma_{H,s}^{R_\alpha}$ 
for the soft function (\ref{eq:soft-coulomb-2}). 
In~\cite{Beneke:2009rj} the required expression has been related to
the constant coefficient in the anomalous dimension of the 
heavy-heavy current in heavy-quark effective theory 
in the limit where the cusp angle goes to infinity. At the two-loop 
order relevant to NNLL resummation, the soft anomalous dimension 
exhibits Casimir scaling, $\gamma_{H,s}^{R_\alpha}=C_{R_\alpha}\gamma_{H,s}$. 
From the expression for the two-loop anomalous 
dimension of the heavy-heavy 
current~\cite{Korchemsky:1991zp, Kidonakis:2009ev}, one extracts 
\begin{equation}
\gamma_{H,s}^{(1)}=-C_A\left(\frac{98}{9}-\frac{2\pi^2}{3}+4\zeta_3\right)
+\frac{40}{9} T_F n_f
\end{equation}
for the coefficient of $(\alpha_s/(4\pi))^2$.
This result was confirmed by an explicit, independent 
calculation~\cite{Czakon:2009zw}.

\subsection{Top-pair invariant mass distribution near threshold}

A Coulomb enhancement of the cross section for the production of top-quark 
pairs close to threshold, which has been discussed in 
Section~\ref{sec:3} in the
context of $e^+e^-$ collisions, can also be observed at hadron colliders, since
it is possible to produce the top quarks in the colour-singlet state.  Indeed,
at LHC, the cross section close to threshold in dominated by the process $gg\to
t\bar{t}$ where the top-quark pair is in the ${}^1S_0$ colour-singlet 
state.  In
contrast to a linear collider, where the physical observable is the total
cross section as a function of energy, at the hadron collider  one considers
the invariant-mass distribution of the top-quark pairs.

The calculation of the cross section within the NRQCD framework contains as
building blocks the hard production cross section for a top-quark pair at
threshold~\cite{Kuhn:1992qw,Petrelli:1997ge} and the non-relativistic Green
function governing the dynamics of the would-be toponium bound-state.  
In~\cite{Kuhn:1992qw} the NLO formulae were derived for quark or gluon
initial states and a quarkonium in a $\mbox{J}^{\mbox{\scriptsize PC}}=0^{-+}$
colour-singlet state, plus possibly a parton. The general case, with the heavy
quark system $(Q\bar Q)$ in S-wave singlet/triplet spin state, and 
colour-singlet/octet configuration is given in~\cite{Petrelli:1997ge}, together
with the corresponding results for P-waves.  The results 
of~\cite{Kuhn:1992qw,Petrelli:1997ge} were presented for stable
bound states. For wide resonances it is convenient to describe the
bound-state dynamics through a Green function.

NLO calculations have been performed in~\cite{Kiyo:2008bv,Hagiwara:2008df}, 
where slightly different
approaches have been applied. Whereas in~\cite{Hagiwara:2008df} the
matching has been performed in the strict threshold limit where the partonic
centre-of-mass energy $\hat{s}$ approaches twice the top-quark mass, the
complete dependence on $\hat{s}$ as given 
in~\cite{Kuhn:1992qw, Petrelli:1997ge} is included in~\cite{Kiyo:2008bv}. 
Thus, formally, the result of~\cite{Hagiwara:2008df} is only valid 
for top-quark production where the velocity of both quarks is small. 
In the approach of~\cite{Kiyo:2008bv} on the other hand, 
the relative velocity still has to 
be small but the combined top-antitop quark system can move with high 
velocity. Furthermore, Ref.~\cite{Kiyo:2008bv} includes all NLO subprocesses,
i.e.~also those which appear for the first time in ${\cal
  O}(\alpha_s^3)$ and performs a soft gluon resummation at the 
NLL order using the Mellin-space approach.

\begin{figure}[t]
  \begin{center}
    \includegraphics[width=\linewidth]{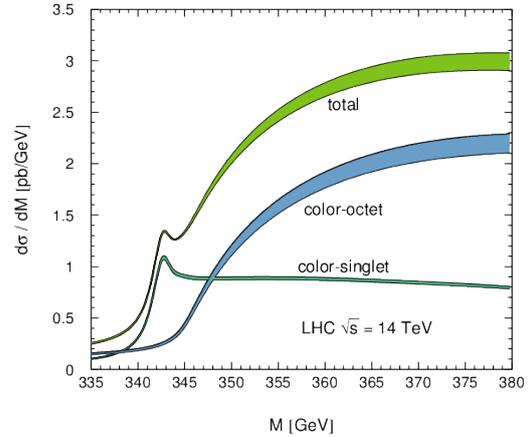}
    \caption{
      \label{fig:xs2}
      Invariant-mass distribution for top-quark pair production. The width of
      the bands reflect the scale dependence of the hard scattering parts. 
Figure from~\cite{Kiyo:2008bv}.}
  \end{center}
\end{figure}

In Figure~\ref{fig:xs2} the invariant-mass distributions for LHC with
$\sqrt{s}=14$~TeV centre-of-mass energy is shown for the sum of all
contributing channels and separately for the colour-octet and colour-singlet
contribution.  The width of the bands is obtained from varying renormalization
and factorization scales in the hard cross section as described above. The
additional uncertainty from the Green function, which is estimated to about
20\% for the singlet and below 5\% for the octet case~\cite{Kiyo:2008bv}, is
not included.

As expected, for invariant mass 
$M<2 m_t$ the production of $t{\bar{t}}$ pairs is dominated
by the singlet contribution. However, for $M>2 m_t$ one observes a strong
rise of the octet contributions, in particular of the 
gluon-induced subprocess, 
which for $M\gsim 2m_t+5$~GeV becomes even larger than the corresponding
singlet contribution.  For the colour-octet case the scale dependence of the
hard scattering amounts to $\pm 7\%$.  Considering the threshold behaviour as
shown in Figure~\ref{fig:xs2} it is clear  
that the location of the threshold is
entirely governed by the behaviour of the colour-singlet (S-wave)
contribution. Thus, as a matter of principle, determining the location of this
step experimentally would allow for a top-quark mass measurement, which is
conceptually very different from the one based on the reconstruction of a
(coloured) single quark in the decay chain $t\to Wb$. In fact, much of the
detailed investigations of $t\bar t$ threshold production at a linear collider
were performed for this situation by establishing the relation 
between the location of the colour
singlet top-antitop resonance and the top-quark
$\overline{\mbox{MS}}$-mass.

%%%%%%%%%%%%%%%%%%%%%%%%%%%%%%%%%%%%%%%%%%%%%%%%%%%%%%%%%%%%%%%%%%%%
%%%%%%%%%%%%%%%%%%%%%%%%%%%%%%%%%%%%%%%%%%%%%%%%%%%%%%%%%%%%%%%%%%%%

Considering the threshold region (say up to $M_{t\bar t}=350~{\rm GeV}$)
separately, an integrated cross section of 15~pb is obtained, which should be
compared to 5~pb as derived from the NLO predictions using a stable top quark
and neglecting the binding Coulomb-force correction.  
Within this relatively narrow region
the enhancement amounts to roughly a factor three and a significant shift of
the threshold. Compared to the total cross section for $t\bar t$ production of
about 900~pb, the increase is relatively small, about 1\%. However, in view 
of the present and future experimental precision
these effects should not be ignored.

\subsection{Total top pair production cross section}

The ability of LHC to produce top-quark pairs in large numbers has 
triggered a large amount of theoretical work devoted to improving 
the accuracy of the prediction of the total cross section. The first 
results referred to approximate NNLO (``NNLO$_{\rm app}$'') 
calculations that aimed at including the singular terms 
in the partonic cross section as $\beta\to 0$ 
\cite{Moch:2008qy,Beneke:2009ye,Aliev:2010zk,Beneke:2010fm}. 
Soft-gluon resummation of the total cross section 
with NNLL accuracy was completed in 
\cite{Beneke:2011mq,Cacciari:2011hy} with two independent calculations,  
one done in the joint soft-Coulomb resummation approach described above, 
the other in the Mellin-space soft-gluon resummation formalism. 
In yet another approach the total cross section is computed from 
NNLL-resummed or approximated NNLO calculations of certain differential 
distributions  \cite{Ahrens:2010zv,Ahrens:2011mw,Kidonakis:2010dk}. 

It should be noted that threshold resummation for the top-pair 
inclusive cross section does not have a clear parametric justification. 
In order to obtain the total hadronic cross section, the partonic
cross section is convoluted with the parton distribution 
functions.  Both at Tevatron and LHC, the top-antitop invariant-mass 
distribution peaks at about $400\,$GeV, which corresponds 
(in the absence of radiation) to $\beta \approx 0.5$. The convolution of 
the partonic cross section with the parton luminosity is therefore 
dominated by the region $\beta>0.3$, where the threshold approximation 
is no longer valid. Nevertheless, one often 
finds that the threshold expansion provides a reasonable approximation 
even outside its domain of validity. At the very least, the approximation 
is better than the one not using this piece of information.

Much of the ambiguity in resummed calculations has been removed 
through the fixed-order 
NNLO calculation \cite{Baernreuther:2012ws,Czakon:2013goa}. The present 
state-of-the-art prediction therefore consists of NNLL resummation 
matched to the full NNLO result, and is available in the 
programs {\sc Topixs}\footnote{http://users.ph.tum.de/t31software/topixs/}~\cite{Beneke:2012wb} 
(based on \cite{Beneke:2011mq}) and 
top++ \cite{Czakon:2011xx} (based on \cite{Cacciari:2011hy}).
{\sc Topixs} includes soft and Coulomb resummation and is therefore 
technically the most complete theoretical prediction. In particular, 
it includes the bound-state effects in the threshold region discussed 
in the previous subsection. However, the 
effect of Coulomb resummation beyond the terms already included 
in the fixed-order NNLO result is rather small for 
the total cross section.

%%%%%%%%%%%%%%%%%%%%%%%%%%%%%%%%%%%%%%%%%%%%%%%%%%%%%%%%%%%%%%%%%%%%
\begin{table}[t]
\begin{center}
\scriptsize
\vspace*{0.2cm}
\begin{tabular}{|l|c|c|c|}
\hline
$\sigma_{t \bar{t}}$[pb]&  Tevatron
& LHC ($\sqrt{s}=$7 TeV) & LHC ($\sqrt{s}=$8 TeV)  \\
\hline
\hline
&&&\\[-0.1cm]
NLO & $6.68^{+0.36+0.23}_{-0.75-0.22}$ & 
$158.1^{+19.5+6.8}_{-21.2-6.2}$ & 
$226.2^{+27.8+9.2}_{-29.7-8.3}$\\[0.2cm]
NNLO$_{\rm app}$ & $7.06^{+0.26+0.29}_{-0.34-0.24}$ & 
$161.1^{+12.3+7.3}_{-11.9-6.7}$&  
$230.0^{+16.7+9.7}_{-15.7-9.0}$\\[0.2cm]
NNLO & $7.01^{+0.27+0.29}_{-0.37-0.24}$ & 
$167.1^{+\phantom{0}6.7+7.7}_{-10.7-7.1}$&  
$239.1^{+9.2+10.4}_{-14.8-9.6}$\\[0.2cm]
NNLL & $7.15^{+0.24+0.30}_{-0.10-0.25}$ & 
$168.5^{+\phantom{0}6.3+7.7}_{-\phantom{0}7.5-7.2}$&  
$241.0^{+8.7+10.5}_{-11.1-9.7}$\\[0.2cm]
\hline
\end{tabular}
\end{center}
\caption{
\label{toptable} Top-quark pair production cross section at 
Tevatron and LHC for
 $m_t=173.3\,$GeV, $\alpha_s(M_Z)=0.1171\pm 0.0014$, 
(N)NLO MSTW08 PDFs~\cite{Martin:2009iq}. The first error represents to 
theoretical uncertainty from independent soft/hard/Coulomb scale 
variations and resummation ambiguities in the partonic cross section, 
the second PDF+$\alpha_s$  at 68\% CL. }
\end{table}
%%%%%%%%%%%%%%%%%%%%%%%%%%%%%%%%%%%%%%%%%%%%%%%%%%%%%%%%%%%%%%%%%%%%

In Table~\ref{toptable} we summarize successive approximations to the 
top-quark pair production cross section at 
Tevatron and LHC from 
NLO to NNLL, where NNLL means resummed matched to full NNLO, 
generated with {\sc Topixs 2.0}. We note that in production at the 
Tevatron, which is dominated by the quark-antiquark initial state, 
the resummation/NNLO effect is significant ($+8\%$, NNLL vs. NLO in the 
table), 
and the threshold approximation to the full result 
worked well (NNLO vs. NNLO$_{\rm app}$). On the other hand, in 
gluon-gluon initiated production at the LHC, resummation is a small 
correction ($+1\%$) and underestimates the full NNLO correction  
($+4\%$). However, in both cases, resummation without the full NNLO 
correction already leads to a significant 
reduction of the theoretical uncertainty 
($8\% \to 3\%$ at Tevatron, $13\% \to 4.5\%$ at LHC, 
excluding the PDF + $\alpha_s$ error), while still including the 
full NNLO+NNLL result in the uncertainty estimate.

%%%%%%%%%%%%%%%%%%%%%%%%%%%%%%%%%%%%%%%%%%%%%%%%%%%%%%%%%%%%%%%%%%%%
\begin{figure}[t]
  \begin{center}
  \includegraphics[width=6cm]{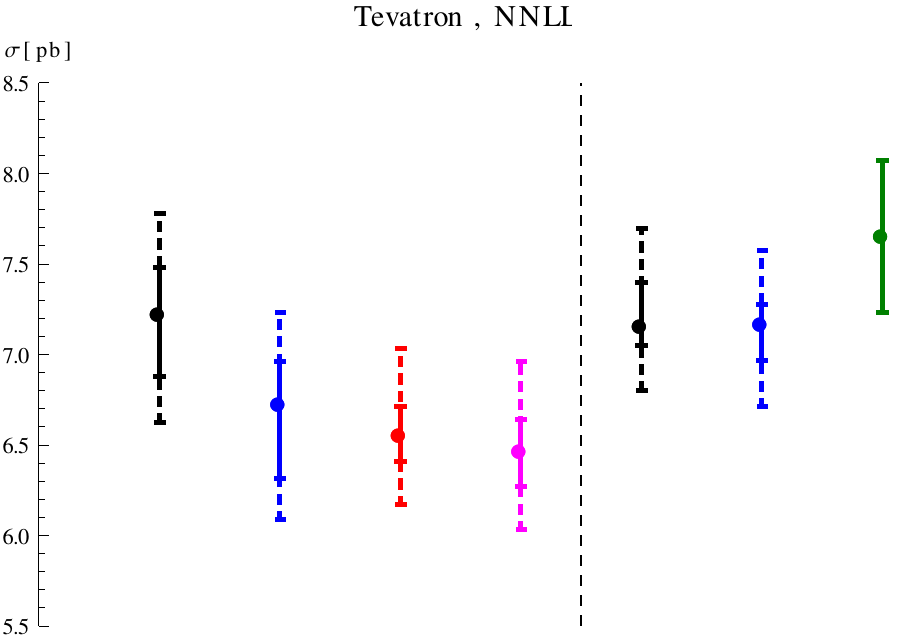}
  \\[-12.7em]
  \fcolorbox{white}{white}{Tevatron, NNLL}
  \\[-0.8em]
  \mbox{}\hspace*{2em}\fcolorbox{white}{white}{$\sigma[pb]$} \hfill \mbox{}
  \\[12em]
  \end{center}
\vspace*{-0.2cm}
\caption{
\label{fig:topcomparison}
Top-pair production cross section calculations with NNLL accuracy 
in $p \bar {p}$ collisions at $\sqrt{s} = 1.96\,$TeV (Tevatron).
To the left of the dashed line: NNLO$_{\rm app}$ + NNLL results.
Black: Beneke et al. \cite{Beneke:2011mq} and {\sc Topixs 1.0} 
($m_t=173.3$~GeV); 
Blue: Cacciari et al. \cite{Cacciari:2011hy} ($m_t=173.3$~GeV)
Red: Ahrens et al. \cite{Ahrens:2011mw} (1PI, $m_t=173.1$~GeV)
Magenta: Ahrens et al. \cite{Ahrens:2011mw} (PIM, $m_t=173.1$~GeV)
To the right of the dashed line: full NNLO + NNLL results.
Black: Update of Beneke et al. \cite{Beneke:2012wb} and 
{\sc Topixs 2.0} ($m_t=173.3$~GeV)
Blue: Czakon et al. \cite{Czakon:2013goa} and {\tt top++2.0}
($m_t=173.3$~GeV)
Darkgreen: CDF/D0 combined measurement quoted from 
CMS/ATLAS-CONF-2012-149.
Error bars: inner solid -- theory uncertainty of the 
partonic cross section excluding $\alpha_s$, 
outer dashed -- PDF (+$\alpha_s$) added.
PDF set: MSTW2008 NNLO with  $\alpha_s(M_Z)=0.1171\pm 0.0014$.
}
\end{figure}

Figure~\ref{fig:topcomparison} summarizes the results for the total 
top-pair cross section at the Tevatron from different 
theoretical calculations. By comparing the black and blue bars 
to the left and right of the vertical dashed line, we observe 
that the predictions from joint soft-Coulomb resummation 
(first bar [black], \cite{Beneke:2012wb}) and the Mellin-space approach 
(second bar [blue], \cite{Czakon:2013goa}) are in very good agreement, once the 
full NNLO result is included and resolves some of the resummation 
ambiguities (in favour of the former). The NNLO+NNLL computations 
are in good agreement with the measurement for the adopted 
top-quark pole mass $m_t=173.3$~GeV, while the NLO result would 
underestimate the measurement significantly (compare 
Table~\ref{toptable}).

The top-quark pair cross section has become a precisely predicted and 
measured quantity, which depends essentially only on the 
fundamental parameters $m_t$ and $\alpha_s$, and the parton 
distributions. Assuming standard physics the measurement constrains 
these parameters. The possibility to determine the top-quark mass 
in a theoretically clean way, though less precisely than from 
reconstruction of top decay products, has been explored, for instance, in 
\cite{Beneke:2011mq,Beneke:2012wb,Langenfeld:2009wd,Abazov:2011pta}. 
The determination of the strong coupling has been considered in
\cite{Chatrchyan:2013haa}. The impact of the top-quark cross section 
at LHC on the gluon distribution in the proton has been investigated 
first in \cite{Beneke:2012wb}, and in more detail in \cite{Czakon:2013tha} 
within the NNPDF framework. See also \cite{Alekhin:2013nda}.

%%%%%%%%%%%%%%%%%%%%%%%%%%%%%%%%%%%%%%%%%%%%%%%%%%%%%%%%%%%%%%%%%%%%

\subsection{Pair production of supersymmetric particles}

The search for the partners of the SM particles predicted by 
supersymmetric extensions of the SM is an integral part of the 
LHC physics programme. Present exclusions already imply that the 
masses of the strongly interacting squarks and gluinos are most 
likely in the TeV range. For such heavy particles, threshold 
resummation is expected to be important, and, contrary to the case of 
top quarks, even parametrically relevant, as a larger fraction 
is produced close to the partonic threshold due to the fall-off 
of the parton distributions at large momentum fraction. 

The study of soft-gluon resummation for pair production of supersymmetric
particles was initiated by \cite{Kulesza:2009kq,Kulesza:2008jb}. The NLL
analysis of the squark-antisquark, squark-squark, (anti) squark-gluino and
gluino-gluino final states \cite{Beenakker:2009ha} finds significant
corrections of several 10\% beyond the fixed-order NLO calculation, especially
for the gluino-gluino final state, which involves the largest colour charges.
Note, that NLO corrections to gluino bound-state production has been
considered in~\cite{Hagiwara:2009hq,Kauth:2009ud} and gluino pair production
close to threshold is presented in~\cite{Kauth:2011vg}.

The resummation formalism for soft and Coulomb gluons was in fact first 
used to predict squark-antisquark production~\cite{Beneke:2010da} 
in the NLL approximation as defined in (\ref{eq:syst}) and then 
extended to all superparticle pair final states in~\cite{Falgari:2012hx}, 
which also generalized the factorization formula to a particular case of 
P-wave production relevant to stop-antistop production. A scenario with 
superparticles with substantial decay widths (for instance, gluinos 
decaying further into squarks and quarks) was also 
investigated~\cite{Falgari:2012sq}, which adds the complication of 
unstable-particle effects and non-resonant production. 

\begin{figure}[t]
\vspace*{0.2cm}
  \begin{center}
  \includegraphics[width=6cm]{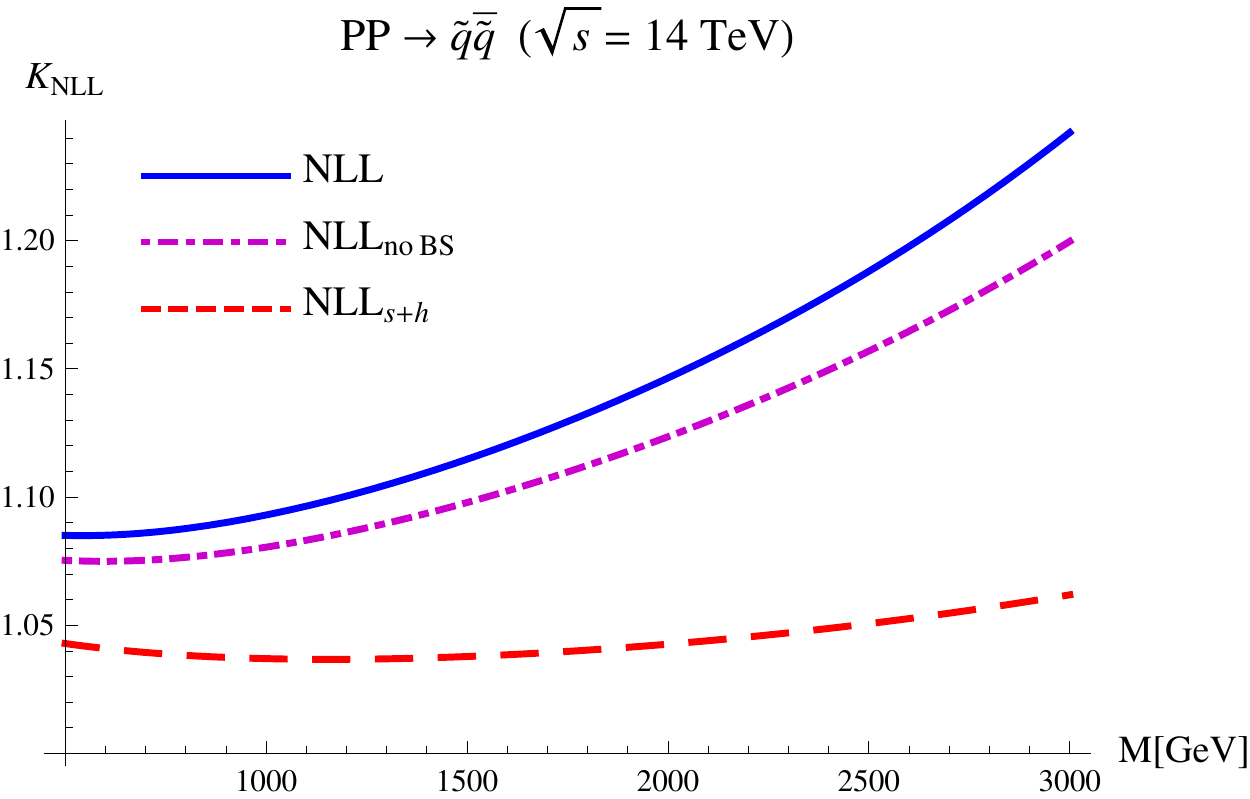}
  \includegraphics[width=6cm]{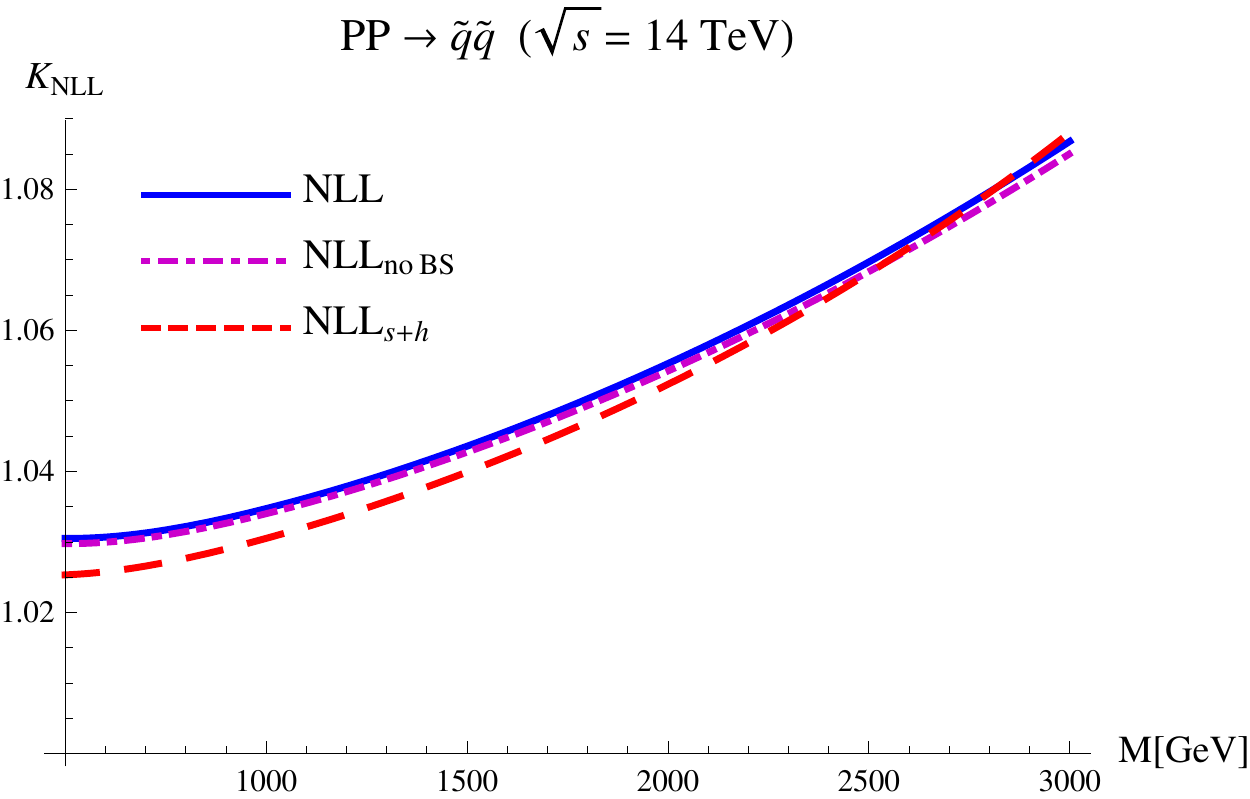}
  \includegraphics[width=6cm]{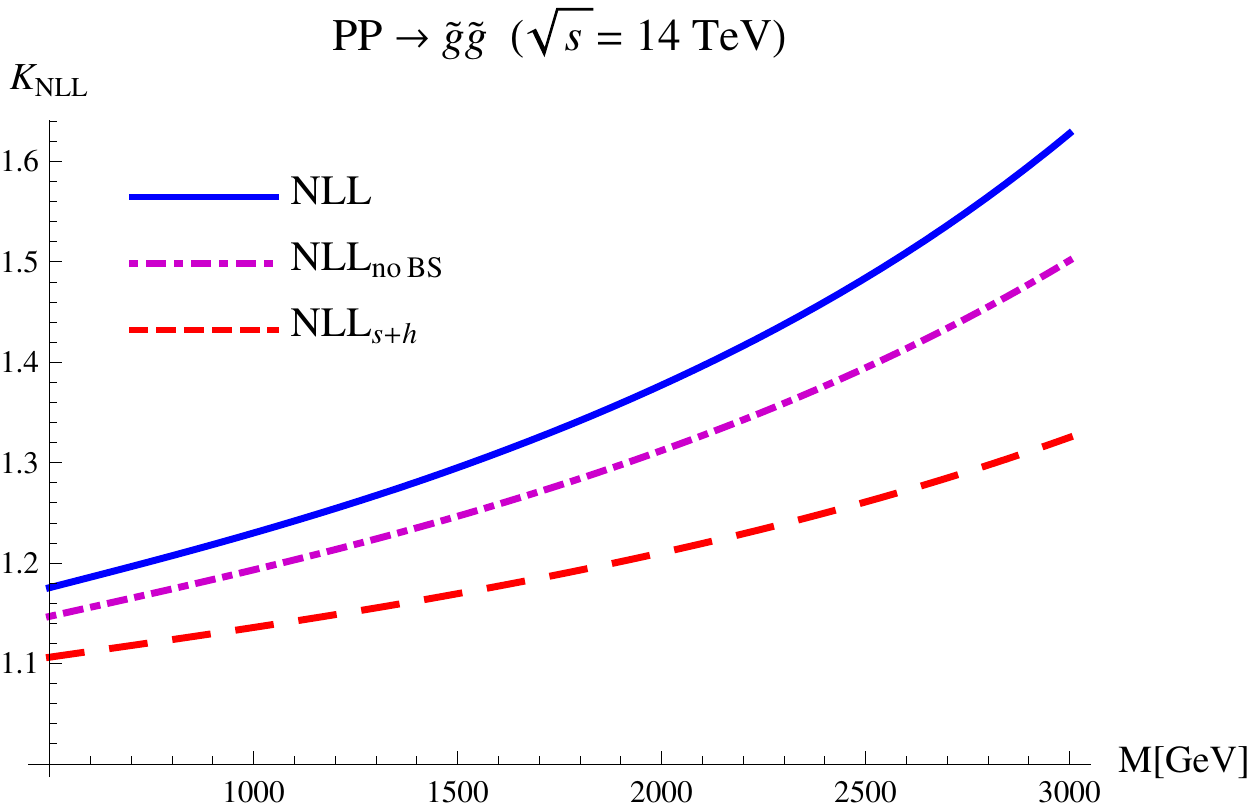}
  \end{center}
\vspace*{-0.2cm}
\caption{
\label{fig:SUSYres}
NLL K-factor for squark-antisquark (top), squark-squark (middle), 
and gluino-gluino (bottom) production at LHC with 
$\sqrt{s} = 14\,$TeV. The plots show $K_{\rm NLL}$ as a function of
$M=m_{\tilde g} = m_{\tilde q}$ 
 for different NLL approximations: NLL (solid blue), NLL$_{\rm no BS}$ 
(bound-state contributions excluded, dot-dashed purple) and 
NLL$_{\rm s+h}$ (dashed red). 
See the text for explanation. Figure from \cite{Falgari:2012hx}.
}
\end{figure}

Different from top-pair production, the summation of Coulomb effects 
is important for some final states, which becomes apparent in large 
differences between the NLL results of \cite{Beenakker:2009ha} and 
\cite{Beneke:2010da,Falgari:2012hx}. The size of the resummation effects 
can be quantified by 
\begin{equation}
K_{\rm NLL} = \frac{\sigma_{\rm NLL, \,matched}}{\sigma_{\rm NLO}},
\end{equation}
where $\sigma_{\rm NLL, \,matched}$ is the resummed result, properly 
matched to the full NLO calculation. The NLL soft- and Coulomb-resummed 
result is shown as solid line in Figure~\ref{fig:SUSYres}, the one 
without Coulomb summation is the dashed NLL$_{\rm s+h}$ line. 
In squark-antisquark production (upper panel), for large squark masses, 
the resummation effect is more than a factor of four larger in the 
former treatment, which highlights the effect of Coulomb attraction. 
No such effect is observed in the squark-squark production process 
(middle panel), due to a cancellation between the numerically dominant, 
repulsive colour-sextet channel for same-flavour squark production, and 
an attractive colour-triplet channel in different-flavour
squark production \cite{Falgari:2012hx}. Gluino-gluino production lies 
in between these extreme cases, but shows the largest resummation effects 
as already mentioned above. Figure~\ref{fig:SUSYres} also demonstrates 
that bound-state production (or rather, the resonant enhancement in 
and below the nominal threshold region) provides a significant further 
enhancement of the total cross section, given 
by the difference between the dot-dashed and solid lines. 

%%%%%%%%%%%%%%%%%%%%%%%%%%%%%%%%%%%%%%%%%%%%%%%%%%%%%%%%%%%%%%%%%%%%
%%%%%%%%%%%%%%%%%%%%%%%%%%%%%%%%%%%%%%%%%%%%%%%%%%%%%%%%%%%%%%%%%%%%

The main difference between  \cite{Beenakker:2009ha} and 
\cite{Beneke:2010da,Falgari:2012hx} can be traced to the 
soft and Coulomb interference term at NNLO, which is already included 
in NLL soft-Coulomb resummation, but not in NLL soft-gluon resummation alone. 
The difference should therefore 
be reduced, when soft-gluon without Coulomb resummation is extended 
to NNLL as has indeed been confirmed by \cite{Beenakker:2011sf}. 
NNLL results in both frameworks have meanwhile been presented 
\cite{Beneke:2013opa,Beenakker:2014sma} for all final states, and 
an additional result is available for gluino pairs \cite{Pfoh:2013iia}.

\section{Sommerfeld enhancement of dark matter 
pair annihilation}
\label{sec:5}

In this last section we turn to a potentially important non-relativistic 
effect in the pair annihilation of dark matter (DM) particles. While the 
nature and origin of dark matter are still unknown it is intriguing 
that the observed abundance can be explained rather naturally as thermal 
relic of a TeV scale particle with weak interaction strength. The relic 
density is determined by the total annihilation cross section, 
$\langle \sigma v\rangle$, averaged over the velocity distribution of 
the particles. Since the abundance froze out when the temperature of 
the Universe was about $M_\chi/25$, where $M_\chi$ is the DM particle mass, 
the typical velocities $v \sim 0.2$ are non-relativistic. 
When DM particles annihilate in the present Universe, 
potentially revealing themselves in cosmic ray signatures,
the typical velocities are $v \sim 10^{-3}$, and the 
annihilation occurs even deeper in the non-relativistic regime.

For heavy, weakly interacting, 
TeV-scale dark matter the exchange of the electroweak gauge 
(and Higgs) bosons between the slowly moving DM particles generates a 
Yukawa potential, which leads to an enhancement of ladder diagrams. 
Unlike the case of the long-range Coulomb potential due to 
gluon or photon exchange discussed up to now, the enhancement 
does not increase as $\alpha_{\rm EW}/v$ at very small velocities, but 
is cut off at $\alpha_{\rm EW} M_\chi/M_W$ by the mass of the 
mediator particle. Nevertheless, additional particle exchange
is not suppressed by the size of the 
electroweak coupling $\alpha_{\rm EW}$ when the 
DM mass is much larger than the mediator mass. This so-called Sommerfeld 
effect can exceed the lowest-order annihilation cross section by orders 
of magnitude. The Yukawa potential has only a finite number of bound 
states of order $\mathcal{O}(\alpha_{\rm EW} M_\chi/M_W)$. When the 
dark matter mass is varied, resonant enhancements of the DM pair 
scattering wave-functions occur whenever a bound state approaches 
the threshold, which can yield pronounced Sommerfeld enhancements 
stronger than in the Coulomb case.

The relevance of the Sommerfeld effect was first pointed out for 
(wino- or higgsino-like) neutralino DM annihilation  
into two photons~\cite{Hisano:2004ds}, and subsequently for relic-density 
calculations~\cite{Hisano:2006nn}, although it was not until 2008, 
when an anomalous positron excess was measured by PAMELA, that Sommerfeld 
enhanced DM models attracted more attention as a mechanism to boost the 
DM annihilation rates~\cite{ArkaniHamed:2008qn}. We note  
that for heavy dark matter with mass far above the electroweak scale, 
the mass splittings between the states of the DM electroweak 
multiplet are naturally in the GeV or sub-GeV range. At freeze-out 
a multitude of co-annihilation processes, including charged states 
are still active, and Sommerfeld enhancements are generic. This 
occurs in particular in the minimal supersymmetric standard model (MSSM). 
Sommerfeld enhancements in cosmic ray 
signatures and in the thermal relic abundance have therefore been discussed 
extensively for relevant MSSM scenarios 
with neutralino DM~\cite{Hryczuk:2010zi,Drees:2009gt,Hryczuk:2011tq,Hryczuk:2011vi,Fan:2013faa,Cohen:2013ama,Hryczuk:2014hpa}, 
but also for generic multi-state 
dark matter models~\cite{Cirelli:2007xd,Cirelli:2010nh,Finkbeiner:2010sm}. 

In the following we summarize the work of 
\cite{Beneke:2012tg,Hellmann:2013jxa,Beneke:2014gja,Beneke:2014hja}, 
which aims at improving the calculation of the dark-matter 
annihilation cross section and relic abundance by including the Sommerfeld 
radiative corrections in the general MSSM, beyond the previously 
considered wino- and Higgsino-limit, where the neutralino can be 
an arbitrary admixture of the electroweak wino, Higgsino and bino 
states. As in previous sections a non-relativistic effective theory 
(of the MSSM) is constructed to separate 
the short-distance annihilation process from the long-distance 
Sommerfeld effect, which is encoded in the matrix elements of local 
four-fermion operators. The approach is very similar to the 
NRQCD treatment of quarkonium annihilation \cite{Bodwin:1994jh}, except 
that we deal with scattering states of several species of  
particles interacting through the electroweak Yukawa force.

%%%%%%%%%%%%%%%%%%%%%%%%%%%%%%%%%%%%%%%%%%%%%%%%%%%%%%%%%%%%%%%%%%%%%%
\begin{figure}[t]
\vskip0.2cm
\begin{center}
\includegraphics[width=6.5cm]{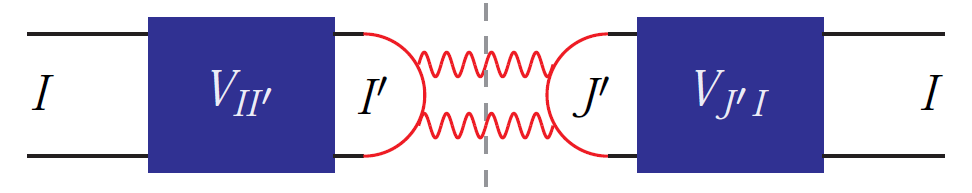}
\end{center}
\vspace*{-0.1cm}
\caption{\label{fig:dmannihilation} Graphical representation 
of the $I\to I$ forward-amplitude. The square stand for the action 
of the potential (the sum of ladder diagram) that scatters 
$I$ into another state $I^\prime$ ($J^\prime$ in the conjugated 
amplitude), which then annihilates into a final state of 
Standard Model particles (and MSSM Higgs bosons, if kinematically 
allowed). [Figure: courtesy C. Hellmann]}
\end{figure}
%%%%%%%%%%%%%%%%%%%%%%%%%%%%%%%%%%%%%%%%%%%%%%%%%%%%%%%%%%%%%%%%%%%%%%

The presence of several particle species $\chi_i$ entails many complications. 
Since electroweak gauge boson exchange may change the 
two-particle state (for instance, scatter a neutralino pair into 
a pair of oppositely charged, nearly degenerate, charginos), 
the short-distance annihilation process is described by a matrix in 
the space of 
two-particle states, which is not diagonal. The off-diagonal terms 
cannot be obtained from the tree-level cross sections computed by 
numerical programs. The non-relativistic expansion of the short-distance 
annihilation cross section is therefore computed analytically rather 
than numerically, which is also necessary to separate S- and P-wave, 
and spin-0 and spin-1 annihilation contributions, which receive different 
Sommerfeld enhancement factors. The large number of final and initial 
states in the MSSM imply that several thousand processes have to 
be computed. The long-distance part of the problem then involves the 
solution of a matrix-valued Schr\"odinger equation in the space 
of $\chi_i\chi_j$ two-particle states. The existence of kinematically 
closed channels leads to numerical instabilities that requires a 
new method to determine the scattering wave functions at the origin 
relevant to the Sommerfeld enhancement. These issues together with a few 
results are summarized below. For many further relevant details, we refer 
to \cite{Beneke:2012tg,Hellmann:2013jxa,Beneke:2014gja,Beneke:2014hja}.

\subsection{Construction of the annihilation rates}

The process we are interested in is shown schematically in 
Figure~\ref{fig:dmannihilation} and is described by the non-relativistic 
MSSM Lagrangian
\begin{equation}
 \mathcal L^{\rm  NRMSSM}
	=
   \mathcal L_{\rm  kin} + \mathcal L_{\rm  pot}
 + \delta\mathcal L_{\rm  ann} + \dots
\,
\end{equation}
$\mathcal L_{\rm  kin}$ contains the bilinear terms in 
the two-component spinor fields $\xi_i$ and  $\psi_j=\eta_j, \zeta_j$ 
that represent the $n_0\le 4$ non-relativistic neutralinos ($\chi^0_i$) 
and $n_+\le 2$ charginos ($\chi^-_j$ and $\chi^+_j$), respectively. 
The entire treatment is restricted to the leading-order terms in 
the velocity expansion in the {\em long-distance} part, hence 
$\mathcal L_{\rm  kin}$ is simply given by
\begin{eqnarray}
\mathcal L_{\rm  kin}&\!\!=\!\!&
 \sum\limits_{i=1}^{n_0}
  \xi^\dagger_i \left( i\partial_t - (m_i - m_{{\rm  LSP}}) + 
\frac{\vec\partial^{\,2}}{2 m_{\mathrm{LSP}} } \right) \xi_i\quad
\nonumber\\
&& \hspace*{-1.35cm}+ 
 \sum_{\psi=\eta,\zeta} \sum\limits_{j=1}^{n_+}
   \psi^\dagger_j \left( i\partial_t - (m_j - m_{{\rm  LSP}}) + 
\frac{\vec\partial^{\,2}}{2 m_{\mathrm{LSP}}} \right) \psi_j.
\qquad
\label{eq:kin}
\end{eqnarray}
The non-relativistic 
energy is measured relative to the mass $m_{{\rm  LSP}}$ of the 
lightest neutralino state. 
To have a consistent power-counting, the mass differences
$(m_i - m_{{\rm  LSP}})$ must formally be considered 
of order $m_{\rm LSP}v^2$. 

The short-distance annihilation of the chargino and neutralino pairs
into SM and light Higgs final states is reproduced in the effective field
theory by local 
four-fermion operators. The leading-order contributions to 
$\delta \mathcal L_{\rm ann}$ are given by
dimension-6 four-fermion operators, that describe leading-order S-wave 
neutralino and chargino scattering processes 
$\chi_{e_1}\chi_{e_2}\to \chi_{e_4}\chi_{e_3}$. 
They read
\begin{widetext}
\begin{equation}
\hspace*{3cm}
\delta \mathcal L_{\rm ann}^{d = 6}  = 
\sum\limits_{ \chi \chi \rightarrow \chi \chi}
\sum\limits_{S = 0,1 }~\frac{1}{4}~
f^{ \chi \chi \to \chi \chi }_{ \lbrace e_1 e_2\rbrace \lbrace e_4 e_3\rbrace }
       \left( {}^{2S+1}S_S \right) \
\mathcal O^{\chi \chi \to \chi \chi }_{ \lbrace e_4 e_3\rbrace 
\lbrace e_2 e_1\rbrace }
\left( {}^{2S+1}S_S \right) \,,
\label{eq:deltaL4fermion}
\end{equation}
\end{widetext}
\noindent 
where $f^{ \chi \chi \to \chi \chi }_{ \lbrace e_1 e_2\rbrace \lbrace e_4 e_3\rbrace } \left( {}^{2S+1}L_J \right)$ 
are the corresponding short-distance coefficients.
The explicit form of the dimension-6 S-wave operators with spin $S=0,1$ is
\begin{eqnarray}
\mathcal O^{\chi \chi \to \chi \chi }_{ \lbrace e_4 e_3\rbrace 
\lbrace e_2 e_1\rbrace }\left( {}^{1}S_0 \right)
 \, &= & \
    \chi^\dagger_{e_4} \chi^c_{e_3}
    \ \chi^{c \dagger}_{e_2}\chi^{}_{e_1}
    \ ,
\label{eq:LOSwave_ops2}
\\
\mathcal O^{\chi \chi \to \chi \chi }_{ \lbrace e_4 e_3\rbrace \lbrace e_2 e_1\rbrace }\left( {}^{3}S_1 \right)
 \, &= & \
    \chi^\dagger_{e_4} \sigma^i \chi^c_{e_3}
    \  \chi^{c \dagger}_{e_2} \sigma^i \chi^{}_{e_1}
    \ .
\label{eq:LOSwave_ops}
\end{eqnarray}
The first sum in (\ref{eq:deltaL4fermion}) is taken over all neutralino 
and chargino forward-scattering reactions $\chi_{e_1} \chi_{e_2} \to 
\chi_{e_4} \chi_{e_3}$, including redundant ones where 
the particle species at the first and second and/or third and fourth 
position are interchanged. Thus in the sector of two-particle states 
with electric charge $Q$, $f^{ \chi \chi \to \chi \chi }_{ \lbrace e_1
  e_2\rbrace \lbrace e_4 e_3\rbrace }$ is a square matrix of dimension 
$N_Q = (N_0,N_1,N_2) = (24,16,4)$. In order to reproduce the 
{\em short-distance} tree-level annihilation cross section with 
$\mathcal{O}(v^2)$ accuracy, we add dimension-8 operators with 
two derivatives accounting for P-wave and corrections to S-wave 
annihilation, but also operators proportional to the mass differences 
$\delta m =(m_{e_4} - m_{e_1})/2$, 
$\delta \overline m = (m_{e_3} - m_{e_2})/2$. Of course, the latter have 
non-vanishing coefficients only for the off-diagonal elements of 
forward amplitude.

%%%%%%%%%%%%%%%%%%%%%%%%%%%%%%%%%%%%%%%%%%%%%%%%%%%%%%%%%%%%%%%%%%%%%%
\begin{figure}[t]
\vskip0.2cm
\begin{center}
\includegraphics[width=7.5cm]{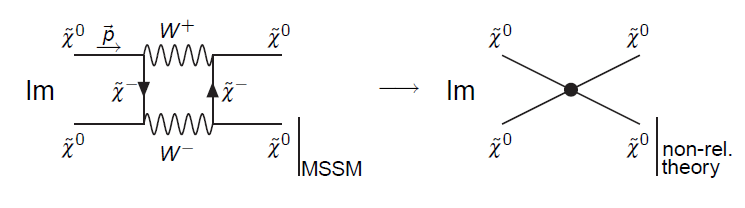}
\end{center}
\vspace*{-0.3cm}
\caption{\label{fig:mssmmatching} 
Matching of the imaginary part of the forward amplitude.
[Figure: courtesy C. Hellmann]}
\end{figure}
%%%%%%%%%%%%%%%%%%%%%%%%%%%%%%%%%%%%%%%%%%%%%%%%%%%%%%%%%%%%%%%%%%%%%%

The short-distance coefficients of all these operators are determined 
by expanding the MSSM amplitudes for the process 
$\chi_{e1}\chi_{e_2} \to X \to \chi_{e_4}\chi_{e_3}$
with SM and light Higgs intermediate states in the relative 
momenta $\vec{p}$, $\vec{p}^{\,\prime}$ and mass differences 
$\delta m$, $\delta \overline m$, and matching them to the tree-level 
matrix element of the four-fermion operators for the same incoming and 
outgoing states to $\mathcal{O}(v^2)$, as illustrated in 
Figure~\ref{fig:mssmmatching}. For the computation of
the neutralino and chargino inclusive annihilation rates, only the 
absorptive part of the short-distance coefficients are required 
according to the optical theorem. At tree-level, it is also possible 
to define the rates to exclusive final states (relevant to indirect 
detection of DM)  in this way. We note that the matching is performed 
for on-shell scattering, which implies that $\vec{p}$ and 
$\vec{p}^{\,\prime}$ are different for the off-diagonal scatterings. 
Energy conservation implies that 
\begin{eqnarray}
\frac{\vec{p}^{\,2}}{2\mu}  &=&  \sqrt{s}-M + 
 ( \,  \delta m +  \delta \overline m \,) + \dots \ ,
\\
\frac{\vec{p}^{\,\prime\,2}}{2\mu}  &=&  \sqrt{s}-M - 
 ( \,  \delta m +  \delta \overline m \,) + \dots,
\end{eqnarray}
where $\sqrt{s}$ is the centre-of mass energy of the scattering, 
$M = (\sum_{i=1}^4 m_{e_i})/2$, $m =(m_{e_1} + m_{e_4})/2$, 
$\overline m = (m_{e_2} + m_{e_3})/2$, and $\mu=m \overline{m}/M$. 
The non-relativistic expansion is strictly valid, when 
$\sqrt{s}-M$, $\delta m$ and $\delta \overline m$ are of order 
$M_\chi v^2$.

The actual calculation is complicated by the presence of many final 
states and interactions in the MSSM. Moreover, for reasons discussed 
in \cite{Beneke:2012tg}, the calculation is most easily done in 
`t~Hooft-Feynman gauge, which, however, adds a large number of unphysical final 
states containing pseudo-Goldstone Higgs and ghost particles. The 
calculation of all S-wave annihilation rates is presented in 
\cite{Beneke:2012tg}, and the coefficients of the $\mathcal{O}(v^2)$ 
terms in \cite{Hellmann:2013jxa}.

%------------------- fig MG5 2 -------------------
\begin{figure}[t]
\vskip0.2cm
\begin{center}
\includegraphics[width=6cm]{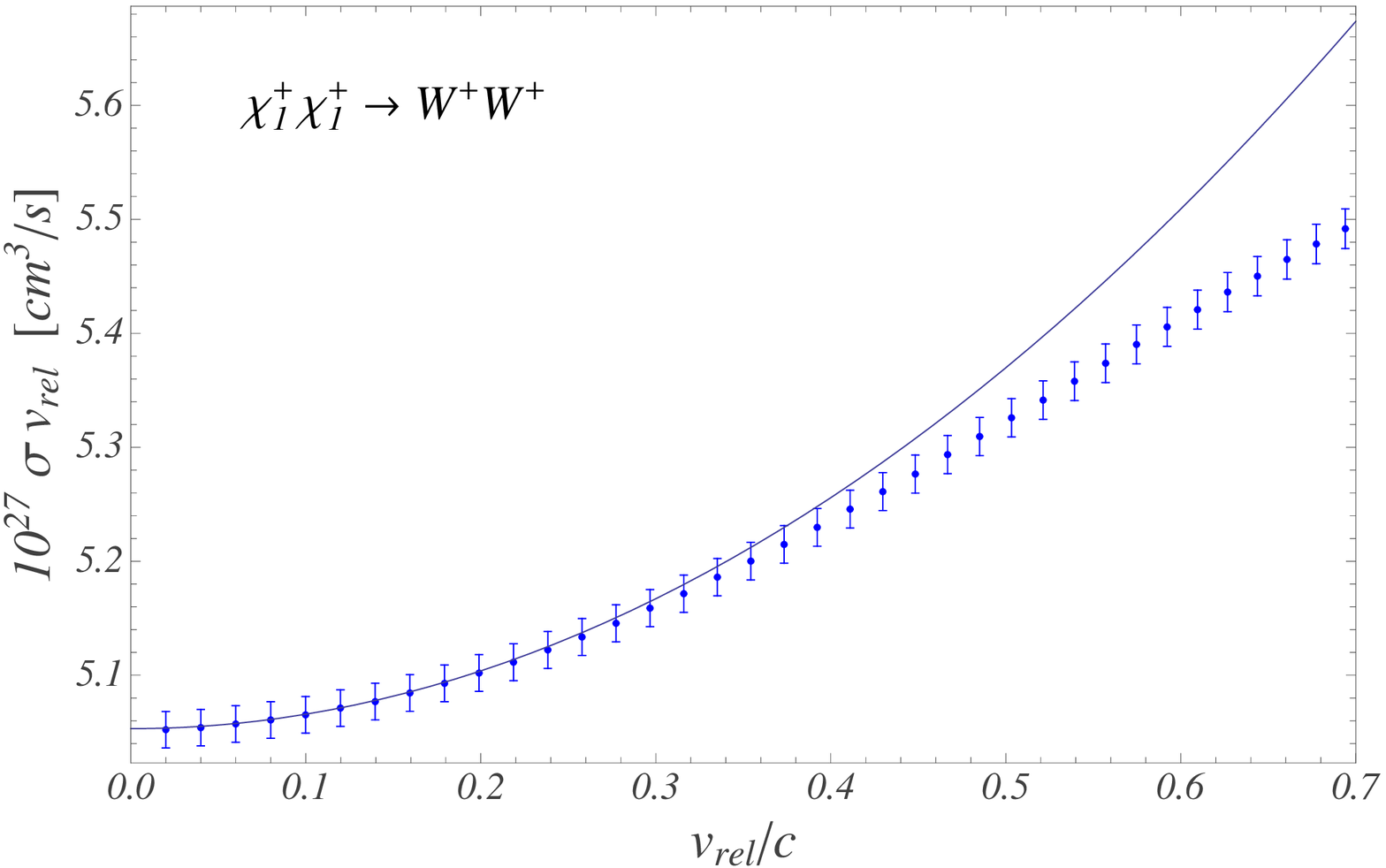} 
\vskip0.2cm
\includegraphics[width=6cm]{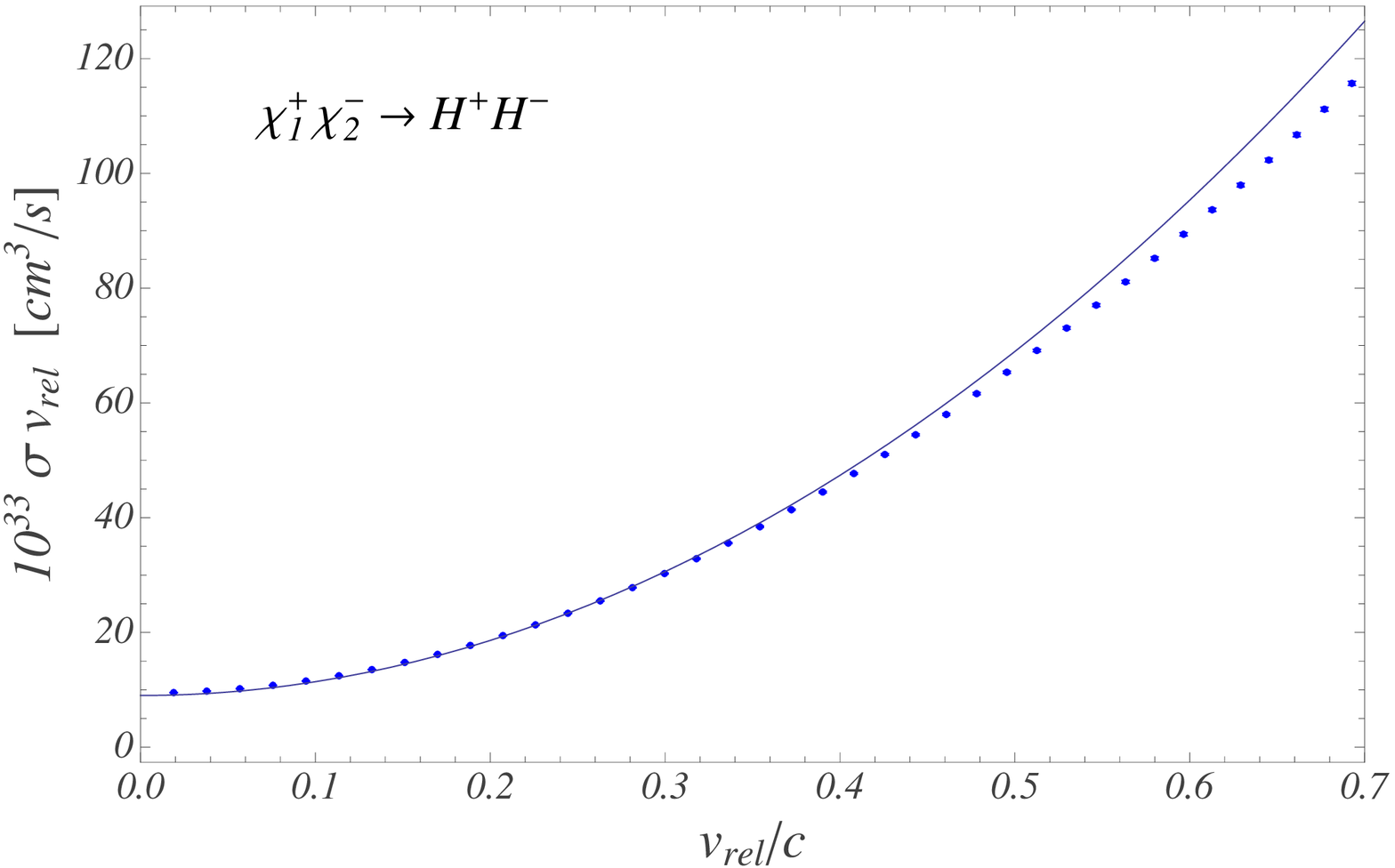}
\end{center}
\caption{Numerical comparison of the non-relativistic approximation 
(solid lines) to the tree-level annihilation cross-section times relative
velocity, $\sigma\, v_{\rm rel}$, for $\chi^+_1 \chi^+_1 \to W^+ W^+$ 
(upper figure) and
$\chi^+_1\chi^-_2 \to H^+ H^-$ (lower figure) with the
corresponding unexpanded annihilation cross section produced with
{\sc MadGraph} \cite{Alwall:2011uj}. The second process is 
dominated by P-wave annihilation.
$v_{\rm rel}$ is given by $v_{\rm rel} = 
\vert \vec v_{e_1} - \vec v_{e_2} \vert$ for
the $\chi_{e_1} \chi_{e_2} \to X_A X_B$ process.
The underlying MSSM spectrum is a wino-like neutralino LSP scenario,
generated with the spectrum calculator {\tt SuSpect} \cite{Djouadi:2002ze}.
The masses of the $\chi^0_1$ and $\chi^+_1$ are given by
$m_{\chi^0_1} = 2748.92\,\mbox{GeV}$, $m_{\chi^+_1} = 2749.13\,\mbox{GeV}$, 
and $m_{\chi^-_2} =  3073.31\,\mbox{GeV}$. The mass of the Higgs particles
$H^\pm$ takes the value $m_{H^\pm} = 167.29\,\mbox{GeV}$. 
Figure from \cite{Beneke:2012tg}.}
\label{fig:xsectionsMG_2}
\end{figure}
%-------------------------------------------------

The non-relativistic expansion to $\mathcal{O}(v^2)$ is usually a 
very good approximation. For the diagonal annihilation rates, 
the result from the above calculation can be compared to the 
unexpanded tree-level cross section. A comparison for two cases is 
shown in Figure~\ref{fig:xsectionsMG_2}. Good agreement is found 
up to relative velocities of about 0.5, which is sufficient for 
all practical purposes. The analytic expressions from 
\cite{Beneke:2012tg,Hellmann:2013jxa} cover all relevant processes, 
when only neutralino and chargino \mbox{(co-)} annihilation processes 
are relevant. MSSM scenarios with sfermion-neutralino co-annihilation 
are presently not covered, neither are resonant 
annihilation models, since in this case the annihilation process is no-longer 
short-distance, and cannot be expanded in the velocity.

\subsection{Computation of the Sommerfeld enhancement}

Once the short-distance coefficients are at hand, the 
Sommerfeld-enhancement is computed from the matrix elements of 
four-fermion operators such as (\ref{eq:LOSwave_ops2}), 
(\ref{eq:LOSwave_ops}). The first step consists in deriving the 
potentials in the expression 
\begin{widetext}
\begin{equation}
\hspace*{1.5cm} \mathcal L_{\rm pot} = 
- \sum\limits_{ \chi \chi \rightarrow \chi \chi}
  \int d^3 \vec{r} ~ V^{\chi\chi \to \chi\chi}_{ \lbrace e_1 e_2\rbrace 
  \lbrace e_4 e_3\rbrace }(r) 
  \, \chi_{e_4}^\dagger (t, \vec{x})   
  \chi_{e_3}^\dagger (t, \vec{x}+\vec{r}\,)  
   \chi_{e_1} (t, \vec{x})   \chi_{e_2} (t, \vec{x}+\vec{r}\,) 
\label{eq:pot}
\end{equation}
%\end{widetext}
from all $\chi_{e_1}\chi_{e_2}\to\chi_{e_4}\chi_{e_3}$ neutral, single-charged
and double-charged scattering reactions through electroweak gauge boson, 
photon and Higgs boson exchange. Assigning mass $m_{\phi}$ to the 
exchanged particle, the potential in coordinate space has the form 
%\begin{widetext}
\begin{equation}
\hspace*{1.5cm} V^{\chi\chi \to \chi\chi}_{ \lbrace e_1 
e_2\rbrace \lbrace e_4 e_3\rbrace }(r) = 
  \left( A_{e_1 e_2 e_4 e_3}\delta_{\alpha_4 \alpha_1}\delta_{\alpha_3\alpha_2}
  +  B_{e_1 e_2 e_4 e_3} \big( \vec{S}^2 \big)_{\alpha_4 \alpha_1,\alpha_3\alpha_2} \right)
  \frac{e^{-m_\phi r}}{r},
\label{eq:potgeneric}
\end{equation}
\end{widetext}
\noindent which contains spin-independent and 
spin-dependent terms. The indices $\alpha_i$ are contracted with the 
(unwritten) spin indices of the field operators in (\ref{eq:pot}), and the 
total spin operator $\vec{S}$ is 
$\vec{S}_{\alpha_4 \alpha_1,\alpha_3\alpha_2}=
 \vec{\sigma}_{\alpha_4\alpha_1}/2\,\delta_{\alpha_3\alpha_2} 
+\delta_{\alpha_4 \alpha_1}\vec{\sigma}_{\alpha_3\alpha_1}/2$. Since 
the total spin is not changed at this order, the two-particle states 
$\chi_{e_1}\chi_{e_2}$ and $\chi_{e_4}\chi_{e_3}$ undergoing potential 
interactions can be decomposed into ${}^{2S+1}L_J$ partial-wave states 
with defined spin $S=0,1$, and the Sommerfeld factors can be determined 
separately for each spin. The MSSM neutralino-chargino potentials are 
given in \cite{Beneke:2014gja}.

The resummation of ladder diagrams carrying the Sommerfeld enhancement 
is achieved by including the interaction (\ref{eq:pot}) into the unperturbed 
Lagrangian. The task is then to evaluate the matrix elements of the 
annihilation operators in this theory. For the general treatment we 
refer to \cite{Beneke:2014gja} and consider here the example of the 
$^1S_0$ operator (\ref{eq:LOSwave_ops2}). Its matrix element can be 
parametrized in the form 
\begin{widetext}
\begin{eqnarray}
\langle \chi_i \chi_j | \, \mathcal O^{ \chi \chi \rightarrow \chi \chi}_{\lbrace e_4 e_3 \rbrace \lbrace e_2 e_1 \rbrace}(^{1}S_0) \, | \chi_i \chi_j \rangle &=& 
\langle \chi_i \chi_j | \, 
    \chi^\dagger_{e_4} \chi^c_{e_3} |0 \rangle \, \langle 0| \chi^{c \dagger}_{e_2}\chi^{}_{e_1}
\, | \chi_i \chi_j \rangle
\nonumber\\
 &=& \Big[\,\langle \xi^{c \dagger}_{j} \xi_{i} \rangle \, 
        \big( \psi^{(0,0)}_{e_4 e_3,\,ij} \, + \,  \psi^{(0,0)}_{e_3 e_4,\,ij} \big) \Big]^*
    \,\langle \xi^{c \dagger}_{j} \xi_{i} \rangle \,
          \big( \psi^{(0,0)}_{e_1 e_2, \,ij} \, + \,  \psi^{(0,0)}_{e_2 e_1, \,ij} \big)
\ ,
\label{eq:wave}
\end{eqnarray}
\end{widetext}
\noindent where (in general) $\psi^{(L,S)}_{e_1 e_2,\,ij}$ is the 
$\chi_{e_1} \chi_{e_2}$-component of the scattering wave function for 
an incoming $\chi_{i} \chi_{j}$ state with centre-of-mass energy $\sqrt{s}$, 
orbital quantum number $L$ and 
total spin $S$, evaluated for zero relative distance and
normalized to the free scattering solution. In the absence of 
potential interactions, the tree-level matrix element of the 
four-fermion operators is obtained by replacing  
$\psi^{(L,S)}_{e_a e_b,\,ij}\to \delta_{e_a i}\,\delta_{e_b j}$. 
It is convenient to introduce a notation where a single index refers to 
a two-particle state (for instance, 
$e={e_1 e_2}, i={ij}$) rather than two indices for 
the individual particle species that make up this state.
The Sommerfeld factor is defined as the annihilation cross section 
including the potential interaction 
relative to the tree cross section. For a given partial-wave 
contribution to the annihilation cross section, it is defined as 
\begin{equation}
S_{i}[\hat f(^{2S+1}L_J) ] = 
\frac{ 
 \left[ \psi^{(L,S)}_{e^\prime i}\right]^*
     \hat f_{ee^\prime}(^{2S+1}L_J)
    \, \psi^{(L,S)}_{ei}}
{ \hat f_{ii}(^{2S+1}L_J)|_{\rm LO}}.
\label{eq:SFdef}
\end{equation}
Here $\hat{f}_{ee^\prime}$ is related to the absorptive part of 
$f^{\chi\chi \to \chi \chi}_{ee^\prime}$. A sum over all two-particle 
states $e$, $e^\prime$ is implicit in (\ref{eq:SFdef}). 

The annihilation cross section including the Sommerfeld corrections 
is constructed by multiplying each spin and partial-wave component 
by its respective Sommerfeld factor (\ref{eq:SFdef}). The tree-level 
cross section expanded to $\mathcal{O}(v^2)$ 
is recovered by setting $S_{i}[\hat f(^{2S+1}L_J) ]\to 1$. 
Defining $\vec{p}_{i}^{\,2}= 2\mu_i \,(\!\sqrt{s}-M_i)+
\mathcal{O}(\vec{p}_{i}^{\,4})$, the relative momentum squared of the two 
annihilating neutralinos, and $M_i$ and $\mu_i$, the total and 
reduced mass, respectively, of the two-particle system,
the master formula for the cross section takes the form 
\begin{widetext}
\begin{eqnarray}
\sigma^{[\chi\chi]_i \to \,{\rm light}} \, v_{\rm rel} 
& =& \, S_{\!i} [\hat f_h(^{1}S_0)] 
     \; \hat  f^{\chi\chi \to \chi \chi}_{ii}(^{1}S_0)
 + \, S_{\!i}[\hat f_h(^{3}S_1)] 
     \; 3 \,\hat  f^{\chi\chi \to \chi \chi}_{ii}(^{3}S_1)
\nonumber\\
&&  \hspace*{-2cm} + \, \frac{\vec{p}_{i}^{\,2}}{M_{i}^2} \, 
     \Big( \, S_{\!i} [\hat g_\kappa(^{1}S_0)]  \; \hat  g^{\chi\chi \to \chi \chi}_{ii}(^{1}S_0)
          +  S_{\!i}[\hat g_\kappa(^{3}S_1)] \; 3 \, \hat  g^{\chi\chi \to \chi \chi}_{ii}(^{3}S_1) + \,S_{\!i} \Big[\frac{\hat f(^{1}P_1)}{M^2}\Big] \; \hat  f^{\chi\chi \to \chi \chi}_{ii}(^{1}P_1)
\nonumber\\
&&  \hspace*{-1.22cm}
        + \,S_{\!i} \Big[\frac{\hat f({}^3P_{\cal J})}{M^2} \Big] \; \hat f^{\chi\chi \to \chi \chi}_{ii}(^{3}P_{\cal J})
     \Big)\,.
\label{eq:SFenhancedsigma}
\end{eqnarray}
\end{widetext}
\noindent 
The precise definition of the annihilation matrices 
appearing in this equation is given in~\cite{Beneke:2014gja}. 

It is well-known that the matrix elements of four-fermion operators 
can be expressed (at this order in the non-relativistic expansion) 
in terms of wave functions at the origin. In the present case  
$\psi^{(0,S)}_{ei} = [\psi_{E}(0)]_{ei}^*$ with $\psi_E(\vec{r}\,)$ 
the matrix-valued solution of the Schr\"odinger equation 
\begin{equation}
\left(\left[-\frac{\vec{\nabla}^{\,2}}{m_{\rm LSP}} - E\right] \delta^{ab}
+ V^{ab}(r)\right) [\psi_E(\vec{r}\,)]_{bi} =0.
\label{eq:schroedinger2}
\end{equation}
Here $E = \sqrt{s} - 2 m_{\rm LSP}$ and 
\begin{equation}
V^{ab}(r) = \hat{V}^{ab}(r) +\delta^{ab} \,\big[M_a-2 m_{\rm LSP}\big]\,.
\label{eq:Vpot2}
\end{equation}
We also use the velocity variable $v$ defined by 
\begin{equation}
E \equiv m_{\rm LSP} v^2\,.
\end{equation}
The solution method proposed in \cite{Slatyer:2009vg} (and reviewed in 
\cite{Beneke:2014gja} in the present notation) relates the Sommerfeld 
enhancement to the asymptotic behaviour of the radial partial-wave function 
$u(r)$ for certain boundary conditions at $r=0$. For the S-wave case, one 
finds  
\begin{equation}
[\psi_{E}(0)]_{ei}^* = [U^{-1}(r_\infty)]_{ei},
\label{eq:Uinv}
\end{equation} 
where $U$ is the matrix
\begin{equation}
U_{ab}(r_\infty) = e^{i k_a r_\infty}\,\left(
[u^\prime(r_\infty)]_{ab} - i k_a [u(r_\infty)]_{ab}\right),\;
\label{eq:Umatrix}
\end{equation}
and $k_a^2 = m_{\rm LSP} (E+i\epsilon-[M_a-2 m_{\rm LSP}])$ 
for a two-particle state $a$ with mass $M_a$. The Schr\"odinger 
equation must be solved from $r=0$ to a value $r_\infty$ large 
enough such that $U_{ab}(r_\infty)$ is close enough to its asymptotic 
value.

%
%---------------------------------------------------------------------------
\begin{figure}[t]
\begin{center}
\includegraphics[width=6.8cm]{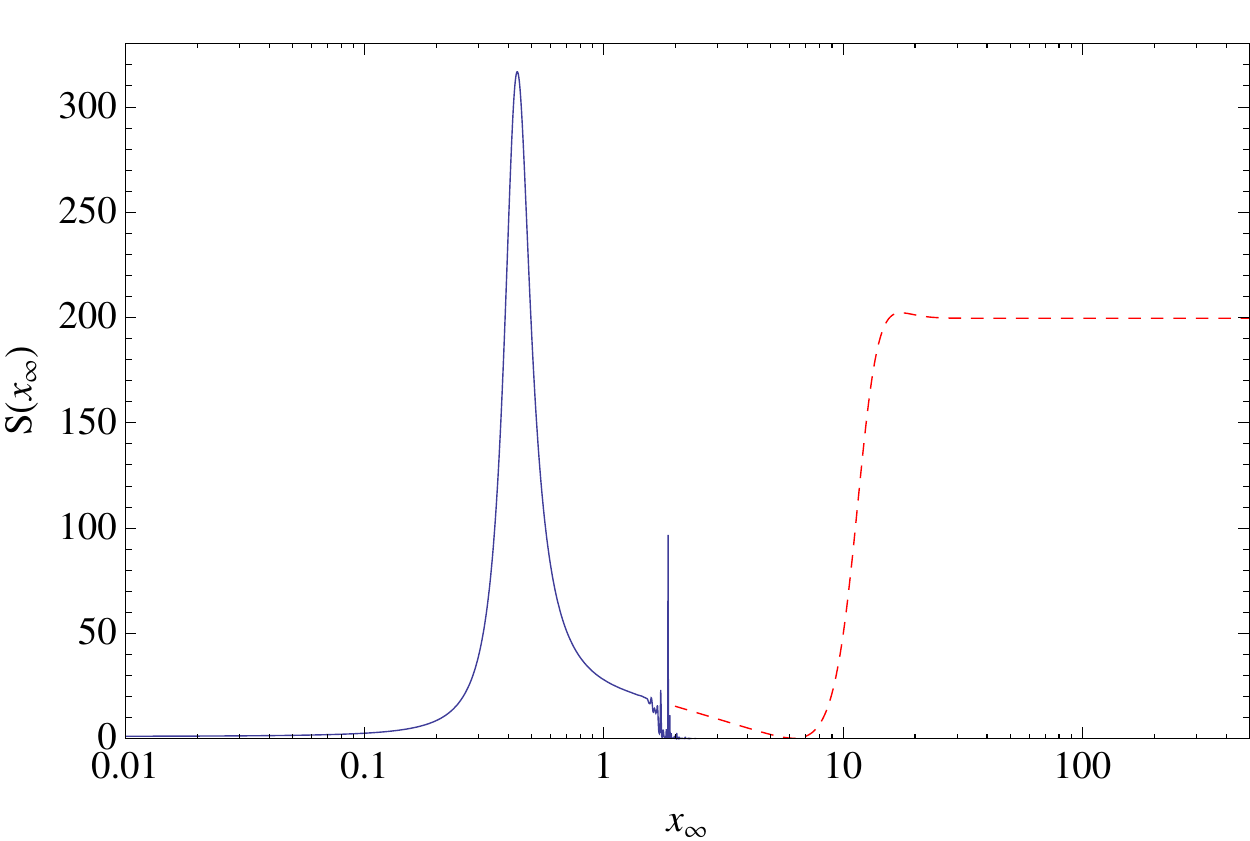}
\caption{S-wave spin-0 Sommerfeld factor for $v=0.012$ in a 
wino-like model as described in the text. 
The light-grey (red) dashed curve shows $S(x_\infty)$, when only 
the $\chi_1^0\chi_1^0$ and $\chi_1^+\chi_1^-$ two-particle 
states in the Schr\"odinger equation and the annihilation 
process are kept and the asymptotic regime is reached for 
$x_\infty>50$. The dark-grey (blue) solid curve shows the result 
when the $\chi_1^0\chi_2^0$ state is included. In this case, the 
evaluation fails for $x_\infty>2$ and no reliable result is 
obtained. Figure from \cite{Beneke:2014gja}.}
\label{fig:Sinfoldmethod}
\end{center}
\end{figure}
%---------------------------------------------------------------------------
%

The procedure described here works well, when {\em all$\,$} $N$ states 
included in the multi-channel Schr\"odinger equation are degenerate to a 
high degree, but fails otherwise. This is illustrated in 
Figure~\ref{fig:Sinfoldmethod}, which shows the Sommerfeld factor 
for the $\chi_1^0\chi_1^0$ $^{1}S_0$ annihilation cross section as 
function of $x_\infty = m_{\rm LSP} v \,r_\infty$ for a MSSM parameter point 
where the lightest neutralino (LSP) 
is wino-like. The relevant masses are $m_{\rm LSP} = 
m_{\chi_1^0} = 2749.4\,$GeV, $m_{\chi_1^+} = 2749.61\,$GeV 
and $m_{\chi_2^0} = 2950.25\,$GeV.  The light-grey (red), 
dashed curve shows $S(x_\infty)$ 
when only the two highly degenerate $\chi_1^0\chi_1^0$ and 
$\chi_1^+\chi_1^-$ two-particle states are included in the Schr\"odinger 
equation and the annihilation process. The velocity is chosen $v=0.012$, 
slightly below the threshold for the $\chi_1^+\chi_1^-$ state. 
After a rapid variation with a peak structure, the Sommerfeld 
factor reaches a plateau and for $x_\infty > 50$ stays at the 
constant value $S(\infty)\approx 199.59$. When the 
$\chi_1^0\chi_2^0$ state is added to the problem, the Schr\"odinger 
system is extended to a $3\times 3$ matrix. Since the new state 
is $200\,$GeV heavier and moreover rather weakly coupled to 
the two lowest, nearly degenerate wino states, it 
should have little effect on the value of the Sommerfeld factor. 
However, now the numerical solution fails when $x_\infty$ is slightly 
larger than 1, as seen from the dark-grey (blue) solid curve in 
Figure~\ref{fig:Sinfoldmethod}, which drops to 0 after a few spikes. 
It is not possible to reach the plateau, where $S(x_\infty)$ 
stabilizes.

The numerical instability originates from the presence of kinematically 
closed two-particle state channels, here the $\chi_1^0\chi_2^0$ state. 
The solution $[u(x)]_{bi}$ for the 
closed channel involves an exponentially growing component 
proportional to $e^{\kappa_b r}$ where $\kappa_b^2 = m_{\rm LSP} 
(M_b - [2 m_{\rm LSP} + m_{\rm LSP} v^2])$. The off-diagonal potentials 
$V^{ab}(r)$ couple the different channels and the 
open-channel solutions  $[u_l(x)]_{ai}$ 
inherit the exponential growth from the closed channels. For the 
two-LSP $\chi_1^0\chi_1^0$ channel, exponential growth occurs 
when at least one of the included kinematically closed channels $b$
satisfies
\begin{equation}  
M_b - [2 m_{\rm LSP} + m_{\rm LSP} v^2] > \frac{M_{\rm EW}^2}{m_{\rm LSP}}
\,.
\end{equation}
Since typically $m_{\rm LSP}\gg M_{\rm EW}$ for the dark-matter scenarios 
of interest, this condition is easily satisfied unless all two-particle 
states included in the computation are degenerate within a 
few GeV or less. In consequence the formally linearly independent 
solutions $[u]_{ai}$ degenerate and the matrix $U_{ai}$ 
becomes ill-conditioned with exponentially growing entries in the 
rows corresponding to open channels $a$. The matrix inversion required for 
(\ref{eq:Uinv}) can no longer be done in practice for $r_\infty$ 
large enough such that the asymptotic regime is reached, 
which causes the instability seen in Figure~\ref{fig:Sinfoldmethod}.

The solution to the problem provided in \cite{Beneke:2014gja} is based on 
an adaptation of the modification of the variable-phase method for 
the Schr\"odinger problem~\cite{Ershov:2011zz}. The idea is to write 
the solution as a linear combination of the linearly independent 
Bessel function solutions $ f_a(x)$, $g_a(x)$ of the free 
Schr\"odinger problem:
\begin{equation}
[u(x)]_{ai} = f_a(x)\alpha_{ai}(x)-g_a(x)\beta_{ai}(x)
\end{equation}
(no sum over $a$). Defining $\tilde\alpha_{ai}=\alpha_{ai}/g_a$, it 
can be shown that the matrix $U$ defined in (\ref{eq:Umatrix}) is 
asymptotically related to $\tilde\alpha_{ai}$ by 
\begin{equation} 
U_{ai}(x) \stackrel{\;x\to\infty\;}{=}
e^{i \hat{k}_a x}\,\tilde{\alpha}_{ai}(x).
\label{eq:Ualpharelation}
\end{equation} 
One then derives a differential equation system for the matrix 
$\tilde{\alpha}^{-1}_{ia}(x)$ from which $[U^{-1}]_{ia}$ follows 
without having to explicitly invert the matrix $U$.

Leaving aside limitations related to the CPU time needed to solve a 
system of many coupled differential equations, this method allows to 
compute the Sommerfeld factors reliably also when many non-degenerate 
two-particle channels are present. This makes the Sommerfeld enhancement 
accessible in a larger part of the MSSM parameter space, away from 
the wino or Higgsino limit, where the Sommerfeld effect is less dramatic 
but potentially still a large radiative correction. To speed up the 
numerical calculation, Ref.~\cite{Beneke:2014gja} also discusses an 
approximation to the treatment of heavier channels, and a relation 
between the $\mathcal{O}(v^2)$ suppressed S-wave operator matrix 
elements and the leading-power ones.

\subsection{Results}

We briefly discuss a selection of results from \cite{Beneke:2014hja} to 
which we refer for further details. 

\subsubsection{Wino-like lightest neutralino}

Wino-like $\chi^0_1$ dark matter arranges into an approximate SU(2)$_L$ fermion
triplet together with the two chargino states $\chi^\pm_1$.
A phenomenological MSSM (pMSSM)
scenario with wino-like $\chi^0_1$ is provided by the SUSY spectrum with
model ID $2392587$ in \cite{Cahill-Rowley:2013gca}. A measure for the 
wino fraction of a given neutralino LSP state is the square of
the modulus of the neutralino mixing-matrix entry $Z_{N\,21}$. For pMSSM scenario
$2392587$ the $\chi^0_1$ constitutes a rather pure wino,
$\vert Z_{N\,21}\vert^2 = 0.999$, with a mass $m_{\rm LSP}\equiv 
m_{\chi^0_1} = 1650.664\,$GeV.
The mass of the chargino partner $\chi^\pm_1$ is given by
$m_{\chi^+_1} = 1650.819\,$GeV, such that $\delta m  = m_{\chi^+_1} - m_{\chi^0_1} = 0.155\,$GeV. For comparison purposes these values are taken 
without any modification from the spectrum card provided
by \cite{Cahill-Rowley:2013gca}, where the mass parameters
refer to the $\overline{{\rm DR}}$-scheme. Eventually, the analysis should 
be done with one-loop renormalized on-shell masses, since the Sommerfeld 
effect is sensitive to the precise value of the mass splitting.

%---------------------------------------------------------------------------
\begin{figure}[t]
\begin{center}
\includegraphics[width=6.5cm]{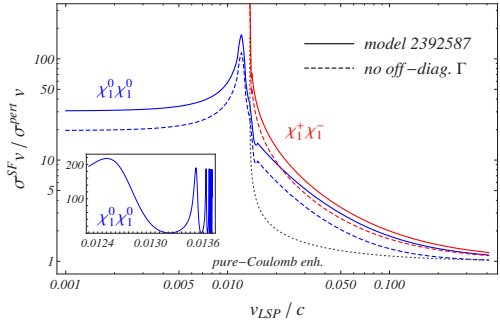}
\vspace*{0.0cm}
\caption{ The enhancement of the $\chi^0_1 \chi^0_1$ and
          $\chi^+_1 \chi^-_1$ annihilation cross sections for Snowmass model
          $2392587$ relative to the perturbative tree-level rate,
          $(\sigma^{\rm SF} v)/ (\sigma^{\rm pert} v)$.
          The solid lines refer to the calculation of the Sommerfeld-enhanced
          rates with off-diagonal entries in the annihilation
          matrices included. The dashed curves show 
the enhancement with respect to the perturbative cross sections when 
off-diagonal annihilation rates are not considered. The dotted curve 
labelled ``pure--Coulomb enh.'' shows the enhancement from photon exchange 
only in the $\chi^+_1\chi^-_1$ channel. Figure from \cite{Beneke:2014hja}.
}
\label{fig:pMSSM_2392587_sigmavoff_Coulomb}
\end{center}
\end{figure}
%---------------------------------------------------------------------------

In Figure~\ref{fig:pMSSM_2392587_sigmavoff_Coulomb} we plot the ratio
$(\sigma^{\rm SF} v)/(\sigma^{\rm pert} v)$ of annihilation rates 
including long-range  interactions, $\sigma^{\rm SF} v$, over the
perturbative tree-level result, $\sigma^{\rm pert} v$, for the two-particle
states $\chi^0_1\chi^0_1$ and $\chi^+_1\chi^-_1$ in the
neutral sector of the model as a function of the velocity
$v_{\rm LSP}$  of the incoming $\chi^0_1$'s in their centre-of-mass frame. 
The enhancement peaks in the vicinity of the threshold of the 
heavier neutral state $\chi^+_1\chi^-_1$ at $v_{\rm LSP}\simeq 0.014$. 
Well below this threshold, the enhancement for the $\chi^0_1\chi^0_1$ system 
is velocity-independent and of $\mathcal O(10)$.

%---------------------------------------------------------------------------
\begin{figure}[t]
\vspace*{1.6cm}
\begin{center}
\vspace*{-1.5cm}
\includegraphics[width=6cm]{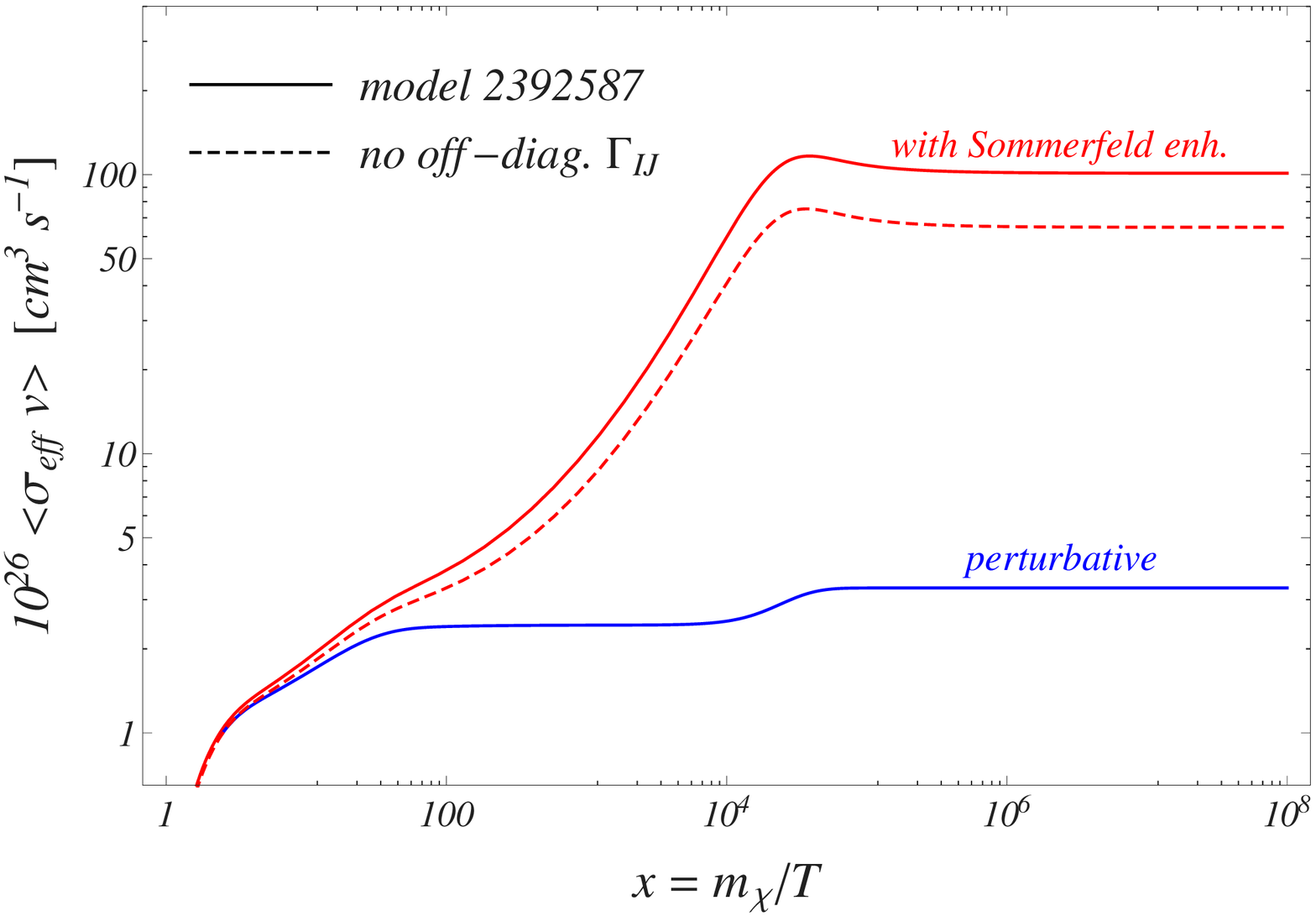}
\includegraphics[width=6cm]{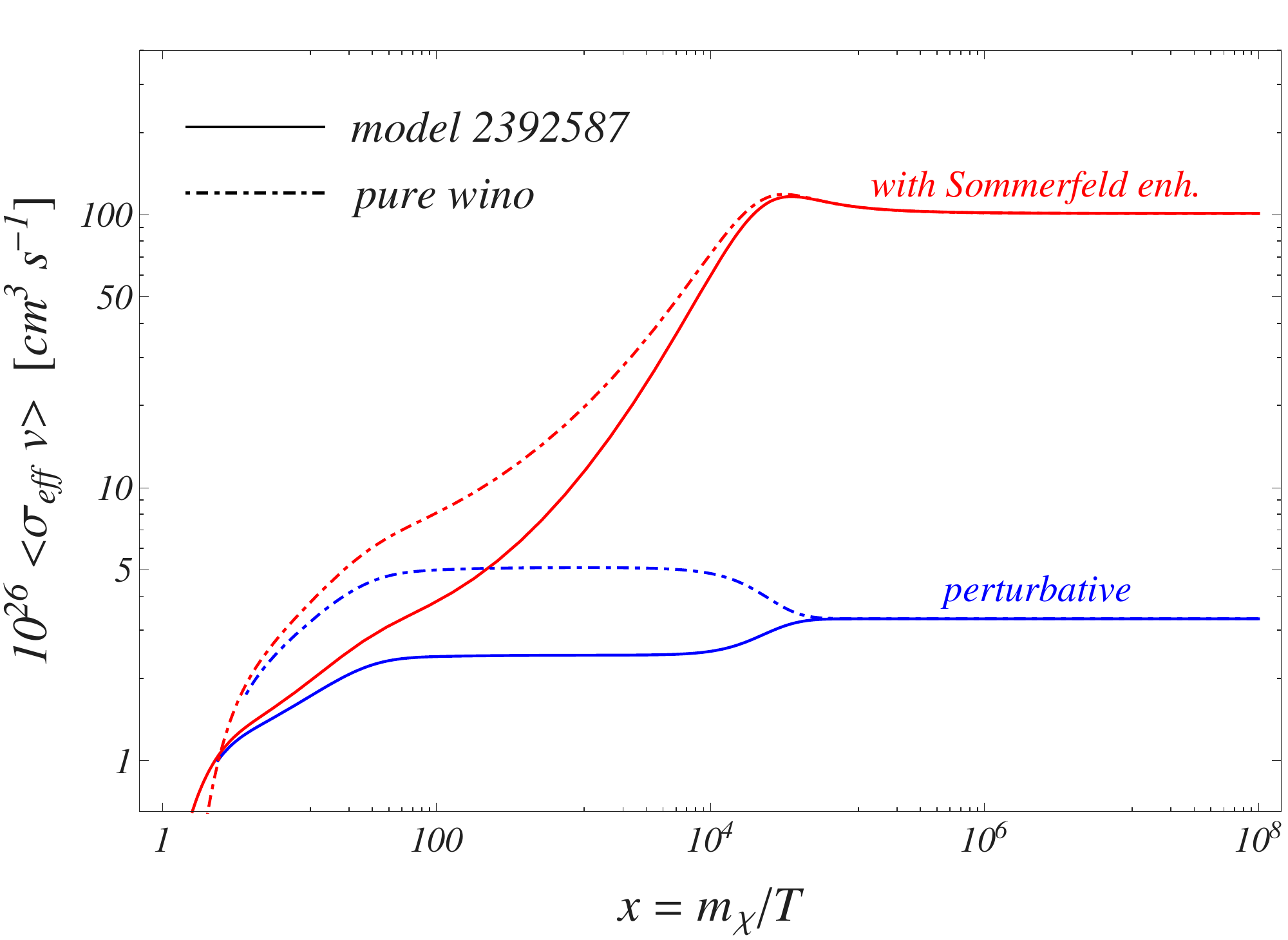}
\end{center}
\caption{ Upper panel: The thermally averaged effective annihilation rate
          $\langle \sigma_{\rm eff} v \rangle$ as a function of the scaled
          inverse temperature $x = m_{\chi}/T$ for Snowmass model
          $2392587$. The two upper (red) curves correspond to the
          Sommerfeld-enhanced annihilation cross sections including
          (solid line) or neglecting (dashed line) the off-diagonals in the annihilation matrices.
          The lower (blue) curve represents $\langle \sigma_{\rm eff} v \rangle$
          obtained from perturbative (tree-level) cross sections.
Lower panel: Comparison of $\langle \sigma_{\rm eff} v \rangle$ for 
model $2392587$ (solid) with the corresponding pure-wino scenario (dot-dashed).
Figures from \cite{Beneke:2014hja}.       }
\label{fig:pMSSM_2392587_sigmaeffoff}
\end{figure}
%---------------------------------------------------------------------------

The quantity that enters the Boltzmann equation for the neutralino 
number density is the thermally averaged
effective annihilation rate summed over all co-annihilation 
channels, $\langle \sigma_{\rm eff} v\rangle$. The upper panel of 
Figure~\ref{fig:pMSSM_2392587_sigmaeffoff} shows 
$\langle \sigma_{\rm eff} v\rangle$ 
as a function of the inverse scaled temperature $x = m_{\chi^0_1}/T$.
As the temperature decreases, the Sommerfeld enhancement increases 
and reaches two orders of magnitude. Around $x\gtrsim10^4$ the number densities
of the $\chi^\pm_1$ are so strongly Boltzmann-suppressed with respect to the
$\chi^0_1$ number density despite the small mass splitting 
that the rates of the charginos basically play no
role in the effective rate $\langle \sigma_{\rm eff} v \rangle$, which is then
essentially given by $\chi^0_1\chi^0_1$ annihilations. After $\chi^\pm_1$ decoupling, $\langle \sigma_{\rm eff} v\rangle$ including the
Sommerfeld enhancements becomes constant, which we can infer from the constant
enhancement factor for the $\chi^0_1\chi^0_1$ system for very low velocities
shown in Figure~\ref{fig:pMSSM_2392587_sigmavoff_Coulomb}. The difference 
between the dashed and solid lines in 
Figure~\ref{fig:pMSSM_2392587_sigmavoff_Coulomb} and the 
upper part of Figure~\ref{fig:pMSSM_2392587_sigmaeffoff} demonstrates 
that it is important to include correctly the off-diagonal annihilation 
reactions.

Since the lightest neutralino is almost a pure wino, the qualitative 
features of model $2392587$ are similar to the pure-wino studied 
already in~\cite{Hisano:2006nn}. There are, however, quantitative 
differences. A pure-wino model contains only an SU$(2)_L$ triplet of 
$\chi$ states in addition to the SM particles, while the MSSM 
model $2392587$ features non-decoupled sfermion states at the $2-3\,$TeV scale
with non-vanishing couplings of the $\chi^0_1$ and $\chi^\pm_1$ to 
sfermions and to the (heavier) Higgs states, which reduce some of the  
annihilation rates relative to the pure-wino dark matter case. The 
difference of the thermally averaged effective annihilation rate for 
the two models, chosen to have the same $\chi_1^0$ mass, is shown in 
the lower panel of Figure~\ref{fig:pMSSM_2392587_sigmaeffoff}. 

\begin{figure}[t]
\begin{center}
\includegraphics[width=7cm]{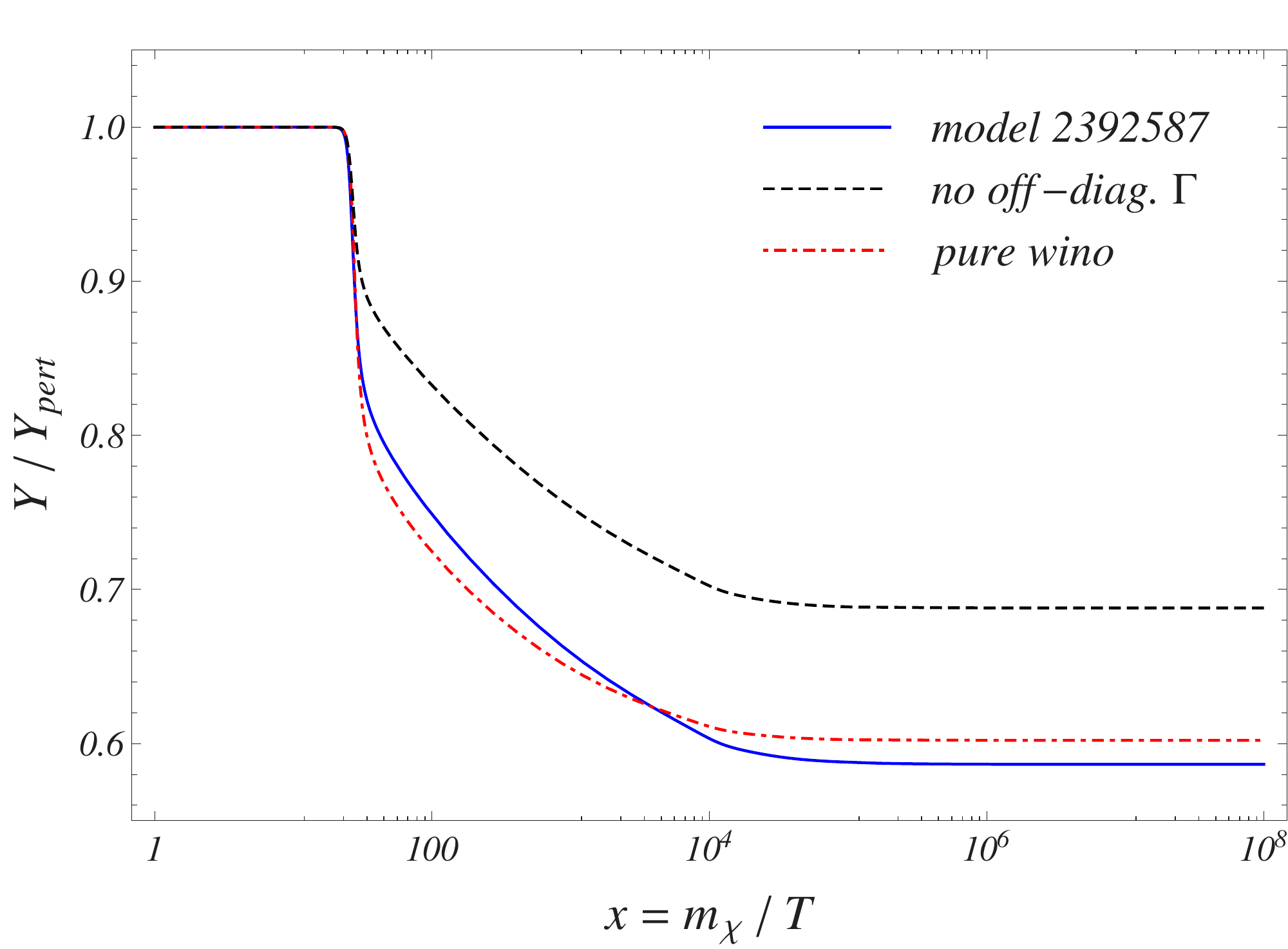}
\caption{ The ratios of the yield $Y/Y_{\rm pert}$ as a function of
          $x = m_{\chi^0_1} / T$, where $Y$ is calculated including the
          Sommerfeld enhancement on the $\chi\chi$ annihilation rates while
          $Y_{\rm pert}$ just uses the perturbative ones.
          The solid (blue) and dashed (black) curves give the results for the
          Snowmass model $2392587$ including and neglecting off-diagonal
          annihilation rates, respectively. The dot-dashed (red) curve
         corresponds to $Y/Y_{\rm pert}(x)$ in the pure-wino model.
 Figure from \cite{Beneke:2014hja}.      }
\label{fig:pMSSM_2392587_yield}
\end{center}
\end{figure}

The $\mathcal{O}(10^2)$ effect seen in 
Figure~\ref{fig:pMSSM_2392587_sigmaeffoff} at large $x$ 
is not relevant to the dark-matter relic density computation, since 
freeze-out occurs already at $x\sim 20$. Nevertheless, the abundance
is significantly modified. It is customary to solve the Boltzmann 
equation that determines the relic density for the 
yield $Y=n/s$, defined as the ratio of the number
density $n$ of all co-annihilating particle species divided by the entropy
density $s$ in the cosmic co-moving frame.
Figure~\ref{fig:pMSSM_2392587_yield} shows the ratio of the yield $Y$
calculated from Sommerfeld-enhanced cross sections in both the pMSSM and the
pure-wino model to the corresponding results using perturbative 
cross sections, $Y_{\rm pert}$, as a function of $x$.
Around $x\sim20$ the yields including Sommerfeld enhancements start to
depart from the corresponding perturbative results.
The enhanced annihilation rates delay the freeze-out of interactions, which
leads to a reduction of the yield $Y$ compared to the perturbative result
$Y_{\rm pert}$. The most drastic reduction in $Y/Y_{\rm pert}$ occurs 
between $x\sim 20$ and $x\sim 10^3$. In this region the
enhancement factors on the cross sections are of $\mathcal O(10)$ (and not yet
$\mathcal O(10^2)$ as for very large $x$). Eventually, 
for $x \gtrsim10^5$ the particle abundances in both the perturbative and 
Sommerfeld-enhanced calculation are frozen in, and the 
fraction $Y/Y_{\rm pert}$ remains constant. In case of the wino-like model 
we find that the relic densities calculated from 
the yield today read $\Omega^{\rm pert} h^2 = 0.112$ and
$\Omega^{\rm SF} h^2 = 0.066$. Hence, the Sommerfeld effect leads to a 
reduction of the calculated relic abundance of around $40\%$ in this 
model. Neglecting the off-diagonal annihilations in the calculation 
would underestimate the effect considerably (dashed curve in 
Figure~\ref{fig:pMSSM_2392587_yield}).

%---------------------------------------------------------------------------
%---------------------------------------------------------------------------

\subsubsection{Higgsino-to-wino trajectory}

\begin{figure}[t]
\vspace*{0.2cm}
\begin{center}
\hskip-0.5cm
\includegraphics[width=6cm]{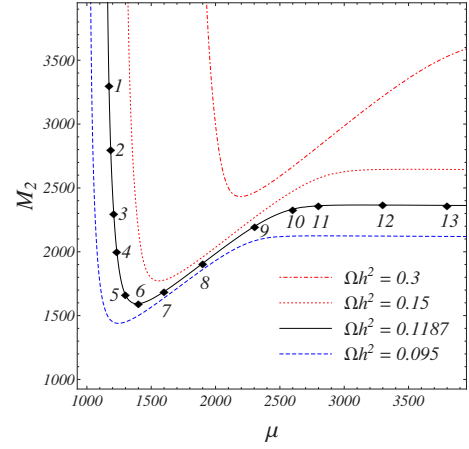}
\caption{ The $\mu-M_2$ plane with the $13$ models defining the higgsino-to-wino
         trajectory, indicated with diamonds.
         All trajectory models lie on the iso-contour for constant
         relic density $\Omega^{\rm DS} h^2 = 0.1187$ calculated with DarkSUSY.
         As reference we also show the iso-contours of constant relic densities
         $\Omega^{\rm DS} h^2 = 0.095$ (lowermost contour-line)
         $0.15$ and $0.3$ (uppermost iso-contour).
 Figure from \cite{Beneke:2014hja}.      }
\label{fig:res_muM2plane}
\end{center}
\end{figure}

The formalism described above allows us to consider lightest neutralino 
states, which are arbitrary admixtures of the electroweak gauginos and 
Higgsinos. Ref.~\cite{Beneke:2014hja} defines a ``trajectory'' in the 
$\mu$ and $M_2$ MSSM parameter space, which interpolates between 
an almost pure-Higgsino ($M_2\gg\mu$) and almost pure wino ($\mu\gg M_2$). 
A large bino fraction is excluded by choosing $M_1=10 M_2$. 
The trajectory is then chosen such that the {\em perturbative} 
relic density computed with 
the program DarkSUSY \cite{Gondolo:2004sc} agrees with the most accurate 
determination obtained
from the combination of PLANCK, WMAP, BAO and high resolution CMB data,
$\Omega_{\rm cdm} h^2 = 0.1187\pm0.0017$. The position of 13 models 
on this trajectory in the $\mu-M_2$ plane is shown in
Figure~\ref{fig:res_muM2plane}. For details on the other MSSM parameters 
and the one-loop mass renormalization scheme, see~\cite{Beneke:2014hja}. 
The models can be categorized as higgsino-like with a wino fraction of 
$\chi^0_1$ below $10\%$ but a higgsino fraction
$\vert Z_{N\, 31}\vert^2 + \vert Z_{N\, 41}\vert^2$ above $0.9$ (models $1-6$),
mixed wino-higgsino $\chi^0_1$ where both the wino and the higgsino fraction
lie within $0.1 - 0.9$ (models $7-9$), or predominantly wino-like $\chi^0_1$
with wino fraction above $0.9$ (models $10-13$).

%---------------------------------------------------------------------------
%---------------------------------------------------------------------------

\begin{figure}[t]
\vspace*{-5.5cm}
\begin{center}
\includegraphics[width=7.8cm]{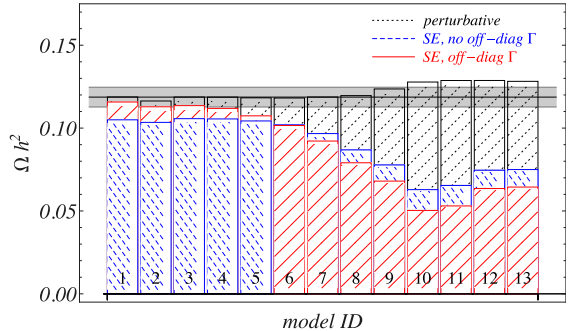}
\caption{ Relic densities $\Omega h^2$ for models $1-13$ on
          the higgsino-to-wino trajectory.
          The charts with dotted (black) hatching are the perturbative results
          $\Omega^{\rm pert} h^2$. Bars with dashed (blue) and solid-line (red)
          hatching refer to a calculation with Sommerfeld-enhanced
          cross sections neglecting and properly including off-diagonal rates,
          respectively.
          The grey shaded band comprises $\Omega h^2$ values within
          $5\%$ around the mean experimental value
          $\Omega_{\rm cdm} h^2 = 0.1187$.
          The latter value is indicated by the black horizontal line and
          agrees with the DarkSUSY result for all $13$ MSSM models on the
          trajectory.
 Figure from \cite{Beneke:2014hja}.      }
\label{fig:trajectory}
\end{center}
\end{figure}

For each of these models the relic density is computed perturbatively 
as well as with the Sommerfeld enhancement included in the non-relativistic 
effective theory approach discussed above. The result of this study is 
shown in Figure~\ref{fig:trajectory}. For the higgsino-like and mixed 
models, the perturbative relic densities $\Omega^{\rm pert} h^2$ 
agree very well with the ones calculated with DarkSUSY for the
same set of input parameters. For the wino-like models a difference of 
up to $8\%$ is observed. The Sommerfeld effect is measured by the height 
of the solid-hatched (red) bars, which give $\Omega^{\rm SF} h^2$, 
relative to the full height, which represents $\Omega^{\rm pert} h^2$. 
The difference between the two is generally 
below 10\% for the most Higgsino-like model $1-5$, in agreement 
with the findings for pure-Higgsino models discussed in the 
context of Minimal Dark Matter \cite{Cirelli:2007xd}. The effect becomes 
more and more significant as the wino fraction increases, reaches 
a maximum for model 10, and then decreases again. For model 10, 
$\Omega^{\rm SF} h^2/\Omega^{\rm pert} h^2 = 0.394$, which implies 
that the relic density is overestimated by a factor 2.5, when the 
Sommerfeld effect is neglected. The occurrence of the maximal effect 
at parameter point 10, which features a lightest neutralino mass 
$m_{\rm LSP} = 2320.986\,$GeV, can be attributed to the above-mentioned 
zero-energy resonance~\cite{Hisano:2004ds} in the Yukawa potential 
of the $\chi^0_1\chi^0_1$ annihilation channel. 

These results demonstrate that it will be necessary to systematically 
include the Sommerfeld effect when MSSM parameter space constraints 
on heavy-neutralino dark matter from direct and indirect searches 
as well as from collider physics are combined with the requirement to 
reproduce, or at least not exceed, the observed abundance of dark matter. 
The formalism and tools developed 
in \cite{Beneke:2012tg,Hellmann:2013jxa,Beneke:2014gja,Beneke:2014hja} 
make it possible to investigate the parameter space of the general
MSSM, and to identify regions where the Sommerfeld
effect is not necessarily as pronounced as in the previously studied wino
limit but still constitutes the dominant radiative correction. 

%---------------------------------------------------------------------------
%---------------------------------------------------------------------------

\section{\label{sec:sum}Summary}

Non-relativistic physics and its effective description play 
an important role in many
different areas of modern particle physics both in the electroweak and in the
strong sector. The field of application reaches from precision determination of
SM parameters, to the calculation of cross sections near thresholds 
for lepton or hadron
collider reactions to high order in perturbation theory, and to the
description of DM annihilation processes in the early Universe.

In this article we reviewed the production of heavy particles close to
threshold within the framework of perturbative QCD. Special emphasis was put on
top-quark pair production both at a future electron-positron collider but also
at hadron colliders like the Tevatron and the LHC.  We explained in detail the
required methods and described the construction of the effective Lagrangians
for the effective theories NRQCD and PNRQCD.

The envisaged experimental accuracy for the measurement of the total cross
section $\sigma(e^+ e^-\to t\bar{t})$ requires perturbative calculations up to
the third order. We discussed several ingredients in detail and described the
construction of the cross section. Due to the broad spectrum of required
techniques, which range from three-loop vertex corrections in full QCD to the
application of non-relativistic perturbation theory to third order including
the corresponding ultrasoft effects, this project, now completed, 
can be considered as a
benchmark calculation in the area of perturbative quantum field theory.

The hadroproduction of top quarks discussed in this review relies to a
large extent on a factorization formula which separates hard and soft scales
and the contributions from the potential region. It is utilized to perform a
simultaneous resummation of Coulomb effects and soft-gluon radiation, which,
in combination with the fixed-order NNLO result, leads to precise predictions
for the total cross section. The same formalism is also applied to the
pair production of supersymmetric particles.

Pair annihilation of heavy, weakly interacting dark matter particles opens 
a new and fascinating area of non-relativistic physics. Non-relativistic 
enhancements of the annihilation rate can be very large despite the fact 
that the force is generated by the exchange of electroweak gauge bosons. 
With TeV scale dark matter particles, degeneracies within the electroweak 
multiplet of DM are generic, leading to a complicated multi-channel 
Schr\"odinger problem. In this article we described the formulation and 
solution of this problem in the non-relativistic effective field theory 
framework with applications to neutralino dark matter of the minimal 
supersymmetric standard model.

\section*{Acknowledgements}

This review summarizes work performed within and
supported by the DFG Sonderforschungsbereich/Transregio 9 ``Computational
Theoretical Particle Physics''. 
We would like to thank our collaborators who contributed to the success of
this project: 
C.~Anzai, D.~Eiras, 
T.~Ewerth, P.~Falgari, C.~Hellmann, B.~Jantzen, 
Y.~Kiyo, J.H.~K\"uhn, P.~Marquard,
S.~Moch, A.~Penin,
J.~Piclum, M.~Prausa, T.~Rauh,
C.~Reisser, P.~Ruiz-Femenia, C.~Schwinn, K.~Schuller, D.~Seidel, A.Smirnov,
V.~Smirnov, P.~Uwer, N.~Zerf.

%\appendix

\section*{References}

%% References

%\nocite{*}
\bibliographystyle{elsarticle-num}
%\bibliography{c3}

\end{document}